\documentclass{CUPLStd}

\usepackage{graphicx}
\usepackage{epstopdf}
\usepackage{amsmath}
\usepackage{amssymb}
\usepackage{color}
\usepackage{icomma}
\usepackage{hyperref}
\usepackage{mathrsfs}

\def\ttotal{{\cal T}_H}
\def\Npresent{N_{\rm past}}
\def\Ntotal{N_{\rm total}}
\def\Nlarge{N_L}
\def\Nsmall{N_S}
\def\Nfuture{N_{\rm future}}

\def\Nworld{{\cal N}_w}

\def\xday{$x$-day~}

\def\MeasureCore{z}
\def\zUN{\MeasureCore}
\def\zpresent{\MeasureCore_{\rm present}}
\def\ztotal{\MeasureCore_{\rm total}}

\def\POS{\mathrm{pos}}
\def\PRIOR{\mathrm{prior}}

\def\MacrotCore{{\Theta}}
\def\MacrotONE{\MacrotCore_1}
\def\MacrotTWO{\MacrotCore_2}
\def\MacrotK{\MacrotCore_k}
\def\MacrotNM{\MacrotCore_M}

\def\MicrohCore{\mu}
\def\MicrohKONE{\MicrohCore_{k,1}}
\def\MicrohKTWO{\MicrohCore_{k,2}}
\def\MicrohKNUMK{\MicrohCore_{k,m_k}}
\def\MicrohKJ{\MicrohCore_{k,j}}
\def\MicrohXY{\MicrohCore_{x,y}}

\def\Data{D}

\def\ProdCore{\mathscr{P}}
\def\Prod{\ProdCore}

\def\ProbCore{P}
\def\Prob{\ProbCore}
\def\ProbKJ{\ProbCore_{k,j}}
\def\ProbPrior{\ProbCore^{\PRIOR}}
\def\ProbPriorK{\ProbCore_k^{\PRIOR}}
\def\ProbPriorKJ{\ProbCore_{k,j}^{\PRIOR}}
\def\ProbPriorXY{\ProbCore_{x,y}^{\PRIOR}}
\def\ProbPos{\ProbCore^{\POS}}
\def\ProbPosK{\ProbCore_k^{\POS}}
\def\ProbPosKJ{\ProbCore_{k,j}^{\POS}}

\def\ProvCore{p}

\def\ProvKJI{\ProvCore_{k,j;i}}
\def\ProvKJZ{\ProvCore_{k,j;z}}
\def\ProvXYI{\ProvCore_{x,y;i}}
\def\ProvXYZ{\ProvCore_{x,y;z}}
\def\ProvPrior{\ProvCore^{\PRIOR}}
\def\ProvPriorKJI{\ProvCore_{k,j;i}^{\PRIOR}}
\def\ProvPriorKJZ{\ProvCore_{k,j;z}^{\PRIOR}}

\def\ProvPos{\ProvCore^{\POS}}
\def\ProvPosKJI{\ProvCore_{k,j;i}^{\POS}}

\def\OSetCore{\mathcal{O}}
\def\OSet{\OSetCore}
\def\OSetPRIME{\OSet^{\prime}}
\def\OSetK{\OSetCore_{k}}
\def\OSetKJ{\OSetCore_{k,j}}
\def\OSetXY{\OSetCore_{x,y}}

\def\WeightCore{{\Xi}}
\def\Weight{\WeightCore}
\def\WeightI{\WeightCore_i}
\def\WeightZ{\WeightCore_z}
\def\WeightPrior{\WeightCore^{\PRIOR}}
\def\WeightPriorI{\WeightCore^{\PRIOR}_i}
\def\WeightPriorONE{\WeightCore^{\PRIOR}_1}
\def\WeightPriorTWO{\WeightCore^{\PRIOR}_2}
\def\WeightPos{\WeightCore^{\POS}}
\def\WeightPosI{\WeightCore^{\POS}_i}
\def\WeightPosONE{\WeightCore^{\POS}_1}
\def\WeightPosTWO{\WeightCore^{\POS}_2}
\def\WeightKI{\WeightCore_{k;i}}
\def\WeightKJI{\WeightCore_{k,j;i}}

\def\WeightNormCore{{\hat{\Xi}}}
\def\WeightNormKJ{\WeightNormCore_{k,j}}
\def\WeightNormXY{\WeightNormCore_{x,y}}

\def\IndexDistCore{\xi}
\def\IndexDistKJI{\IndexDistCore_{k,j;i}}

\def\HumanSetCore{\mathcal{H}}
\def\HumanSet{\HumanSetCore}
\def\HumanSetA{\HumanSetCore_A}
\def\HumanSetB{\HumanSetCore_B}
\def\HumanSetR{\HumanSetCore_r}
\def\HumanSetONE{\HumanSetCore_1}
\def\HumanSetTWO{\HumanSetCore_2}

\def\Cyclicranknoon{N_{1/2}^{\prime}}
\def\xrank{N_x}

\def\MaxRanks{{\cal R}}

\def\objectiveName{physical}

\def\SBInfty{SB-B$^n$}
\def\SBBCompound{SB-B$^{\prime}$}

\def\xrankText{$x$-rank}

\def\WorldONE{\mathrm{i}}
\def\WorldTWO{\mathrm{ii}}

\def\ga{\gtrsim}

\newcommand{\mean}[1]{\ensuremath{\langle #1 \rangle}}

\newcommand{\obsd}[1]{\overline{#1}}

\def\endash{\text{--}}

\def\apj{\emph{Astrophysical Journal}}
\def\apjl{\emph{Astrophysical Journal Letters}}

\def\araa{\emph{Annual Review of Astronomy and Astrophysics}}
\def\icarus{\emph{Icarus}}
\def\mnras{\emph{Monthly Notices of the Royal Astronomical Society}}
\def\nat{\emph{Nature}}
\def\prd{\emph{Physical Review D}}
\def\prl{\emph{Physical Review Letters}}
\def\qjras{\emph{Quarterly Journal of the Royal Astronomical Society}}
\def\sovast{\emph{Soviet Astronomy}}

\jname{International Journal of Astrobiology}

\begin{document}

\supertitle{Research Paper}

\title[Noonday: Fine-graining as Copernicanism]{The Noonday Argument: Fine-Graining, Indexicals, and the Nature of Copernican Reasoning}

\author[Lacki]{Brian C. Lacki$^{1}$}

\corres{\name{Brian C. Lacki} \email{astrobrianlacki@gmail.com}}

\address{\auadd{1}{Breakthrough Listen, Astronomy Department, University of California, Berkeley, CA, USA}}

\begin{abstract}
Typicality arguments attempt to use the Copernican Principle to draw conclusions about the cosmos and presently unknown conscious beings within it, including extraterrestrial intelligences (ETI). The most notorious is the Doomsday Argument, which purports to constrain humanity's future from its current lifespan alone. These arguments rest on a likelihood calculation that penalizes models in proportion to the number of distinguishable observers. I argue that such reasoning leads to solipsism, the belief that one is the only being in the world, and is therefore unacceptable. Using variants of the ``Sleeping Beauty'' thought experiment as a guide, I present a framework for evaluating observations in a large cosmos: Weighted Fine Graining (WFG). WFG requires the construction of specific models of physical outcomes and observations. Valid typicality arguments then emerge from the combinatorial properties of third-person physical microhypotheses. Indexical (observer-relative) facts do not directly constrain physical theories, but instead weight different provisional evaluations of credence. As indexical knowledge changes, the weights shift. I show that the self-applied Doomsday Argument fails in WFG, even though it can work for an external observer. I argue that the Copernican Principle does not let us apply self-observations to constrain ETIs.
\end{abstract}


\Abbreviations{ETI: extraterrestrial intelligence, FGAI: Fine Graining with Auxiliary Indexicals, SETI: Search for Extraterrestrial Intelligence, SIA: Self-Indication Assumption, SSA: Self-Sampling Assumption, SSSA: Strong Self-Sampling Assumption{; WFG: Weighted Fine-Graining}}

\maketitle

\setlength\parindent{1em} 

\section{Background}
\label{sec:Intro}
What can we learn from the simple fact that we exist where and when we do?  The answer may bear on many profound questions, including the nature and size of the cosmos, the existence and types of extraterrestrial intelligence, and the future of humanity.

Attempts to reason about the cosmos from the fact of our existence have been called anthropic reasoning \citep[e.g.,][]{Leslie96,Bostrom13}. The anthropic principle argues that our existence is expected even if the events that lead up to it are rare \citep{Carter74}. In its weakest form, it simply asserts that humanity's existence is probabilistically likely in a large enough universe \citep{Carter83}. The strongest versions of the anthropic principle have an opposite premise at heart \citep{Bostrom13}: our existence is not a fluke, but somehow necessary in a logical sense \citep{Barrow83}. Anthropic reasoning has been extended to include Copernican principles, which emphasize the typicality of our environment. Weak forms point out our evolution was not a special rupture in the laws of physics, but one possible outcome that can be repeated if given enough ``trials'' in sufficiently many cosmic environments like ours.  Strong forms state that conscious beings (``observers'') like ourselves are common.  

Anthropic reasoning frequently tries to synthesize the anthropic and Copernican principles: we should regard our circumstances as typical of observers like ourselves.  \citet{Bostrom13} has formalized this notion as the Self-Sampling Assumption (SSA). The group of observers considered similar enough to us for Copernican reasoning to be valid is our ``reference class''.  It may be as wide as all possible sentient beings or as narrow as people exactly identical to your current self.  There is no consensus on a single reference class, or indeed whether we might use a multitude \citep[e.g.,][]{Neal06,Garriga08,Bostrom13}, although the more extreme Copernican formulations apply the universal reference class of all observers.  Typicality arguments are often justified in normal experiments to derive conclusions when unusual outcomes are expected given enough ``trials''. In fact, some kind of typicality assumption seems necessary to reason about large cosmologies, where thanks to the anthropic principle there will exist observers like ourselves with certainty even in Universes distinct from ours -- otherwise observations have no power to constrain the nature of the world \citep{Bostrom02,Bousso08}. Typicality is commonly invoked in discussions of cosmology as the ``principle of mediocrity'' \citep{Vilenkin95}.

\Fpagebreak

The seemingly reasonable Copernican statement of the SSA has led to controversy, as it can be applied to constrain the cosmological contexts of as-of-yet unobservable intelligences in the Universe. Few applications of the Copernican principle are more contentious than the Doomsday Argument. In its most popular form as presented by \citet{Gott93}, we are most likely ``typical'' humans and therefore are unlikely to be near humanity's beginning or end \citep[see also][]{Nielsen89}. The Bayesian Doomsday Argument, most strongly defended by \citet{Leslie96}, has a more robust basis: our birthrank (the number of humans before us) is treated as a uniform random variable drawn from the set of all birthranks of the final human population. A larger human population has more ``outcomes'', resulting in a smaller likelihood of ``drawing'' your specific birthrank and a Bayesian shift favoring a short future for humanity \citep[see also][]{Knobe06,Bostrom13}. A generalized non-ranked variant has been brought to bear to evaluate the existence of beings unlike ourselves in some way, as in the Cosmic Doomsday argument \citep{Olum03}.

The Doomsday Argument and similar typicality arguments would have profound implications for many fields. For example, in the Search for Extraterrestrial Intelligences \citep[SETI][]{Tarter01} the prevalence of technological societies is critically dependent on the lifespan of societies like our own \citep[e.g.,][]{Bracewell60,Sagan73,Forgan11}. Doomsday would be an extraordinarily powerful argument against prevalent interstellar travel, much less more exotic possibilities of astronomical-scale ``megastructures'' \citep{Dyson60,Kardashev64}. The instantaneous population of a galaxy-spanning society could be $\sim 10^{20}$ \citep{Gott93,Olum03} while predictions of intergalactic travel and astro-engineering suggest populations greater than $10^{50}$ \citep{Bostrom03-Pop,Cirkovic04}. Yet the Doomsday Argument applies huge Bayes Factors against the viability of these possibilities or indeed any long future, essentially closing off the entire field of SETI as difficult to futile \citep{Gott93}. Indeed, this is a straightforward implication of the Cosmic Doomsday argument, which is more well known in cosmology \citep{Olum03}. Similar Bayesian shifts might drastically cut across theories in other fields.

This sheer power cannot be stressed enough. These likelihood ratios for broad classes of theories ($>10^{40}$ in some cases!) imply more than mere improbability, they are far more powerful than those resulting from normal scientific observation. If we have any realistic uncertainty in such futures, even the slightest possibility of data being hoaxed or mistaken (say $10^{-9}$) results in epistemic closure. If we discovered a galaxy-spanning society, or if we made calculations implying that the majority of observers in the standard cosmology live in realms where the physical constants are different from ours, then the evidence would force us to conclude that scientists are engaged in a diabolical worldwide conspiracy to fake these data. The SSA even can lead to paradoxes where we gain eerie ``retrocausal'' influence over the probabilities of \emph{past} events by prolonging humanity's lifespan \citep{Bostrom01}.

Given the unrealistic confidence of the Doomsday Argument's assertions, it is not surprising that there have been many attempts to cut down its power and either tame or refute typicality \citep[e.g.,][]{Dieks92,Kopf94,Korb98,Monton01,Olum02,Neal06,Bostrom13,BenetreauDupin15,Garisto20}. These include disputing that our self-observation can be compared to a uniformly drawn random sampling \citep{Dieks92,Korb98,Garisto20}, arguing for the use of much narrower reference classes \citep{Neal06,Bostrom13,Friederich17,Cirkovic20}, or rejecting the use of a single Bayesian credence distribution \citep{Srednicki10,BenetreauDupin15}. The most common attack is the Self-Indication Assumption (SIA): if we really are drawn randomly from the set of possible observers, then any given individual is more likely to exist in a world with a larger population. The SIA demands that we adopt a prior in which the credence placed on a hypothesis is directly proportional to the number of observers in it, which is then cut down by the SSA using our self-observation \citep[][see also \citealt{Neal06} for further discussion]{Dieks92,Kopf94,Olum02}. 

The pitfall of these typicality arguments is the Presumptuous Philosopher problem where philosophical arguments giving us ridiculous levels of confidence about otherwise plausible beings in the absence of observations. The problem was first stated for the SIA: the prior posits absurd levels of confidence (say $10^{100}$ to $1$) in models with a very large number of observers (for example, favoring large universe cosmologies over small universe cosmologies; \citealt{Cirkovic01,Bostrom03-SIA}). This confidence is ``corrected'' in the Doomsday Argument, but \emph{a prior} favoring a galactic future $\ga 10^{20}$ to 1 is unlike how we actually reason, and there is no correction when comparing different cosmologies for example \citep{Cirkovic01,Bostrom03-SIA}. But the Doomsday Argument is a Presumptuous argument too, just in the opposite direction -- one develops extreme certainty about far-off locations without ever observing them.

In this paper, I critically examine and deconstruct the Copernican typicality assumptions used in the Doomsday Argument and present a framework for understanding them, Fine Graining with Auxiliary Indexicals (FGAI). 

\section{{Statement of the central problem}}

{It is important to specify the issue at stake -- what is being selected and what we hope to learn. The Doomsday Argument is sometimes presented as a way to prognosticate about \emph{our} future specifically \citep{Gott93,Leslie96,Bostrom13}. This question can become muddled if there are multiple ``copies'' of humanity because the universe is large. Now, if we already know the fraction of societies that are long-lived versus short-lived, the fraction of observers at the ``start'' of any society's history follows from a simple counting argument, with no Doomsday-like argument applying \citep{Knobe06,Garisto20}.

The real problem is that we have nearly no idea what this probability distribution is, aside from it allowing our existence. What typicality arguments actually do is attempt to constrain the probability distribution of observers throughout the universe. This is implicit even in \cite{Gott93}, which argues the negative results of SETI are an expected result, because the Doomsday Argument implies long-lived societies are rare and the vast megastructure-building ETIs are nearly non-existent. \citet{Knobe06} makes this argument explicit with the ``universal Doomsday argument'', suggesting a large fraction of observers are planetbound like us and not in galaxy-spanning societies ETIs. More generally, typicality arguments have been purported to constrain the distribution of other properties of ETIs, like whether they are concentrated around G dwarfs like the Sun instead of the more numerous red dwarfs \citep{HaqqMisra18} or if ETIs in elliptical galaxies can greatly outnumber those in spiral galaxies \citep{Zackrisson16,Whitmire20}. These all are questions about the distribution of observers.

Even questions about \emph{our} future can be recast into a question about lifespan distributions. Specifically, what is the lifespan distribution of societies that are exactly like our own at this particular point in our history? There presumably is only one such distribution, even if there are many instantiations of humanity in the Universe. If our future is deterministic, then this distribution is monovalued, with a single lifespan for all humanities out there including ours. More generally, we can imagine there is some mechanism (e.g., nuclear war or environmental collapse) that could cut short the life of societies like ours, and we try to evaluate the probability that it strikes a random society in our reference class.

Thus, what we are comparing are different theories about this distribution, with two important qualities. First, the theories are mutually exclusive in the \citet{Garisto20} sense -- only one distribution can exist. Even if we suppose there is a multiverse with a panoply of domains, each with their own distribution (e.g., because large-scale travel and expansion is easier under certain physical laws than others), there is presumably only one distribution of these pockets, which in turn gives only one cosmic distribution for the observers themselves. Second, however, there is no ``draw'' by an external observer when we are doing self-observation. All the observers are realized, none more real than the others. This makes it unlike other cases of exclusive selection.

The central problem, then, is whether given mutually exclusive theories about the distribution of observers, should we automatically prefer theories with narrower distributions, such that we are more typical, closer to the bulk? Should we do this even if the broader distributions have more total observers, so that both predict similar numbers of observers like us? Throughout this paper, I use the terms ``Small'' (S) and ``Large'' (L) to broadly group theories about unknown observers \citep[as in][]{Leslie96}. In Small theories, the majority of actually existing observers are similar to us and make similar observations of their environment, while Large theories propose additional, numerically dominant populations very dissimilar to us. The version of the Doomsday Argument that I focus on applies this to the distribution of the total final population (or lifespan). In this context, Small theories are ``Short'', while Large theories are considered ``Long'' (e.g., $\Ntotal \gg 10^{11}$ for human history). 

I illustrate the Argument with some simple toy models. In the prototypical model, I consider only two distributions, both degenerate:
\begin{itemize}
\item In the Small theory, all ``worlds'' have the same small $\Ntotal$ value of $1$.
\item In the Large theory, all ``worlds'' have the same large $\Ntotal$ value, taken to be $2$.
\end{itemize}
Each ``world'' consists of an actually existing sequence of observers, ordered by birthrank, the time of their creation (note that world does not mean theory; there can be many ``worlds'' in this sense but only one correct theory). The observers may be undifferentiated, or may have a distinct label that individuates them. The data each observer can acquire is their own birthrank, and their label if they have one. Numerous variants will be considered, however, including ones where additional data provide additional distinctions in the theory, and ones where $\Ntotal$ is fixed but not the distribution. I also consider models where there may be a mixture of actually existing worlds of different sizes, with or without a distribution known to the observer. To distinguish this variance from Small and Large theories, I denote small and large worlds with lowercase letters (s and $\ell$), reserving uppercase S and L for mutually exclusive theories.

}

\section{The Doomsday Argument and its terrible conclusion}
\label{sec:Doomsday}
The Doomsday Argument is arguably the most far-reaching and contentious of the arguments from typicality. It generates enormous Bayes factors against its Large models, despite relatively plausible (though still very uncertain) routes to Large futures. By comparison, we have no specific theory or forecast implying that {non-artificial} inorganic lifeforms \citep[c.f.,][]{Neal06} or observers living under very different physical constants elsewhere in a landscape dominate the observer population by a factor of $\gg 10^{10}$. Cases where we might test typicality of astrophysical environments, like the habitability of planets around the numerically dominant red dwarfs \citep{HaqqMisra18}, generate relatively tame Bayes factors of $\sim 10 \endash 100$ that seem plausible (a relatively extreme value being $\sim 10^4$ for habitable planets in elliptical galaxies from \citealt{Dayal15,Whitmire20}). For this reason, it is worth considering Doomsday as a stringent test of the ``Copernican Princple''.

\subsection{Noonday, a parable}
\label{sec:Noonday}

\begin{quotation}
A student asked their teacher, how many are yet to be born?

The teacher contemplated, and said: ``If the number yet to be born equaled the number already born, then we would be in the center of history.  Now, remember the principle of Copernicus: as we are not in the center of the Universe, we must not be in such a special time. We must evaluate the $p$-value of a possible final population as the fraction of people who would be closer to the center of history.

``Given that 109 billion humans have been born so far,'' continued the teacher, citing \citet{Kaneda20}, ``The number of humans yet to be born is not between 98 and 119 billion with 95\% confidence.  There may be countless trillions yet to be born, or none at all, but if you truly believe in Copernicus's wisdom, you must be sure that the number remaining is \textit{not} 109 billion.''

And so all who spoke of the future from then on minded the teacher's Noonday Argument.
\end{quotation}

\subsection{\texorpdfstring{Noonday, \xday, and the frequentist Doomsday Argument}{Noonday, \textit{x}-day, and the frequentist Doomsday Argument}}
\label{sec:Frequentist}
\label{sec:IntervalFlaw}
In its popular frequentist form, the Doomsday Argument asks \textit{Wouldn't it be strange if we happened to live at the very beginning or end of history?} Given some measure of history $\zUN$, like humanity's lifespan or population, let $F \in [0, 1]$ be the fraction $\zpresent/\ztotal$ that has passed so far. If we regard $F$ as a uniform random variable, we construct confidence intervals: $F$ is then between $[(1 - p)/2, (1 + p)/2]$ with probability $p$. Thus for a current measure $\zpresent$, our confidence limit on the total measure $\ztotal$ is $[2 \zpresent / p, 2 \zpresent/(1-p)]$, allowing us to estimate the likely future lifespan of humanity \citep{Gott93}. This form does not specifically invoke the SSA: $F$ can parametrize measures like lifespan that have nothing to do with ``observers'' and require no reference class. It is motivated by analogy to similar reasoning applied to external phenomena. Unlike the Bayesian Doomsday Argument, it penalizes models where $F = 1$. {A more proper Doomsday Argument would constrain the $\ztotal$ distribution, from which the expected $F$ distribution can be derived, although the basic idea of finding the $[(1-p)/2, (1+p)/2]$ quantile still applies.}

The parable of the Noonday Argument demonstrates the weakness of the frequentist Doomsday Argument: it is not the only confidence interval we can draw.  The Noonday Argument gives another, one arguably even more motivated by the Copernican notion of us not being in the ``center'' \citep[c.f.,][]{Monton01}. An infinite or zero future is maximally compatible with the Noonday Argument. But there is no reason to stop with Noonday either. Wouldn't it be strange if our $f$ happened to be a simple fraction like $1/3$, or some other mathematically significant quantity like $1/e$, or indeed any random number we pick? 

Doomsday and Noonday are just two members of a broad class of \emph{\xday Arguments}. Given any $x$ in the range $[0, 1]$, the \xday Argument is the observation that it is extremely unlikely that a randomly drawn $F$ will just happen to lie very near $x$. A confidence interval with probability $p$ is constructed by excluding values of $F$ in between $[x - (1-p)/2] \mod 1$ and $[x + (1-p)/2] \mod 1$, wrapping $x$ as in the frequentist Doomsday Argument. The Doomsday Argument is simply the $0$-day Argument and the $1$-day Argument, whereas the Noonday Argument is the \textonehalf-day Argument. 

By construction, all frequentist \xday Arguments are equally valid frequentist statements if $F$ is truly a uniformly distributed random variable. Any notion that being near the beginning, the end, or the center is ``strange'' is just a subjective impression. The confidence intervals are disjoint but their union covers the entire range of possibilities, with an equal density for any $p$ covering any single $\ztotal$ value \citep{Monton01}.

The Noonday Argument demonstrates that not every plausible-sounding Copernican argument is useful, even when technically correct. Confidence intervals merely summarize the effects of likelihood -- which is small for Large models, as reflected by the relative compactness of the 0/1-day intervals. A substantive Doomsday-style Argument therefore requires something more than the probability of being near a ``special'' time.

\subsection{Bayesian Doomsday and Bayesian Noonday}
\label{sec:Bayesian}
\label{sec:BayesianDoomsday}

Bayesian statistics is a model of how our levels of belief, or credences, are treated as probabilities conditionalized by observations.  We have a set of some models we wish to constrain, and we start with some prior credence distribution over them. The choice of prior is subjective, but a useful prior when considering a single positive parameter $\lambda$ of unknown scale is the flat log prior: $\ProbPrior (\lambda) \propto 1/\lambda$. An updated posterior credence distribution is calculated by multiplying prior credences by the likelihood of the observed data $\Data$ in each model $\boldsymbol{\lambda_i}$, which is simply the probability $\Prod(\Data|\boldsymbol{\lambda_i})$ in that model that we observe $y$:
\begin{equation}
\label{eqn:Bayes}
\ProbPos (\boldsymbol{\lambda_j} | \Data) = \displaystyle \frac{\ProbPrior (\boldsymbol{\lambda_j}) \Prod(\Data | \boldsymbol{\lambda_j})}{\sum_{\boldsymbol{\lambda_x} \in \Lambda} \ProbPrior (\boldsymbol{\lambda_x}) \Prod(\Data | \boldsymbol{\lambda_x})},
\end{equation}
where $\Lambda$ is the set of possible $\boldsymbol{\lambda}$.\footnote{Bayes' theorem can be adapted to continuous parameters by replacing $\ProbPrior$ and $\ProbPos$ with probability distributions and the sum in the denominator with an integral over all possible parameter values.} 

Bayesian probability provides a more robust basis for the Doomsday Argument, and an understanding of how it supposedly works. In the Bayesian Doomsday argument, {we seek to constrain the distribution of $\Ntotal$}, the final total population of humanity and its inheritors, using birthranks as the observable. The key assumption is to apply the SSA with all of humanity and its inheritors as our reference class. If we view ourselves as randomly selected {from a society with a final total population of $\Ntotal$}, our birth rank $\Npresent$ is drawn from a uniform distribution over $[0,\Ntotal-1]$, with a likelihood of {
\begin{equation}
\label{eqn:FixedNtotalLikelihood}
\Prod (\Npresent | \Ntotal = n) = \begin{cases}
                                   1/n              & n \ge \Npresent\\
																	 0                & n < \Npresent .
															\end{cases}
\end{equation}
Since we are deciding between different distributions of $\Ntotal$, we must average this over the probability that a randomly selected observer is from a society with $\Ntotal$,
\begin{equation}
\Prod_o (\Ntotal = n) = \frac{n f(\Ntotal = n)}{\sum_{n^{\prime} = 1}^{\infty} n^{\prime} f(\Ntotal = n^{\prime})},
\end{equation}
where $f(\Ntotal = n)$ is the probability that a randomly selected \emph{society} has a final population $\Ntotal$. This gives a likelihood for the distribution $f$ of
\begin{align}
\nonumber \Prob (\Npresent | f) & = \sum_{n = \Npresent}^{\infty} \Prod (\Npresent | \Ntotal = n) \Prod_o (\Ntotal = n) \\
                                & = \frac{F(\Ntotal \ge \Npresent)}{\mean{\Ntotal}} .
\end{align}
Here, $F$ is the complementary cumulative mass function.} Starting from the uninformative flat prior $\ProbPrior(\mean{\Ntotal}) \propto 1/\mean{\Ntotal}$, applying Bayes' theorem results in the posterior:
{
\begin{equation}
\label{eqn:DoomsdayPosterior}
\ProbPos (\mean{\Ntotal}) \propto \frac{F(\Ntotal \ge \Npresent)}{\mean{\Ntotal}^2} .
\end{equation}
}
The posterior in equation~\ref{eqn:DoomsdayPosterior} is strongly biased against Large models, with $\ProbPos(\Ntotal \ge N) \approx \Npresent / N$. 

The parable's trick of excluding $\Ntotal$ values near $2 \Npresent$ is irrelevant here. Bayesian statistics does define credible intervals containing a fraction $p$ of posterior credence and we could draw Noonday-like intervals that include Large models but exclude a narrow range of Small models. But this is simply sleight-of-hand, using well-chosen integration bounds to hide the fundamental issue that $\ProbPos (\Ntotal \gg \Npresent) \ll 1$.

\label{sec:BayesianNoonday}

But could there be a deeper \xday Argument beyond simply choosing different credible intervals?  Perhaps not in our world, but we can construct a thought experiment where there is one.  Define an $x$-ranking as $\xrank = \Npresent - x \Ntotal$; allowed values are in the range of $[-x\Ntotal, (1 - x)\Ntotal]$.  If through some quirk of physics we only knew our $\xrank$ \footnote{\citet{Neal06} briefly discusses a case related to $x = 1$, with a deathrank motivated by a hypothetical asteroid impact.}, we would treat $\xrank$ as a uniform random variable, calculate likelihoods proportional to $1/\Ntotal$ and derive a posterior of $1/\Ntotal^2$ for allowed values of $\Ntotal$ ($\Ntotal \ge -\xrank / x~\text{and}~\xrank < 0 ~\text{or}~ \Ntotal \ge \xrank / (1 - x)~\text{and}~\xrank > 0$). Thus all Bayesian \xday Arguments rule out Large worlds, including the Bayesian Noonday Argument, even though the parable's Noonday Argument implies we should be perfectly fine with an infinite history. This is true even if $x$ is not ``special'' at all: most of the \xday Arguments are perfectly consistent with a location in the beginning, the middle, or the end of history -- anything is better than having $F \approx x$, even if the maximum likelihood is for a ``special'' location. For example, the Bayesian Doomsday Argument, which uses our $0$-ranking, predicts a maximum likelihood for us being at the very end of history. The oddity is also clear if we knew a cyclic Noonday rank $\Cyclicranknoon$ that wraps from the last to the first human: the most likely $\Ntotal$ value would be $\Cyclicranknoon + 1$, a seemingly anti-Copernican conclusion favoring us being adjacent to the center of history.

The Bayesian \xday Arguments provide insight into the heart of the Bayesian Doomsday Argument, which has nothing to do with any particular point in history being ``special''.  What powers all of these arguments is the SSA applied with a broad reference class: the $1/\Ntotal$ factor in the likelihood {of equation~\ref{eqn:FixedNtotalLikelihood}}.  According to these SSA-based arguments, the reason Large worlds are unlikely is because measuring \emph{any} particular value of $\xrank$ whatsoever is more unlikely as $\Ntotal$ increases.  This is a generic property when one actually is randomly drawing from a population; it does not even depend on numerical rankings at all. Ultimately, the result follows from the Bayesian Occam's Razor effect, in which Bayesian probability punishes hypotheses with many possible outcomes \citep{Jefferys91}.  

\subsection{The Presumptuous Solipsist}
\label{sec:Solipism}
The SSA is doing all of the work in the \xday Arguments, but nothing restricts its applications solely to birth rankings or $\xrank$ in general. In fact, the unrestricted SSA argues against Large models of all kinds. It can be used to derive ``constraints'' on all kinds of intelligences. One example is the Cosmic Doomsday argument against the existence of interstellar societies, solely by virtue of their large populations regardless of any birthrank \citep{Olum03}. If thoroughly applied, however, it would lead to shockingly strong evidence on a variety of matters, from cosmology to astrobiology to psychology \citep[as noted by][]{Neal06,Hartle07}. 

The existence of non-human observers is widely debated, and a central question in SETI. The unrestricted SSA rules out large sentient populations that are not humanlike in some way, like having a non-carbon based biology or living in the sea. The Cosmic Doomsday argument notes how ``unlikely'' it is to be in a planetbound society if some ETIs have founded galaxy-wide societies \citep{Olum03}. In fact it more generally obliterates the case for SETI even without the Doomsday Argument, for if many aliens existed in the Universe, what would the probability be of being born a human instead any of the panoply of extraterrestrial species that are out there?\footnote{\citet{Hartle07} argue the unrestricted SSA is absurd because we could rule out aliens on Jupiter this way. But {a proponent could argue that this example counts as a success: our modern evidence is consistent with Jupiter being uninhabited, the majority view now} despite the many astronomers of past centuries who believed other planets were similar to Earth on typicality grounds \citep{Crowe99}. The real problem is that the unrestricted SSA rules against intelligent life everywhere else, even the nearest $10^{100}$ Hubble volumes. Animal consciousness may be a better historical example as we have reason to suspect many animals are indeed conscious in some way \citep[e.g.,][]{Griffin04}.}

The consequences extend to other fields, where the consequences become harder to accept. Several theories of physics predict a universe with infinite extent (in space, time, or otherwise), including the open and flat universes of conventional cosmology, eternal inflation, cyclic cosmology, and the many-world interpretation of quantum mechanics, with a nearly endless panoply of Earth at least slightly different from ours. But according to the unrestricted SSA, the probability that a randomly selected observer would find themselves in \emph{this} version of Earth might as well be zero in a sufficiently big universe. It also makes short work of the question of the question of animal consciousness \citep[c.f.,][]{Griffin04}. Isn't it strange that, of all the creatures on the Earth, you happen to be human?  The SSA would be a strong argument against the typical mammal or bird being a conscious observer, to say nothing of the trillions of other animals and other organisms \citep[][]{Neal06}.

Unrestrained application of the SSA leads to a far more radical, and ominous conclusion, however.  Why not apply the SSA \emph{to other humans}?  We can do this even if we restrict our reference class to humanity and remain agnostic about aliens, multiverses and animal consciousness. Solipsism, the idea that one is the only conscious being in existence and everything else is an illusion, is an age-old speculation. Obviously, most people do not favor solipsism \emph{a priori}, but it cannot actually be disproved, only ignored as untenable. {Solipsism would imply that scientific investigation is pointless, however. Any method that leads to near certainty in solipsism is self-defeating, indicating a flaw in its use.}

What happens when we apply the SSA to our sliver of solipsistic doubts?  Let $\epsilon$ be the credence you assign to all solipsistic ideas that are constrained by SSA. Although surely small, perhaps $10^{-9}$ being reasonable, it should not have to be zero -- if only because cosmology predicts scenarios like solipsism could happen (like being a Boltzmann brain in a tiny collapsing cosmos). What are the odds that you are \emph{you}, according to the SSA? The principle assigns a likelihood of $\le 1/10^{11}$ for a realist worldview; only the normalization factor in Bayes' theorem preserves its viability. But the SSA indicates that, according to solipsism, the likelihood that an observer has your data instead of the ``data'' of one of the hallucinatory ``observers'' you are imagining is 1. According to Bayes' theorem, your posterior credence in solipsism is $\epsilon/[1/\Npresent + \epsilon(1 - 1/\Npresent)]$, or $1 - 1/(\Npresent \epsilon)$. If your prior credence is solipsism was above $\sim 10^{-11}$, the SSA magnifies it into a virtual certainty. In fact, the problem is arguably much worse than that: \citet{Bostrom13} further proposes a Strong Self-Sampling Assumption (SSSA), wherein individual conscious experiences are the fundamental unit of observations.  What are the odds that you happen to make this observation at \emph{this} point of your life instead of any other? Your credence in extreme solipsism, where only your current observer-moment exists, should be amplified by a factor $\ga 10^{17}$, and more if one believes there is good evidence for an interstellar future, animal consciousness, alien intelligences, or the existence of a multiverse. Unless one is unduly prejudiced against it -- Presumptuously invoking the absurdly small prior probabilities that are the problem in the SIA -- the principle behind the Doomsday Argument impels one into believing nobody else in existence is conscious, not even your past and future selves.\footnote{The SSSA does allow there to be multiple copies of you, exactly identical to yourself, since you have no way of telling which of these selves you are, but that is hardly any better.}

One might object that ``me, now'' is a superficial class, that everyone in the real world is just as unique and unrepresentative, so there is no surprise that you are \emph{you} instead of everything else. This is invalid according to unrestricted Copernican reasoning. You are forced to grapple with the fundamental problem that your solipsistic model has only one possible ``outcome'' but realist models have many because there really do exist many different people -- in the same way that drawing a royal flush from a stack of cards on the first try is very strong evidence it is rigged even if every possible draw is equally unlikely with a randomly shuffled deck. Indeed, we can use trivial identifying details to validly make inferences about external phenomena, like in the urn problems used as analogies for the Doomsday Argument. The \xday Argument prevent us from saving the Doomsday Argument by appealing to the unlikelihood of being born near a ``special time'' in the future. {T}here exists an $\xrank$ that is small for each individual in any population.

A later section will provide a way out of solipsism even if one accepts the SSA. This is to fine-grain the solipsism hypothesis by making distinctions about what the sole observer hallucinates, or which of all humans is the ``real'' one.  The distinctions between specific observers are thus important and cannot be ignored. In the context of the Doomsday Argument, we can ask what is the likelihood that the sole solipsist observer imagines themself to have a birthrank of $10^{11}$ instead of $1$ or $10^{50}$. Indeed, I will argue that fine-graining is an important part of understanding the role of typicality: it imposes constraints that limit its use in Doomsday Arguments.

\subsection{The anti-Copernican conclusion of the maximal Copernican Principle}
The Copernican Principle is inherently unstable when adopted uncritically.  Even a small perturbation to one's initial prior, a sliver of doubt about there \emph{really} being 109 billion people born so far, is magnified to the point where it can completely dominate one's views about the existence of other beings.  This in turn leads to epistemic instability, as everything one has learned from the external world is thrown into doubt.  Although the SSA started out as a way of formalizing the Copernican Principle, it has led to what may be considered an anti-Copernican conclusion in spirit. Weaker Copernican principles suggest that you consider yourself one of many possible minds, not considering yourself favored, just as the Earth is one of many planets and not the pivot of the Universe.  But this strong formulation suggests that you are the \emph{only} kind of mind, unique in all of existence.  Instead of a panoply of intelligences, we get at most an endless procession of copies of you and no one else.  Like the anthropic principle, the most extreme versions of the Copernican principle presumptuously tell you that \emph{you} are fundamental.

\section{Deconstructing typicality}

\label{sec:Deconstructing}

\subsection{Why is typicality invoked at all?}
\label{sec:Typicality}
Typicality, the principle behind the SSA, has been invoked to explain how we can conclude anything at all in a large Universe.  Since the Universe appears to be infinite, all possible observations with nonzero probability will almost surely be made by some observers somewhere in the cosmos by the anthropic principle. That is, the mere fact that there exists an observer who makes some observation has likelihood $1$ in every cosmology where it is possible. Furthermore, a wide range of observations is possible in any given world model. Quantum theories grant small but nonzero probabilities for measurements that diverge wildly from the expected value: that a photometer will detect no photons if it is pointed at the Sun, for example, or that every uranium nucleus in the Earth will spontaneously decay within the next ten seconds. Most extreme are Boltzmann brains: any possible observer (for a set of physical laws) with any possible memory can be generated wherever there is a thermal bath. Thus we expect there exist observers who make any possible observation in infinite cosmologies that can sustain cognition. Without an additional principle to evaluate likelihoods, no evidence can ever favor one theory over another and science is impossible \citep{Bostrom02,Bousso08,Srednicki10,Bostrom13}.

The common solution has been to include indexical information in our distributions.  Indexicals are statements relating your first-person experience to the outside world. They are not meaningful for a third-person observer standing outside the world and perceiving its entire physical content and structure. The SSA, and the SIA in reply, attempt to harness indexicals to learn things about the world: they convert the first-person statement into a probabilistic objective statement about the world, by treating you as a ``typical'' observer.  Frequently, the physical distinctions between observers are left unspecified, as if they are intrinsically identical and only their environments are different.   

When we make an observation, we do at least learn the indexical information that we are an observer with our data $\Data$.  The idea of these arguments, then, is to construct a single joint distribution for physical theories about the third-person nature of the cosmos and indexical theories about which observer we are.  Often this is implicit -- rather than having specified theories about which observer we are, indexical information is evaluated with an indexical likelihood, the fraction of observers in our reference class with data $\Data$.  While all possible observations are consistent with a theory in an infinite Universe, most observations will be clustered around a typical value that indicates the true cosmology \citep{Bostrom13}.  Thus the indexical likelihoods result in us favoring models that predict most observers have data similar to ours over those where our observations are a fluke.  

Typicality is usually fine for most actual cosmological observations, but it yields problematic conclusions when attempting to choose between theories with different population sizes.  These problems will motivate the development of Fine-Graining with Attached Indexicals (FGAI) approach to typicality over the next sections.

\subsection{Sleeping Beauty as three different thought experiments}
\label{sec:SleepingBeauty}
\label{sec:SBC}

\begin{figure*}
\includegraphics[width=18cm]{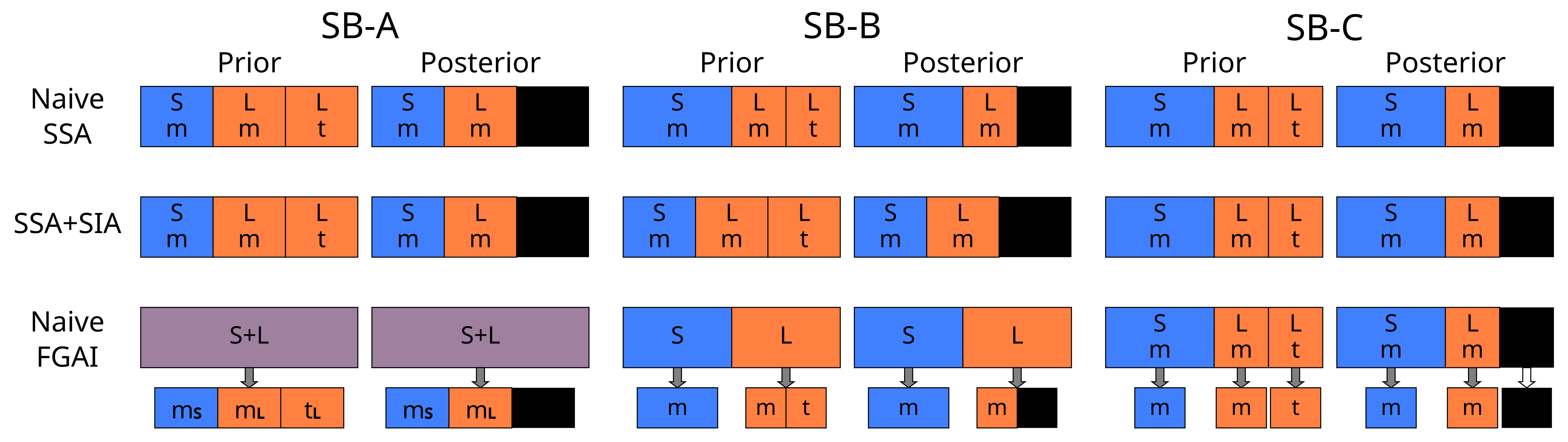}
\caption{The Sleeping Beauty variants SB-A, SB-B, and SB-C, illustrating how different theories of typicality handle Bayesian credence, before and after learning it is Monday ({m}) instead of Tuesday ({t}). Ruled out hypotheses are colored in black and do not count towards the normalization. The SSA, with or without the SIA, leads to presumptuous conclusions in SB-B. In FGAI, indexical and physical distributions are not mixed.  Instead, there is an overarching physical distribution, and each model has an associated indexical probability distribution (indicated by the arrows). \label{fig:SB_ABC}}
\end{figure*}

The core of the Doomsday Argument, and many similar ``Copernican'' arguments, can be modeled with a thought experiment known as the Sleeping Beauty problem. Imagine that you are participating in experiment in which you wake up in a room either on just Monday or on both Monday and Tuesday.  Each day your memory of any previous days has been wiped, so you have no sense of which day it is. Now, suppose you knew that the experiment was either Short, lasting for just Monday, or Long, lasting for Monday and Tuesday. In the original formulation of the thought experiment (\textbf{SB-O}), the experimenters flip a coin, running the Short version if it came up Heads and the Long version if it came up Tails \citep{Elga00}.  You wake up in the room, not knowing how the coin landed, ignorant of whether you are in a Short run or a Long run.  What probability should you assign to the possibility that the coin landed heads and the run is Short, $1/2$ or $1/3$ (Figure~\ref{fig:SB_ABC})?\footnote{To emphasize the potential for unrealistic Bayes factors, consider the case where you are immortal and the Long version lasts for a trillion days. If you use the SIA, could you ever be convinced the experimenters are truthful if they come in and tell you the coin landed on Heads?}

{According to the ``halfer'' camp, i}t is obvious that you have absolutely no basis to choose between Short and Long because the coin is fair, and that you should have an uninformative prior assigning weight $1/2$ to each possibility.\footnote{This paper uses thought experiments where there is a simple binary choice between Small and Large models, for which $1/2$ is the uninformative prior probability. If there are many choices for $\Ntotal$ spanning a wide range of values, a flat log prior in $\Ntotal$ is more appropriate.}

{The ``thirder'' camp instead argues that it is} obvious that you are more likely to wake up on any particular day in the Long experiment.  That is, there are three possibilities -- Short and Monday, Long and Monday, and Long are Tuesday -- so the odds of Short and Monday are $1/3$.  Treating these hypothetical days with equal weight is not as facetious as it sounds: if the experiment was run a vast number of times with random lengths, $1/3$ of all awakenings really would be in Short variations.  One could even construct bets that favor odds of $1/3$.

And what if the experimenters then tell you it is Monday?  In the Short theory, there is a 100\% chance that today is Monday, but in the Long theory, there is just a 50\% chance.  If you initially adopted $1/2$ as the probability in Short, you now favor Short with a credence of $2/3$.  In fact, this the basic principle of the Doomsday Argument.  On the other hand, if you initially adopted $1/3$ as the probability in Long, you now have even credences in Short and Long.  And indeed, if many such experiments were being run, half of the Monday awakenings occur in the Long runs. Still, this leads to the odd situation where we start out fairly confident that we are in a Long experiment. It devolves into the Presumptuous Philosopher problem, forcing us to start out virtually certain that we live in a universe with many inhabitants. Just as in the Doomsday Argument, we are drawing confident conclusions about the Universe without actually looking at anything but ourselves.

The conventional Sleeping Beauty thought experiment has been used to compare the SSA and competing principles \citep{Neal06}, but it conceals some very different situations.  This is why it is not immediately obvious the probability is $1/2$ or $1/3$.  The following versions of the experiment clarify this distinction (see Figure~\ref{fig:SB_ABC}):
\begin{itemize}
\item[] \textbf{(SB-A)} You know that the experiments proceed with a Short run followed by a Long run of the experiment, and you are participating in both.  Today you wake up in the room with no memories of yesterday.  With what probability is today one the day you awaken during the Short run?
\item[] \textbf{(SB-B)} The experimenters have decided, through some unknown deterministic process, to run only either the Short or the Long version of the experiment.  You have absolutely no idea which one they have decided upon. Today you wake up in the room with no memories of yesterday. What credence should you assign to the belief that the experimenters have chosen to do the Short run?
\item[] \textbf{(SB-C)} The experimenters have decided, through some unknown deterministic process, to run only either the Short or a variant of the Long version of the experiment.  In the modified Long experiment, they run the experiment on both Monday and Tuesday, but only wake you on one of the two, chosen by another unknown deterministic process.  Today you wake up in the room.   What credence should you assign to the belief that the experimenters have chosen to do the Short run?
\end{itemize}

In the Doomsday Argument, we essentially are in SB-B: we know that we are ``early'' in the possible history and want to know if we can conclude anything about conscious observers at ``later'' times. Invocations of typicality then presume a similarity between either SB-A or SB-C to SB-B. Yet these analogies are deeply flawed.  Both SB-A and SB-C have obvious uninformative priors yielding the same result with or without the SIA, but they point to different resolutions of the Sleeping Beauty problem: $1/3$ for SB-A and $1/2$ for SB-C. \footnote{\citet{Garisto20} notes the distinction between an ``inclusiverse'' like SB-A where all possibilities exist and an ``exclusiverse'' like SB-B and SB-C where only some do, arguing there is a weighting factor in inclusive selections that changes the prior.} Thus thought experiments lead to ambiguous conclusions (for example, \citealt{Leslie96}'s ``emerald'' thought experiment motivates typicality by noting that we should \emph{a priori} consider it more likely we are in the ``Long'' group in a situation like SB-A because they are more typical, but this could be viewed as support for the SIA).

SB-A and SB-B leave us with \emph{indexical} uncertainty. In SB-A, this is the only uncertainty, with all relevant objective facts of the world known with complete certainty. Because only an indexical is at stake, \emph{there can be no Presumptuous Philosopher problem} in SB-A -- you are \emph{already} absolutely certain of the ``cosmology''. But although it has been reduced to triviality in SB-A, there is actually a second set of credences for the third-person physical facts of this world: our 100\% credence in this cosmology, with both a Short and Long run. SB-B instead posits that you are not just trying to figure out where you are in the world, but the nature of the world in the first place.

SB-C and SB-B leave us with \emph{objective} uncertainty about the physical nature of the world itself.  The objective frequentist probability that you are in the Short run with SB-B is neither $1/2$ nor $1/3$ -- it is either $0$ or $1$. Instead the Bayesian prior is solely an internal one, used by you to weigh the relative merits of different theories of the cosmology of the experiment. Thus, it makes no sense to start out implicitly biased against the Short run, so the probability $1/2$ is more appropriate for a Bayesian distribution. The big difference between SB-B and SB-C is that there are two possible physical outcomes in SB-C's Long variant, but only one in SB-B's.  If we knew whether the experiment was Short or Long, we could predict with 100\% certainty what each day's observer will measure.  But in SB-C,  ``a participant awakened on Monday'' is not a determined outcome in the SB-C Long theory.

\subsection{Separating indexical and physical facts}
\label{sec:Separation}

According to typicality arguments, indexical and \objectiveName{} propositions can be mixed, but in this paper I regard them as fundamentally different. Indexicals are like statements about coordinate systems: they can be centered at any arbitrary location, but that freedom does not fundamentally change the way the Universe objectively works.  It follows that \emph{purely indexical facts cannot directly constrain purely \objectiveName{} world models}.  The Presumptuous Philosopher paradoxes are a result of trying to force indexical data into working like \objectiveName{} data.

Since Bayesian updating does work in SB-A and SB-C, we can suppose there are in fact two types of credence distributions, \objectiveName{} and indexical. In FGAI there is an overarching physical distribution describing our credences in \objectiveName{} world models.  Attached to each \objectiveName{} hypothesis is an indexical distribution (Figure~\ref{fig:SB_ABC}). Each indexical distribution is updated in response to indexical information \citep[c.f.,][]{Srednicki10}. For example, in the Sleeping Beauty thought experiment, when the experimenter announces it is the first day of the experiment, you learn both an \objectiveName{} fact (``a participant wakes up on Monday'') and an indexical fact (``I'm the me waking on Monday''). The \objectiveName{} distribution is insulated from changes in the indexical distribution, protecting it from the extremely small probabilities in both the SSA and SIA. Thus, in SB-B, learning ``today is Monday'' only affects the indexical distribution, invalidating a Doomsday-like argument. Within the context of a particular world-model, one may apply typicality assumptions like the SSA/SIA to the indexical assumption.

\begin{figure}
\centerline{\includegraphics[width=8.6cm]{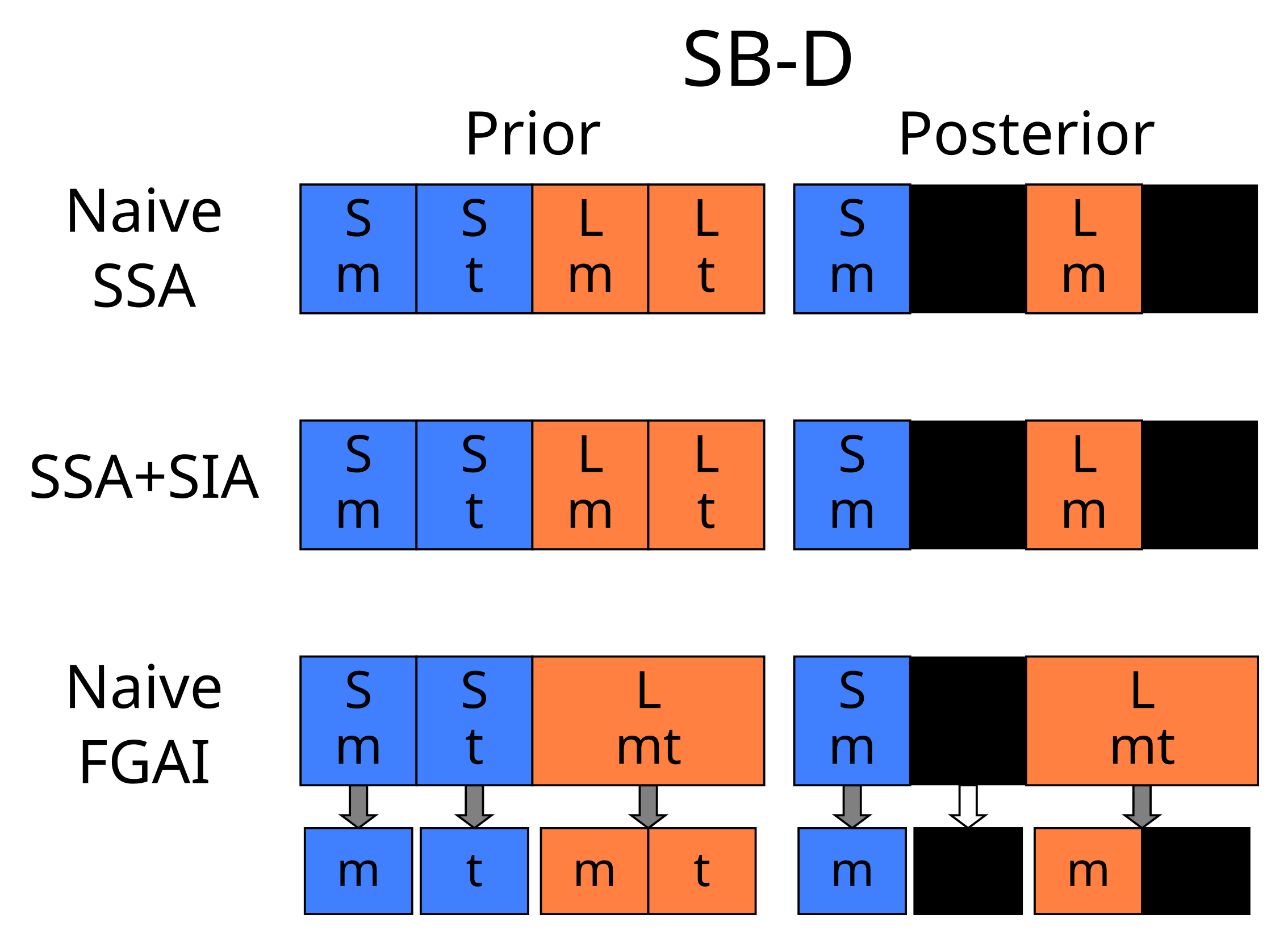}}
\caption{SB-D is a variant of Sleeping Beauty that is challenging for theories with separate indexical and physical distributions. More outcomes are instantiated in the Long theory than in either Short microhypothesis, thus seemingly favoring the Long theory unless the SSA is applied.\label{fig:SBD}}
\end{figure}

A fourth variant of the Sleeping Beauty thought experiment will complicate this attempted reconciliation:
\begin{itemize}
\item[] \textbf{(SB-D)} The experimenters have decided, through some unknown deterministic process, to run only either a modified Short or the standard (SB-B) Long version of the experiment. This version of the Short experiment is identical to the Long variant in SB-C: the experimenters wake you up on one of Monday or Tuesday, chosen through an unknown process. Today you wake up in the room.   What credence should you assign to the belief that the experimenters have chosen to do the Short run?
\end{itemize}
Both the SSA and SIA yield the natural result of equal credences in Short and Long before and after learning it is Monday, while a simple separation of indexical and \objectiveName{}~facts favors Long (Figure~\ref{fig:SBD}). This situation is distinct from the Doomsday Argument because it is not possible for \emph{only} high $\xrank$ humans to exist without the low $\xrank$ ones. An attempt to address this issue will be made in the section on Weighted {Fine Graining (WFG)}.

\subsection{The astrobiological relevance of Sleeping Beauty}
The variants of the Sleeping Beauty thought experiment are models of ``Copernican'' arguments about questions in astrobiology. In all these cases, we are interested in the existence and nature of beings who are unlike us in some way. In these analogies, humanity might be likened to a ``Monday'' observer, and we are considering ``Tuesday'' beings like those in the clouds of Jupiter, around red dwarfs, or in different types of galaxies or different cosmological epochs. With that analogy in mind, consider these hypothetical scenarios in astrobiology:

\begin{itemize}
\item[] \textbf{(AB-A)} We start out already knowing that the Milky Way disk, Milky Way bulge, and M33 disk are inhabited (with M33 lacking a major bulge), and that the number of inhabitants in these three regions is similar. We know we are in one of these three regions, but not which one. Are we more likely to be in the Milky Way or M33? Are we more likely to be in a spiral disk or a bulge? What if we learn we are in a disk?
\item[] \textbf{(AB-B)} We start out knowing that we live in the Milky Way's disk. We come up with two theories: in theory S, intelligence only evolves in galactic disks, but in theory L, intelligence evolves in equal quantities in galactic disks and bulges. Does the Copernican principle let us favor theory S?
\item[] \textbf{(AB-C)} We have two theories: in theory S, intelligence only evolves in galactic disks. In theory L, intelligence either evolves only in galactic disks or only in galactic bulges but not both, with equal credence in either hypothesis. We then discover we live in a galactic disk. Do we favor theory S or theory L?
\item[] \textbf{(AB-D)} We do not know our galactic environment because of our limited observations, but from observations of external galaxies we suspect that galactic disks and bulges are possible habitats for ETIs. Theory S is divided into two hypotheses: in $S_1$, intelligence only evolves in galactic disks and in $S_2$, it evolves only in galactic bulges. Theory L proposes that intelligence is evolves in both disks and bulges. How should we apportion our credence in $S_1$, $S_2$, and L? Once we learn the Earth is in the galactic disk, how do our beliefs change?
\end{itemize}

Put this way, it is clear that neither AB-A nor AB-C are like our current astrobiological questions, and thus neither are SB-A nor SB-C. In AB-A, we already have an answer to the question of whether galactic bulges are inhabited, we are merely uncertain of our address! AB-C, meanwhile, is frankly bizarre: we are seemingly convinced that intelligence cannot evolve in both environments, as if the mere existence of intelligence in galactic bulges always prevents it from evolving in galactic disks. Since we start out knowing there is life on Earth and we wonder if there is life in non-Earthly environments, clearly AB-B -- and SB-B -- is the better model of astrobiology.

AB-D is an interesting case, though, and it too has relevance for astrobiology. In the 18th century, it was not clear whether Earth is in the center of the Milky Way or near its edge, but the existence of ETIs was already a well-known question \citep{Crowe99}. AB-D really did describe our state of knowledge about different regions of the Milky Way being inhabited back then. In other cases, the history is more like AB-B; we discovered red dwarfs and elliptical galaxies relatively late. As the distinction between AB-B and AB-D basically comes down to the order of discoveries in astronomy, it suggests AB-B and AB-D should give similar results in a theory of typicality, with the same true for SB-B and SB-D.

\subsection{The frequentist limit and microhypotheses}
\begin{figure}
\includegraphics[width=8.6cm]{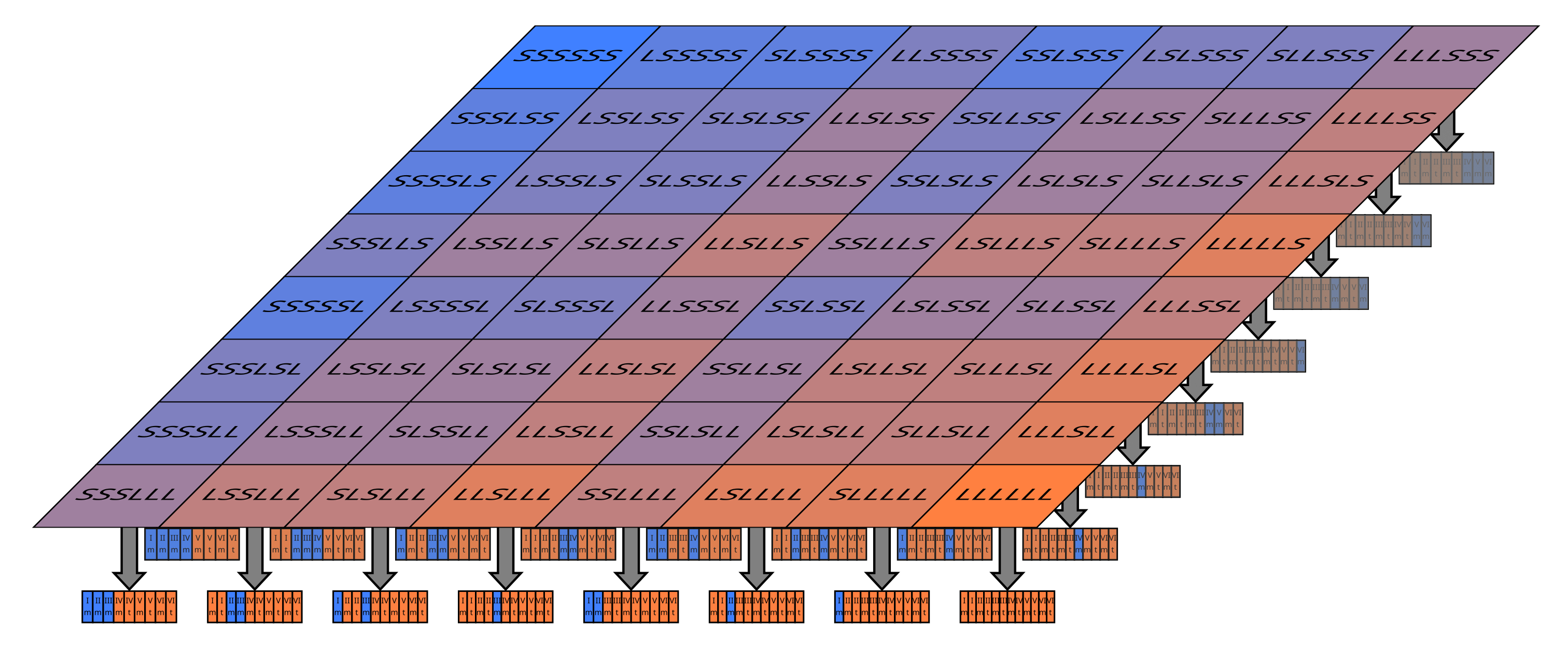}
\caption{If the Sleeping Beauty experiment is run many times, with each possible sequence of Long and Short \emph{a priori} equally likely, a vast number of microhypotheses about the sequence of Short and Long is generated.  Shown here are the 64 microhypotheses when there are $n = 6$ runs.  Attached to every single one of these fine models is an indexical distribution.\label{fig:SB_Many}}
\end{figure}

Finally, it's worth noting that the frequentist $1/3$ probability slips back in for a frequentist variant:
\begin{itemize}
\item[] \textbf{(\SBInfty)} You know for certain that the experiment is being run for $n$ times where $n \gg 1$.  Whether a given run is Short or Long is determined through some deterministic but pseudo-random process, such that any possible sequence of Shorts and Longs is equally credible from your point of view.  Each day, you wake up with an identical psychological state.  Today you wake up in the room.  What credence should you assign to the belief that today is happening during a Short run \citep[c.f., the ``Three Thousand Weeks'' thought experiment of][]{Bostrom07}?
\end{itemize} 
There are now $2^n$ competing \objectiveName{} hypotheses, one for each possible sequence of Shorts and Longs. In most of these hypotheses, however, about half of the runs are Short and Long, and only about one-third of the observers are in the Short runs through simple combinatorics (Figure~\ref{fig:SB_Many}). Thus, as $n$ tends to infinity, the coarse-grained \objectiveName{} distribution converges to one like SB-A, with unbalanced Short/Long runs being a small outlier. We can then say in any likely scenario, the probability we are in a Short run is $\sim 1/3$.

\SBInfty calls to mind statistical mechanics, where a vast number of physical microstates are grouped into a small number of distinguishable macrostates.  By analogy, I call each possible detailed world model a \emph{microhypothesis}, which are then grouped into \emph{macrotheories} defined by statistical properties.

\section{Replacing typicality}
\label{sec:Replacing}

\subsection{The fine-grained approach to typicality}
\label{sec:FGAI_Deterministic}

How are we to make inferences in a large universe, then, without directly mixing indexicals into the credence distribution? I propose that most of the work performed by typicality can instead be performed by \emph{fine-graining}. Fine-graining of \objectiveName{} theories is the first principle of FGAI.

The common practice is to treat observers in a fairly large reference class as interchangeable when discussing their observations, but this is merely a convenience. In fact, we can make fine distinctions between observers -- between Earth and an inhabited planet in Hubble volume \# 239,921, for example, or between me and you, or even between you now and you last Thursday.  Macrotheories often cannot predict exactly which specific observer measures a particular datum. Thus, every theory is resolved into myriads of \emph{microhypotheses}, each of which does make these predictions. Because the distinctions between these observers are \emph{\objectiveName}, statements about a specific observer making a particular measurement are evaluated as purely \objectiveName{} propositions, without invoking indexicals. Some microhypotheses will be consistent with the data, others will not be. The resulting credences in the macrotheories are entirely determined by summing the posterior weight over all microhypotheses, in many cases through simple counting arguments. Typicality then follows from the likelihood values of the microhypotheses -- as it indeed does in conventional probability, where specific ``special'' events like getting a royal flush from a randomly shuffled deck are no more rare than specific mundane outcomes \citep{Laplace1902}. Thus, in most cases, there is no need to invoke any separate Copernican principle, because it is a demonstrable consequence of our theories.

The other main precept of FGAI is that purely indexical facts do not directly affect third-person propositions about the \objectiveName{} world, rather modifying indexical distributions attached to each world model. Observations must be treated as physical third-person events when constraining the \objectiveName{} distribution.  Statements like ``I picked ball 3 out of the urn'' must be recast into third-person statements like ``Brian Lacki picked ball 3 out of the urn''.  Each microhypothesis requires an \emph{observation model}, a list of possible observations that each observer may make. Observation models necessarily impose physical constraints on which observations can be made by whom, forbidding impossible observations like ``Hypatia of Alexandria observed that her peer was Cyborg 550-319447 of Gliese 710'' from being considered as possible outcomes.

The fine-graining is most straightforward when every microhypothesis predicts that all physically indistinguishable experiments lead to the same outcome. This follows when we expect conditionalization to entirely restrict possible observations. For example, the Milky Way could be the product of an indeterministic quantum fluctuation in the Big Bang, but copies of our Earth with its data (e.g., photographs of the Milky Way from inside) do not appear in elliptical galaxies except through inconceivably contrived series of coincidences. More difficult are purely indeterministic cases, when any specific observer can observe any outcome, which is true for most quantum experiments. I will argue that even then we can form microhypotheses by assuming the existence of an appropriate coordinate system.

A more serious difficulty is what to do when different plausible indexical hypotheses would lead to different likelihood evaluations for microhypotheses. That is, we may not know enough about where we are to determine whether a microhypothesis predicts an observation or not. In this section, I will adopt the perspective that I call Naive FGAI: we adopt the maximum possible likelihood over all observers we could be.

\subsection{Naive FGAI}
FGAI constructs the \objectiveName{} probability distributions using the Hierarchical Bayes framework, dividing theories into finer hypotheses about internal parameters, possibly with intermediate levels.  Suppose we have $M$ macrotheories $\MacrotONE$, $\MacrotTWO$, ... $\MacrotNM$.  Each macrotheory $\MacrotK$ has $m_k$ microhypotheses $\MicrohKONE$, $\MicrohKTWO$, ..., $\MicrohKNUMK$. Each microhypothesis inherits some portion of its parent macrotheory's credence or weight. Sometimes, when the microhypotheses correspond to exact configurations resulting from a known probabilistic (e.g., flips of an unfair coin), then the prior credence in each $\ProbPriorKJ$ can be calculated by scaling the macrotheory's total prior probability accordingly. In other cases, we have no reason to favor one microhypothesis over another, and by the Principle of Indifference, we assign each microhypothesis in a macrotheory equal prior probability: $\ProbPriorKJ = \ProbPriorK / m_k$. Some macrotheories are instead naturally split into mesohypotheses describing intermediate-level parameters, which in turn are fine-grained further into microhypotheses. Mesohypotheses are natural when different values of these intermediate-level parameters result in differing numbers of outcomes -- like if a first coin flip determines the number of further coin flips whose results are reported.  Finally, in each $\MicrohKJ$, we might be found  at any of a number of locations. More properly, as observers we follow \emph{trajectories} through time, following a sequence of observations at particular locations, as we change in response to new data \citep[c.f.,][]{Bostrom07}. The set $\OSetKJ (\Data)$ is the set of possible observer-trajectories we could be following allowed by the microhypothesis $\MicrohKJ$ and the data $\Data$.  {In Naive FGAI, t}he set $\OSetKJ$ describes our reference class if $\MicrohKJ$ is true.

In Naive FGAI, prior credences $\ProbPriorKJ$ in $\MicrohKJ$ are updated by data $\Data$ according to:
\begin{equation}
\label{eqn:NaiveFGAI}
\ProbPosKJ = \frac{\displaystyle \ProbPriorKJ \times \max_{i \in \OSetKJ (\Data)} \Prod(o @ i \to \Data | \MicrohKJ)}{\displaystyle \sum_{x = 1}^M \sum_{y = 1}^{m_x} \ProbPriorXY \times \max_{z \in \OSetXY (\Data)} \Prod(o @ z \to \Data | \MicrohXY)},
\end{equation}
where $\Prod(o @ i \to \Data | \MicrohKJ)$ is the likelihood of $\MicrohKJ$ if the observer ($o$) located at position $i$ in $\OSetKJ (\Data)$ observes data $\Data$. Of course if the likelihoods are equal for all observers in the reference class $\OSetKJ (\Data)$, equation~\ref{eqn:NaiveFGAI} reduces to Bayes' formula. The posterior credences in the macrotheories can be found simply as:
\begin{equation}
\label{eqn:NaiveFGAIMacro}
\ProbPosK = \sum_{j = 1}^{m_k} \ProbPosKJ .
\end{equation}
Naive FGAI is sufficient to account for many cases where typicality is invoked. Paradoxes arise when the number of observers itself is in question, as in Doomsday, requiring a more sophisticated treatment.

\subsection{A simple urn experiment}
\label{sec:NaiveFGAI_Urn}
In some cases, microhypotheses and observations models are nearly trivial. Consider the following urn problem: you are drawing a ball from an urn placed before you that contains a well-mixed collection of balls, numbered sequentially starting from $1$.  You know the urn contains either one ball (theory A) or ten (theory B), and start with equal credence in each theory.  How does drawing a ball and observing its number constrain these theories? Both theory A and theory B have microhypotheses of the form ``The urn contains $N$ balls and ball $j$ is drawn at the time of experiment''. In theory A, there is only one microhypothesis, which inherits the full 50\% of Theory A's prior credence.  Theory B has 10 microhypotheses, one for each possible draw and each of equal credence, so its microhypotheses start with 5\% credence each.  Each microhypotheses about drawing ball $j$ has an observation model containing the proposition that you observe exactly ball $j$ -- this observation is a physical event, since you are a physical being.  

Then the likelihoods of an observed draw is either $1$ (if the ball drawn is that predicted in the microhypothesis) or $0$ (if the ball is not the predicted one).  If we draw ball $3$, for example, the credence in all hypotheses except Theory B's ``Ball 3 is drawn'' microhypothesis is zero.  Then the remaining microhypothesis has 100\% credence, and Theory B has 100\% credence as well.  If instead ball $1$ is drawn, Theory A's sole microhypothesis and one of Theory B's microhypotheses survive unscathed, while the other nine microhypotheses of Theory B are completely suppressed.  That is, 100\% of Theory A's credence survives, while only 10\% of Theory B's credence remains; therefore, post-observation, the credence in Theory A is $10/11$ and the credence in Theory B is $1/11$.  This, of course, matches the usual expectation for the thought experiment.

But what if instead you and 99 other attendees at a cosmology conference were drawing from the urn with replacement and you were prevented from telling each other your results? If we regard all one hundred participants as exactly identical observers, the only distinct microhypotheses seem to be the frequency distribution of each ball $j$ being drawn. It would seem that there could be no significant update to Theory B's credence if you draw ball $1$: all you know is that you are a participant-observer and there exists an observer-participant who draws ball $1$, which is nearly certain to be true. Inference would seem to require something like the SSA. Yet this is not necessary in practice because if this experiment were carried out at an actual cosmology conference, the participants \emph{would} be distinguishable. We then can fine-grain Theory B further by listing each attendee by name and specifying which ball they draw for each microhypothesis, and forming an observation model where names are matched to drawn balls. For example, we could order the attendees by alphabetical order and each microhypothesis would be a 100-vector of integers from $1$ to $10$.

With fine-graining, there is only $1^{100} = 1$ microhypothesis in Theory A -- all participants draw ball $1$ -- but $10^{100}$ microhypotheses in Theory B.  In only $10^{99}$ of B's microhypotheses do you specifically draw ball $1$ and observe ball $1$. Thus only 10\% of Theory B's microhypotheses survive your observation that you drew ball $1$. The credence in Theory B is again $1/11$, but is derived without appealing to typicality. Instead, typicality \emph{follows} from the combinatorics. The original one-participant version of this thought experiment can be regarded as a coarse-graining of this 100-participant version, after marginalizing over the unknown observations of the other participants. 

\subsection{Implicit microhypotheses: A thought experiment about life on Proxima b}

In other cases, the microhypotheses can be treated as abstract, implicit entities in a theory. A theory may predict an outcome has some probability, but provide no further insight into which situations actually lead to the outcome. This happens frequently when we are actually trying to constrain the value of a parameter in some overarching theory that describes the workings of unknown physics. Not only do we not know their values, we have no adequate theories to explicitly predict them. Historical examples include the terms of the Drake equation and basic cosmological parameters like the Hubble constant. Yet these parameters are subject to sampling variance; some observers in a big enough Universe should deduce unusual values far from their expectation values. Naive FGAI can be adapted for such cases by positing there are \emph{implicit} microhypotheses that we cannot specify yet. The probability that we observe an outcome is then treated as if it is indicating the fraction of microhypotheses where that outcome occurs.

Suppose we have two models about the origin of life, L-A and L-B, both equally plausible \emph{a priori}.  L-A predicts that all habitable planets around red dwarfs have life. L-B predicts that the probability that a habitable zone planet around a red dwarf has life is $10^{-100}$.\footnote{{This is a toy model. In reality, one would need a very rigidly defined mechanism with no possibility for unanticipated factors (e.g., directed panspermia) to calculate a probability this low. A modified version of L-B may very well yield higher probabilities.}} Despite this, we will suppose that the conditions required for life on the nearest potentially habitable exoplanet Proxima b are \emph{not} independent of our existence on Earth, and that any copy of us in a large Universe will observe the same result. The butterfly effect could impose this conditionalization -- small perturbations induced by or correlated with (un)favorable conditions on Proxima b may have triggered some improbable event on Earth necessary for our evolution.  

We wish to constrain L-A and L-B by observing the nearest habitable exoplanet, Proxima b, and we discover that Proxima b does have life on it. {The measurement is known to be perfectly reliable.} Copernican reasoning suggests that in L-B, only one in $10^{-100}$ inhabited G dwarf planets would observe life around the nearest red dwarf, and that as typical observers, we should assign a likelihood of $10^{-100}$ to L-B.  Thus, L-B is essentially ruled out.

\begin{figure*}
\includegraphics[width=17.2cm]{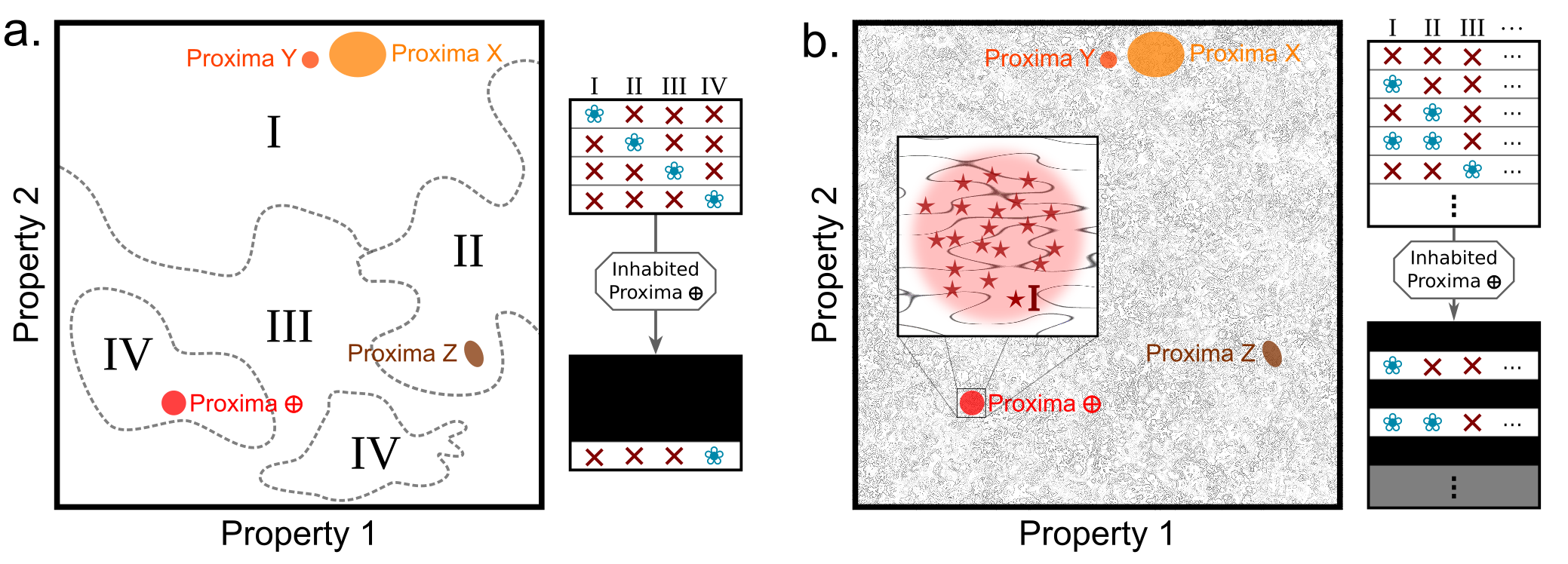}
\caption{Extremely simplified representation of how FGAI treats theories like L-B. Probabilities of observing a particular outcome in a macrotheory may be presumed to result from some unknown fine-scale division of parameter space into regions delineating different equivalence classes. Then microhypotheses would be constructed by considering all possible outcomes for all regions. For L-B, we observe the red dwarf nearest to Earth (Proxima $\oplus$) and see whether it is inhabited (flower symbol). The observation of an inhabited Proxima $\oplus$ reduces the number of allowed microhypotheses. Left: observers on other planets distinguishable from Earth would observe red dwarfs with different properties (Proxima X, Y, and Z) and probe different classes. Right: many identical Earths observe distinct Proxima $\oplus$ (red stars). These are treated by assuming there is some indexing that allows microhypotheses to be constructed, and their likelihoods calculated by symmetry. \label{fig:LB_Conditionalized}\label{fig:LB_NonConditionalized}}
\end{figure*}

L-B does not directly specify which properties of a red dwarf are necessary for life on its planets; it merely implies that the life is the result of some unknown but improbable confluence of properties.  Nonetheless, we can interpret L-B as grouping red dwarfs into $10^{100}$ equivalence classes, based on stellar and planetary characteristics. Proxima Centauri would be a member of only one of these. L-B then would assert that only one equivalence class of $10^{100}$ bears life.  Thus, L-B actually implicitly represents $10^{100}$ microhypotheses, each one an implicit statement about which equivalence class is the one that hosts life (Figure~\ref{fig:LB_Conditionalized}).  In contrast, L-A has only one microhypothesis since the equivalence class contains all red dwarfs.  We then proceed with the calculation \emph{as if} these microhypotheses were known.

If we observe life on Proxima b, then the sole microhypothesis of L-A survives unscathed, but implicitly only one microhypothesis of L-B of the $10^{100}$ would survive.  Thus, after the observation, L-A has a posterior credence of $100\%/(1 + 10^{-100})$, while L-B has a posterior credence of $10^{-100}/(1 + 10^{-100})$.  As we would hope, FGAI predicts that we would be virtually certain that L-A is correct, which is the result we would expect if we assumed we observed a ``typical'' red dwarf.

What of the other inhabited G dwarf planets in an infinite Universe? The nearest red dwarfs to these will have different characteristics and most of them will belong to different equivalence classes (Figure~\ref{fig:LB_Conditionalized}). In principle, we could construct microhypotheses that specify what each type of these observers will around their nearest red dwarf, and implicitly we assume they exist. If we failed to develop the capability to determine whether Proxima b has life but trustworthy aliens from 18 Scorpii broadcast to us that their nearest red dwarf has life, the result would be the same. 

\subsection{Fine-graining and implicit coordinate systems}
\label{sec:FGAI_Random}
\label{sec:Implicit}

The strictest interpretation of the separation of \objectiveName{} and indexical facts is that we cannot constrain \objectiveName{} models if the observed outcome happens to \emph{any} observer physically indistinguishable from us. This is untenable, at least in a large enough universe -- quantum mechanics predicts that all non-zero probability outcomes will happen to our ``copies'' in a large universe. But this would mean no measurement of a quantum mechanical parameter can be constraining. Surely if we do not observe any radiodecays in a gram of material over a century, we should be able to conclude that its half-life is more than a nanosecond, even though a falsely stable sample will be observed by \emph{some} copy of us out there in the infinite universe.\footnote{\citet{Leslie96} takes the position that indeterminism blunts SSA-like arguments, but only if the result has not been decided yet, because it is obvious the probability of an indeterministic future event like a dice throw cannot be affected by who we are. I believe this attempt to soften Doomdsay fails, because we are considering the probability \emph{conditionalized on you being you}{, an individual whose possible location is contingent on those indeterministic events}.} 

Strict indeterminism is not necessary for this to be a problem, either. In the last section, we might have supposed that the existence of life on Proxima b depends on its exact physical microstate five billion years ago, and that these microstates are scattered in phase space.  Yet, Proxima b is more massive than the Earth and has a vaster number of microstates -- by the pigeonhole principle, in an infinite Universe, most Earths exactly identical to ours would neighbor a Proxima b that had a different microstate, opening the possibility of varying outcomes (Figure~\ref{fig:LB_Conditionalized}).

But the separation of indexicals and \objectiveName{} theories need not be so strict. Indexicals are regarded in FGAI as propositions about coordinate systems. In \objectiveName{} theories we can and do use coordinate systems, sometimes arbitrary ones, as long as we do not ascribe undue objective significance. We can treat these situations by imposing an implicit indexing, as long as we do not ascribe undue \objectiveName{} significance to it.  Thus, in an infinite Universe, we label ``our'' Earth as Earth 1.  The next closest Earth is Earth 2, the third closest Earth is Earth 3, and so on.  We might in fact only implicitly use a coordinate system, designating our Earth as Earth 1 without knowing details of all the rest.  Our observations then are translated into third-person propositions about observations of Earth 1. The definition of the coordinate system imposes an indexical distribution where we must be on Earth 1. Of course, this particular labeling is arbitrary, but that hardly matters because we would reach the same conclusions if we permuted the labels. If we came into contact with some other Earth that told us that ``our'' Earth was Earth 3,296 in their coordinate system, that would not change our credence in a theory.

In a Large world, then, the microhypotheses consists of an array listing the outcome observed by each of these implicitly indexed observers. Only those microhypotheses where Earth (or observer) 1 has a matching observation survive. If the observed outcome contradicts the outcome assigned to Earth 1 by the microhypothesis, we cannot then decide we might actually be on Earth 492,155 in that microhypothesis because Earth 492,155 does observe that outcome. The coordinate system's definition has already imposed the indexical distribution on us.

There are limits to the uses of implicit coordinate systems, however, if we wish to avoid the usual Doomsday argument and its descent into solipsism.  In SB-B, could we not assign an implicit coordinate system with today at index 1, and the possible other day of the experiment at index 2?  A simplistic interpretation would then carve the Long run theory into two microhypotheses about which day has index 1, and learning ``today is Monday'' leads us to favor the Short theory. 

There is a very important difference between SB-B and the previous thought experiments. In those, our \objectiveName{} theory did not in any degree predict which observers get a particular outcome -- the likelihood distribution for the outcomes is exactly identical for all observers. This is why we are able to assign likelihoods even though the mapping between our implicit coordinate system and some external coordinate system is unknown.  In SB-B, though, our \objectiveName{} theory does predict which observer gets a particular outcome.  The likelihood of observing ``it is Monday'' is not identical for each day.  Instead, ``it is Monday'' needs to be interpreted as an indexical fact.  We can indeed create an implicit coordinate system with today at index 1 and the other day at index 2.  But we \emph{cannot calculate likelihoods in this coordinate system} -- the referents of the indices in the theory are unknown, and until we can connect the implicit coordinates with the coordinate system used by the Long theory, we cannot update credences either.  All we can say is that if observer 1 is located on Monday, they observe ``it is Monday'' with certainty; if observer 1 is located on Tuesday, they observe ``it is Tuesday'' with certainty. In these kinds of situations, a more sophisticated theory is needed.

\section{Reconstructing typicality}
\label{sec:Reconstructing}

\subsection{The indication paradox in Naive FGAI}

Naive FGAI, supplemented by the use of implicit microhypotheses and implicit coordinate systems, is sufficient to handle many practical cases where typicality is invoked. In these cases, however, it has been possible to calculate a single likelihood for each result because of an underlying symmetry in the problem.

\begin{figure*}
\centerline{\includegraphics[width=17.2cm]{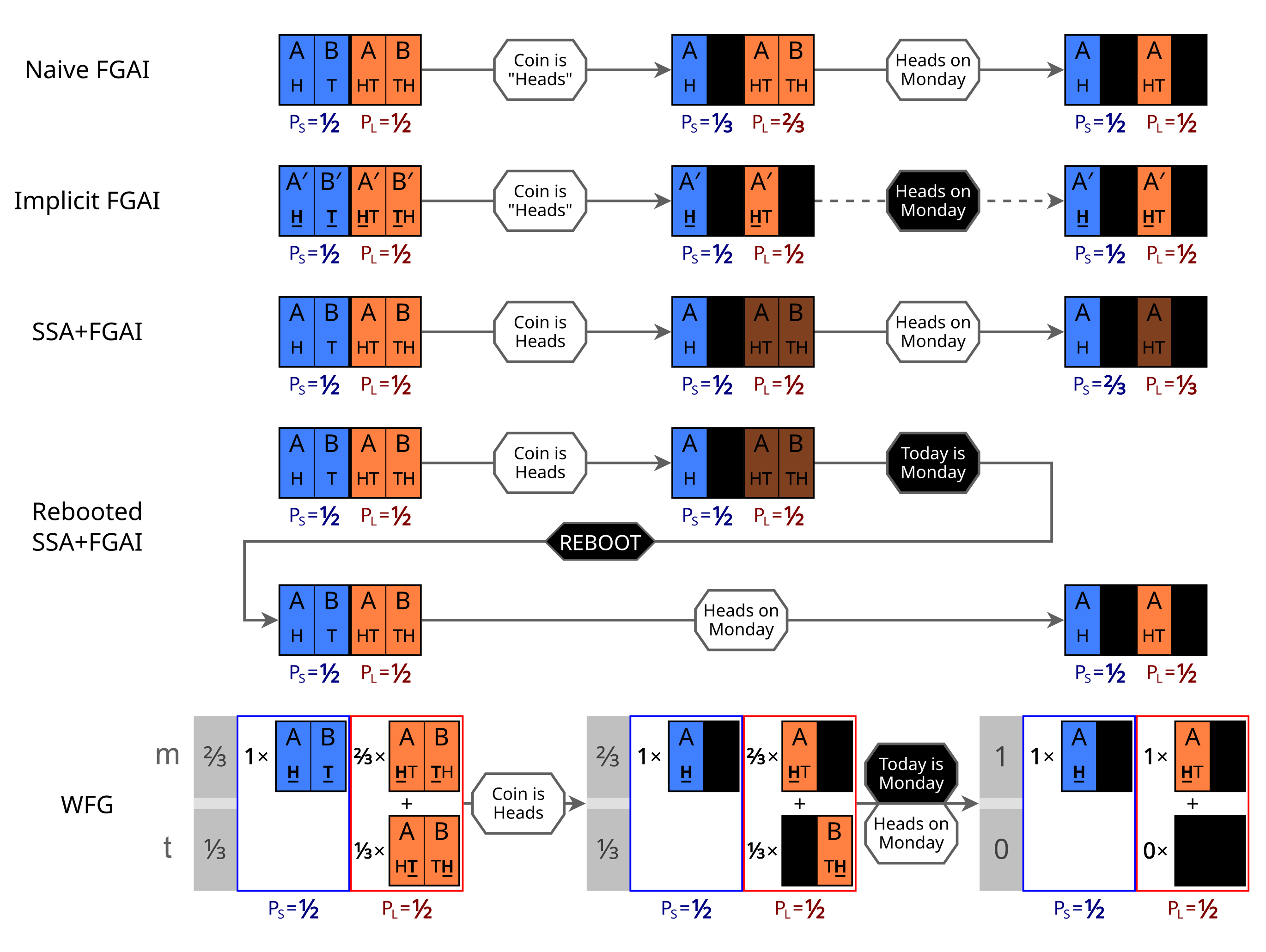}}
\caption{The indication paradox of \SBBCompound, and how it is treated with different modifications of FGAI. Each box, representing the credence in a microhypothesis, lists the observed outcomes, ordered by index in the used coordinate system. An underlined, bold outcome is treated as the one the observer measures for the purposes of calculating likelihoods. In the {Weighted Fine Graining} treatment, the normalized weights applied to each microhypotheses are given to the left of each provisional \objectiveName{} credence.\label{fig:IndicationParadox}}
\end{figure*}

In other cases, however, Naive FGAI leads to a bizarre SIA-like effect that always favors Large models{.} {T}his thought experiment {demonstrates the unwanted effect of an irrelevant detail}:
\begin{itemize}
\item[] \textbf{(\SBBCompound)} You wake up in the room. The experimenters tell you they are running SB-B, but each day they will also flip a fair coin each day the experiment is run and tell you the result.  But a team member gains your trust and tells you they are lying! The other experimenters have either adopted Strategy A or Strategy B. In Strategy A, they tell you the coin lands Heads on Monday, and if you awaken on Tuesday, they will tell you it landed Tails. In Strategy B, the supposed outcomes are opposite Strategy A's. Suspecting your new friend's duplicity, the other experiments exclude them from the decision of which Strategy to go with. Now the experimenters inform you the coin landed Heads. How does that affect your credence in a Short or Long run?
\end{itemize}
It is obvious the coin flip ``result'' should not affect our beliefs in a Short or Long experiment, because of the symmetry between the coin flip outcomes. Yet in Naive FGAI, \emph{either} coin flip ``result'' leads us to favoring a Long experiment (Figure~\ref{fig:IndicationParadox})! This is because either outcome is compatible with at least one of your wakenings in Long, but only in one of the Short microhypotheses. The indication paradox is that \emph{any} datum about the false coin flip favors the Long run. 

So, then, how should we calculate likelihoods when we do not even know who or where we are? In cases like SB-B and \SBBCompound{}, our likelihoods depend on our indexical position. Naive FGAI simply used the maximum likelihood among all possible observers, but there are other ways to approach the problem (Figure~\ref{fig:IndicationParadox}). {These highlight different pitfalls to avoid in reconstructing a theory of typicality.}

\subsubsection{Implicit FGAI{: can we just use observer-relative indexing?}}
{One method to address \SBBCompound{} is to reject its formulation, demanding the observer-relative indexing of ``today'' and the ``other day''.} ``Monday'' and ``Tuesday'' do not have any meaning in this coordinate system except as purely indexical information. Instead, our microhypotheses are mixtures of Strategies A and B, with the coin reported as Heads ``today'' and Tails ``the other day'' in Strategy A$^{\prime}$, and the reverse in B$^{\prime}$. Then the experimenters announcing Heads rules out B$^{\prime}$ in both Short and Long. As before, Implicit FGAI is only practical because of the symmetry. We cannot translate predictions between one coordinate system and the next. {Many situations of interest, like evaluating the merits of theories where we are probably a Boltzmann brain against those where we are probably evolved, break this symmetry and require some way to compare different classes of observers.}

\subsubsection{SSA+FGAI: {can we have Doomsday while avoiding solipsism with fine-graining?}}
A more radical solution is to reintroduce the SSA: instead of applying the maximum possible likelihood for an outcome among all observers we might be, we apply the mean likelihood among these possible observers. Equation~\ref{eqn:NaiveFGAI} is modified to:
\begin{equation}
\label{eqn:SSA_FGAI}
\ProbPosKJ = \frac{\displaystyle \ProbPriorKJ \times \sum_{i \in \OSetKJ (\Data)} \frac{\Prod(o @ i \to \Data | \MicrohKJ)}{|\OSetKJ (\Data)|}}{\displaystyle \sum_{x = 1}^M \sum_{y = 1}^{m_k} \ProbPriorXY \times \sum_{z \in \OSetXY (\Data)} \frac{\Prod(o @ z \to \Data | \MicrohXY)}{|\OSetXY (\Data)|}} .
\end{equation}
Then, in \SBBCompound, the result of the ``coin flip'' does not affect our credences in Short and Long. Both microhypotheses of Long would survive the observation, but they would be penalized by a $1/2$ factor because only one in two participants is given a Heads outcome. But this penalty continues to apply even when it is no longer needed: when we now apply the observation that ``today is Monday'' (and thus Strategy A was adopted) to the resulting distribution, Short is now favored two to one. 

SSA+FGAI also gives the correct answer for the SB-D variant. The Doomsday Argument is hence possible in SSA+FGAI, but it does not lead to solipsism if we start out ignorant of our location and include hypotheses in our prior that we are in the ``wrong'' place. That is the crucial difference between Doomsday and solipsism: it is impossible for high $\xrank$ observers to exist without low $\xrank$ observers, but it is possible that intelligence only exists in gas giants, or only around red dwarfs, or only in galactic bulges. Indeed, we must include these ``wrong'' hypotheses to avoid solipsism, as though we are starting out as a prescientific society, although these possibilities usually are not included when ``Copernican'' arguments are made. Even the Doomsday Argument may be tamed or neutralized in certain fine-grainings (KSO with a restricted reference class, from later in the paper).

Although tamer than the unrestricted SSA, SSA+FGAI is unsatisfactory both in that it allows (potentially unrealistically powerful) Doomsday Arguments at all and because it never ``forgives'' the penalties it imposes. The $1/2$ factor applied to Long's likelihood in \SBBCompound{} arose from our uncertainty in our location. That uncertainty vanishes when one learns ``today is Monday'', but the application of the SSA cannot be undone. Essentially, SSA+FGAI double-counts evidence against Large theories: penalizing all the microhypotheses because we do not know who we are, and then penalizing some of the microhypotheses because we \emph{do} know who we are. SSA+FGAI is unsatisfactory because it either reaches different conclusions based on which order data is learned, or it demands {that we ignore data to avoid this double-penalty effect.}

\subsubsection{Rebooted SSA+FGAI{: the use of shrinking reference classes}}
``Rebooted SSA+FGAI'' is a modification of SSA+FGAI to eliminate the order-dependence of learning data. It is like SSA+FGAI, except that the credence distribution is reset to the original prior whenever new indexical information is learned. Then, all physical and indexical data is re-applied simultaneously. The likelihood of microhypotheses are evaluated according to the SSA applied only among observers with all your current indexical data. Essentially, with each reboot, the reference class shrinks, undoing the now-redundant penalty of the SSA.

{In summary, a theory of typicality should ideally (1) be able to evaluate data when observers at different locations have different likelihoods, (2) avoid universally favoring Large theories no matter the datum by using some kind of typicality assumption, but (3) be able to retroactively update the reference class used in the typicality calculations.}

\subsection{Weighted Fine Graining}
Weighted {Fine Graining}'s premise is that there is no single physical distribution for the microhypotheses.  Instead, there is a set of \emph{provisional physical credences} for each microhypothesis, each associated with a possible observer location. {The credence is essentially a superposition of these provisional credences.} We evaluate our credences in the various microhypotheses by averaging over the provisional physical distributions. \footnote{To use a loose analogy, the combination of indexical weights and provisional physical distributions can be compared to density matrices in quantum mechanics. Density matrices are generally needed to describe the mixed state of a part of a system. Metaphorically speaking, {WFG} is needed when an observer sees only a part of the consequences of a microhypothesis and no single pure distribution can describe credences.} 

The provisional physical credence $\ProvKJI$ is proportional to the credence in $\MicrohKJ$ calculated by an observer known to be along observer-trajectory $i \in \OSetKJ${:
\begin{equation}
\label{eqn:ProvDef}
\frac{\ProvKJI}{\sum_{\MicrohXY} \ProvXYI} \equiv \Prob(\MicrohKJ | o @ i) .
\end{equation}
If $i \notin \OSetKJ$, then $\ProvKJI = 0$. Additionally, the ratio of the provisional credences in $\MicrohKJ$ for different observer-trajectories gives the indexical distribution for $\MicrohKJ$:
\begin{equation}
\label{eqn:WFGAI_IndexicalDist}
\IndexDistKJI = \frac{\ProvKJI}{\sum_{z \in \OSetKJ} \ProvKJZ} .
\end{equation}
The usual case is to start out with an uninformative indexical distribution for all $o \in \OSetKJ$, resulting in $\ProbPriorKJ = \ProvPriorKJI = \ProvPriorKJZ$.}

Prior provisional credences are updated with data $\Data$. The observation model of $\MicrohKJ$ gives a single likelihood for $D$ conditionalized on our location being $i$, which we use to update the provisional physical credences:
\begin{equation}
\label{eqn:ProvUpdate}
\ProvPosKJI = \ProvPriorKJI \Prod(o @ i \to \Data| \MicrohKJ) .
\end{equation}

Our credence in microhypothesis $\MicrohKJ$ is then generated from the provisional physical distributions {using} indexical weights:
\begin{equation}
\label{eqn:WeightedFG}
\ProbKJ = \frac{\displaystyle \frac{1}{\WeightNormKJ} \sum_{i \in \OSetKJ} \WeightI \ProvKJI}{\displaystyle \sum_{\MicrohXY} \frac{1}{\WeightNormXY} \sum_{z \in \OSetXY} \WeightZ \ProvXYZ} .
\end{equation}
where posterior provisional credence and indexical weights are substituted to find $\ProbPosKJ$ and prior provisional credences and weights for $\ProbPriorKJ$. As in Naive FGAI, posterior credence in a macrotheory is found by summing over microhypotheses. The use of the weights means that {WFG} is not a strictly Bayesian theory. We cannot calculate the ``actual'' \objectiveName{} credence until we can assign a single, unambiguous likelihood, and we cannot do this in FGAI. Instead the posterior credence is a construction built from possible \objectiveName{} credences using the indexical weights.

{The weight for $i$ is proportional to the total provisional credence for that observer-trajectory:
\begin{equation}
\WeightI = \frac{\sum_{\MicrohXY} \ProvXYI}{\sum_{\MicrohXY} \sum_{z \in \OSetXY} \ProvXYZ} .
\end{equation}
These weights perform an averaging over all microhypotheses in all theories.\footnote{Additionally, the weights are assigned to individual locations rather than indexical probability distributions. \citet{Srednicki10} first proposed that we test indexical distributions (called xerographical distributions) by comparing predictions in physical theories with evidence. The problem is that an indexical distribution expressing uncertainty in our location can be consistent with evidence even when we know our exact location, as pointed out by \citet{Friederich17}. {WFG} avoids this issue because indexical locations themselves are weighted -- the xerographical distributions associated with each weight are orthogonal.} While not a pure indexical distribution, which in FGAI is considered to be associated with only one microhypothesis, it forms an effective indexical distribution. In fact, it is equal to the indexical distribution according to the SIA prior. The results differ from SIA because of the normalization factor applied to the weights:}
\begin{equation}
\WeightNormKJ \equiv \sum_{i \in \OSetKJ} \WeightI .
\end{equation}
{The normalization factor is critical in that it allows a ``dilution'' of credence in microhypotheses with more observers. It is a reflection of the fact that the set of observers in our reference class differs from one hypothesis to the next, and so will our conclusions for a typical observer. Thus, although the indexical prior reflects the SIA, the evaluated credences in equation~\ref{eqn:WeightedFGAI} lack this indication effect.}

{In summary, the evaluated credences are calculated by (1) constructing the provisional credences using equations~\ref{eqn:ProvDef}, \ref{eqn:WFGAI_IndexicalDist}, and \ref{eqn:ProvDef}; (2) calculating the weights by summing the provisional credences for each observer-trajectory and normalizing; and (3) calculating the credence by taking the weighted average of the provisional credences for each microhypotheses using those weights, and normalizing.}

The indexical weights probabilistically define our current reference class, which evolves as we learn new information. Yet the identity of the observer at each location (trajectory) $i \in \OSet$ depends on the microhypothesis, just as who is born at birthrank $1$ in the Doomsday Argument varies. Thus the indexical weights describe a reference class of locations or contexts (or more properly, trajectories), not observers. The indexical weights tell us \emph{where} we are, not {\emph{what}} we are. {The characteristics of the observer at that location is, if specified at all, given by the microhypothesis.}

{WFG} has several useful limits:
\begin{itemize}
\item {When we already are certain of the physical nature of the world, $\ProvKJI$ is nonzero only for the known world model $\MicrohKJ$. Then $\WeightI = \IndexDistKJI$, with simple Bayesian updating, just as we expect from SB-A.}
\item If $\Prod(o @ i \to \Data | \MicrohKJ)$ is the same for all possible locations{, and this is true for all $\MicrohKJ$}, then Equation~\ref{eqn:WeightedFG} reduces to Bayes' formula {applied to
\begin{equation}
\widetilde{\ProbKJ} \equiv \sum_{i \in \OSetKJ} \frac{\WeightPosI}{\WeightNormKJ} \ProvKJI .
\end{equation}
}
\item When indexical information is entirely irrelevant, $\ProvKJI = \ProbKJ$, and all indexical dependence vanishes from equation~\ref{eqn:WeightedFG}{, which reduces to simple Bayesian updating.}
\item Suppose the data is consistent only with a subset $\OSetPRIME$ of locations, and that the posterior weights are uninformative ({$\WeightPosI = 1 / |\OSet|$}) among this subset. Then $\OSetPRIME$ is effectively our new location reference class, and equation~\ref{eqn:WeightedFG} implements SSA+FGAI over it. Equation~\ref{eqn:WeightedFG} can thus be viewed as {WFG}'s rendition of the Observer Equation proposed by \citet{Bostrom13}.
\item If we become certain of $\MicrohKJ$, then provisional credences for all other microhypotheses are zero, and $\ProbPosKJ = 1$.
\end{itemize}

The provisional \objectiveName{} {credences update solely in response to the \objectiveName{} fact that a specific observer measures the datum, independent of what happens at other locations. This ensures the insulation between the \objectiveName{} distribution and the indexical distribution, which emerges from the ratios of the provisional credences}, albeit the weights have a more active role than in Naive FGAI. The weights shift adaptively in response to indexical data, as we can more accurately assess likelihoods over microhypotheses. The novel feature of {WFG} is the shifting indexical weights' ability to ``revive'' theories that were provisionally disfavored by the SSA. If we had only a single joint probability distribution, evidence can only eat away the prior probability in a theory because likelihoods are always $\le 1$.  The only way a theory survives with a single credence distribution is if the other theories' prior credence is also consumed at a comparable rate.  In {WFG}, the provisional probability invested in disfavored observer trajectories is not {necessarily} lost but re-assigned to the microhypotheses consistent with likely observer trajectories.

This formulation of {WFG} is based on the assumption that there are only a finite number of locations or trajectories, parameterized by a finite set of indexical weights. Infinite volumes are a common feature of modern cosmologies, however, and this issue needs to be addressed. Perhaps the discrete distributions can be replaced with probability densities in a way that recovers conclusions found for a large, finite number of locations. Alternatively, the number of expected configurations of a Hubble volume is expected to be finite \citep[e.g.,][]{Bousso02}, which may be consistent with the discrete distributions, if the observer locations are treated as equivalence classes of identical Hubble volumes at the moment we start the experiment.

\subsection{An example: two urn problems}
{Urn problems give natural cases} where we should favor a ``Small'' theory where our observation is more typical. In Naive FGAI, this can follow from combinatorics when observers are distinguishable. The conclusion also holds up in {WFG}.

{In the first urn problem (\textbf{U-I}), there is one urn. The urn may have a Small set of balls numbered $1$ through $N_S$, or a Large set numbered $1$ through $N_L$; we know that this number follows from some unknown deterministic process. Now, a series of $N$ participants including us draws a ball from the urn in sequence, with replacement. We do not know how many participants there are before or after us,} nor any other identifying details{, nor what ball the other participants draw. We thus have a series of provisional credences for each possible participant, describing microhypotheses about the sequence of balls drawn.} Now, if $N$ is big enough, any possible {draw} will occur with high probability. However, in any theory about which type of urn is being used, the probability of drawing a given ball is equal for all locations. Thus, if we draw a numbered ball that is found in both a Small urn and a Large urn, we favor the Small urn, as expected -- the provisional physical distribution has fewer compatible microhypotheses at each location without a compensating weight shift.

\begin{figure*}
\centerline{\includegraphics[width=12.9cm]{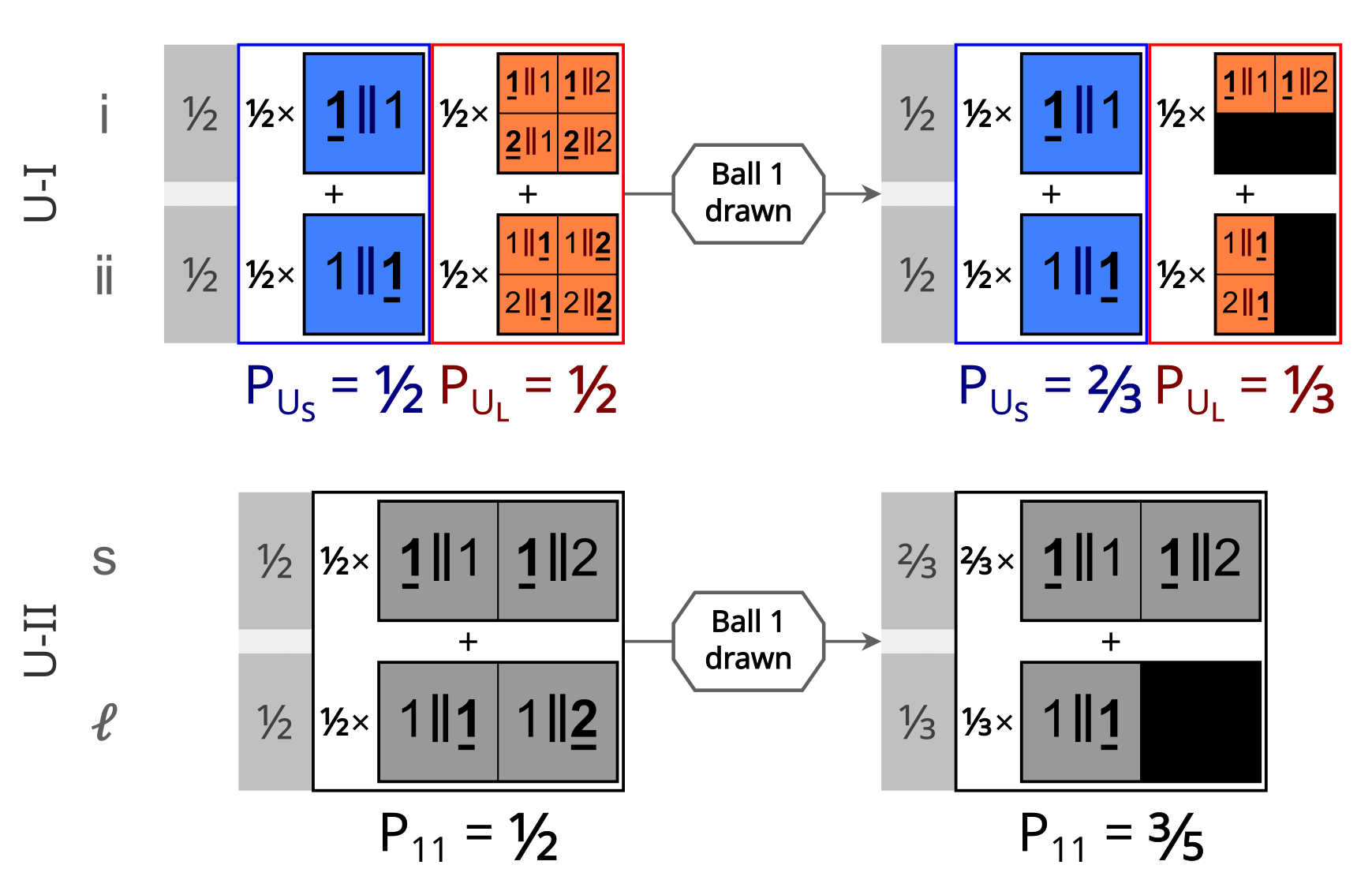}}
\caption{Treatment of the ``urn problems'' U-I and U-II in {WFG}. In U-I (top), there is one urn. We and one other participant try to determine if it is Small or Large by drawing one ball with replacement. The different indices correspond to the order in which the participants draw. In U-II (bottom), there are two urns at different locations. We and one other participant are assigned randomly to the different urns and try to determine which urn location we are at based on that draw. \label{fig:WFGAI_Urn}}
\end{figure*}

{Figure~\ref{fig:WFGAI_Urn} demonstrates a simple worked example. Here, there are two participants. They draw from the urn with replacement at distinct locations $\WorldONE$ and $\WorldTWO$. We are completely agnostic about which of these two we are: $\ProvPrior_{k,j;\WorldONE} = \ProvPrior_{k,j;\WorldTWO}$. If the Small urn theory is true, there is only one ball in the urn, with one microhypothesis inheriting all of the Small theory's credence: $\ProvPrior_{S;\WorldONE} = \ProvPrior_{S;\WorldTWO} = 1/2$. According to the Large urn theory, there are two balls, with four microhypotheses, with $\ProvPrior_{L,11;z} = \ProvPrior_{L,12;z} = \ProvPrior_{L,21;z} = \ProvPrior_{L,22;z} = 1/8$. Because the sums of the provisional credences for position $\WorldONE$ and $\WorldTWO$ are the same, $\Weight_{\WorldONE} = \Weight_{\WorldTWO} = 1/2$. It can then be shown that $\ProbPrior_S = \ProbPrior_L = 1/2$.}

{Now we draw ball one. The probability of this happening if we are an observer who draws ball two is zero, of course, updating some of the provisional credences: $\ProvPos_{L,12;\WorldTWO} = \ProvPos_{L,21;\WorldONE} = \ProvPos_{L,22;\WorldONE} = \ProvPos_{L,22;\WorldTWO} = 0$. The first and second position are equally likely to draw ball two in the Long theory, so the sums of the provisional credences for these positions remains the same, $1/2 + 1/8 + 1/8 + 0 + 0 = 3/4$. Thus, the weights remain the same for both theories, $\WeightPos_{\WorldONE} = \WeightPos_{\WorldTWO} = 1/2$. The evaluated credence in the Small theory is now
\begin{equation}
\ProbPos_S = \frac{(2/2) \times (1/2)}{(2/2) \times (1/2) + [(2/2) + 2 \times (1/2)] \times (1/8)} = 2/3 ,
\end{equation}
as expected from the SSA. 
}

{What if instead we already know the distribution of urn sizes, and we are trying to determine which urn we are drawing from? In \textbf{U-II}, there are only two urns, and two participants including us, one for each urn. At location $s$ is a small urn with a single ball numbered $1$, while at location $\ell$ is a large urn with two balls numbered $1$ and $2$. We wish to know which location we are at, but we have no idea. The overall distribution of urns -- one small and one big -- is known, but there are still two microhypotheses regarding for which ball is drawn at $\ell$. The provisional credences are equal for both locations and microhypotheses, to yield uninformative \objectiveName{} and indexical priors: $\ProvPrior_{11;s} = \ProvPrior_{11;\ell} = \ProvPrior_{12;s} = \ProvPrior_{12;\ell} = 1/2$. Now we draw ball $1$. If we are at the small urn, or if the $\ell$ participant draws ball $1$, then this is consistent with this datum ($\ProvPrior_{11;s} = \ProvPrior_{11;\ell} = \ProvPrior_{12;s} = 1/2$), but not if we are the $\ell$ participant in the microhypothesis where they draw ball $2$ ($\ProvPos_{12;\ell} = 0$). By summing these provisional credences, we find $\WeightPos_{s} = (1/2 + 1/2)/[(1/2 + 1/2) + 1/2] = 2/3$ and $\WeightPos_{\ell} = 1/3$. This indexical formulation of the urn problem gives the same result -- we favor our having a small urn by 2:1.}

{An interesting effect of the updating weights is shown in Figure~\ref{fig:WFGAI_Urn}. What if, in U-II, we want to know whether both participants drew ball $1$? Of course, if we are certain we are at the small urn, then we definitely have no information about what the large urn participant draws and $\ProbPos_{11} = \ProbPrior_{11} = 1/2$. If we remained completely uncertain about our location, as with a naive application of the SSA, we would average the probability that a random participant draws ball $1$: $\ProbPos_{11} = (1/2) \times 1 + (1/2) \times 1/2 = 3/4$. In {WFG}, we favor our being at location $s$ but are not certain about it, resulting in an intermediate credence: $\ProbPos_{11} = (2/3 \times 1 + 1/3 \times 1) / [(2/3 \times 1 + 1/3 \times 1) + (2/3 \times 1 + 1/3 \times 0)] = 3/5$.}

\subsection{WFG and Sleeping Beauty}

\begin{figure*}
\centerline{\includegraphics[width=17.2cm]{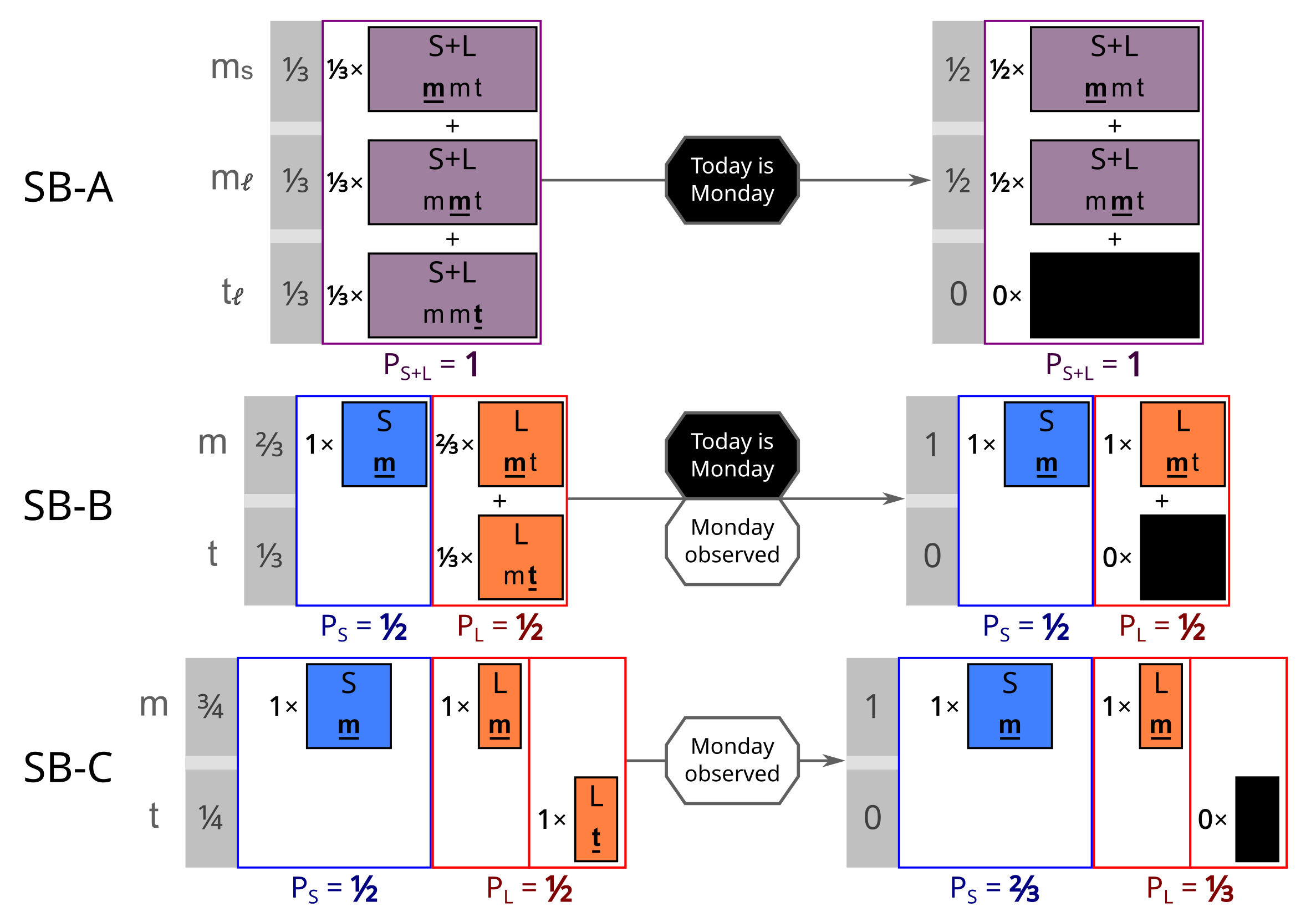}}
\caption{Treatment of the fundamental variations of the Sleeping Beauty problem in {WFG}. In SB-A, only the indexical weights shift, while in SB-C, the physical credence distributions entirely drive the conclusion. In SB-B, the indexical information shifts the indexical weights assigned to each provisional distribution, ensuring that neither the Short nor Long distributions are favored. \label{fig:WeightedSB}}
\end{figure*}

{WFG}'s treatment of the Sleeping Beauty examples is illustrated in Figure~\ref{fig:WeightedSB}.  Because both the physical provisional distributions and indexical weights are updated according to Bayes' equation, we can recover the naive Bayesian predictions of SB-A and SB-C.  

In the SB-B variant, there are only the Short and the Long theories, with one trivial microhypothesis each, and each having $1/2$ prior credence.  The Short theory has a normalized indexical weighting of $1$ for Monday.  The Long theory has a Monday provisional {credence} and a Tuesday provisional {credence}, both still equal to $1/2$. {The prior indexical weights are $2/3$ for Monday and $1/3$ for Tuesday.} Now we learn ``today is Monday''.  The likelihood that this observation is made is $1$ if we are the Monday observer so the Monday provisional {credences} have a likelihood of $1$.  The likelihood this observation is made is $0$ if we are the Tuesday observer, so the Tuesday provisional {credence} in the Long theory is eliminated.  However, {we have} 100\% confidence in our being the Monday observer if today is Monday, and {the} indexical weights shift to $1$ for Monday and $0$ for Tuesday.  As a result, the probability lost in the Tuesday provisional distribution is irrelevant.  Our credence in the Long theory is $(1 \times 1/2 + 0 \times 0) / [(1 \times 1/2) + (1 \times 1/2 + 0 \times 0)] = 1/2$, just as before.

There may be some situations like SB-B when we do want an indication effect: if SB-B is carried out with two already extant observers Alice and Bob who are released after the experiment. If Alice wakes up during the experiment instead of afterwards, she can arguably conclude the Long scenario is more likely. \citet{Garisto20} emphasizes this distinction between situations where one outcome is ``Picked'' and those where each outcome is simply a different location one can ``Be''. Indexical weights representing trajectories instead of static locations allow for this distinction. In the Alice-first Long microhypothesis, the Monday trajectory connects with Alice's Sunday location instead of Bob's Sunday location; in the Bob-first Long microhypothesis, the Monday trajectory connects with Bob-on-Sunday. There are thus four possible indexical weights, and each microhypothesis is compatible with only some of them. 

\subsection{SB-D, virtual observers, and the nature of the provisional distributions}
Explaining SB-D is a challenge for FGAI, and the resolution I will now sketch provides insight into the nature of the provisional distributions.

The issue in SB-D arises when a participant builds the provisional distributions as if they are an external, timeless observer $X$ of the experiment. Then, one may reason that Short and Long are equally likely, so the two Short microhypotheses (waking on Monday and waking on Tuesday) inherit half of that credence. Short-Monday's entire credence of $1/4$ is given to the Monday provisional distribution, Short-Tuesday's entire credence of $1/4$ is given to the Tuesday provisional distribution, and the weights normalize so that Short and Long have equal prior credence.

\begin{figure*}
\centerline{\includegraphics[width=17.2cm]{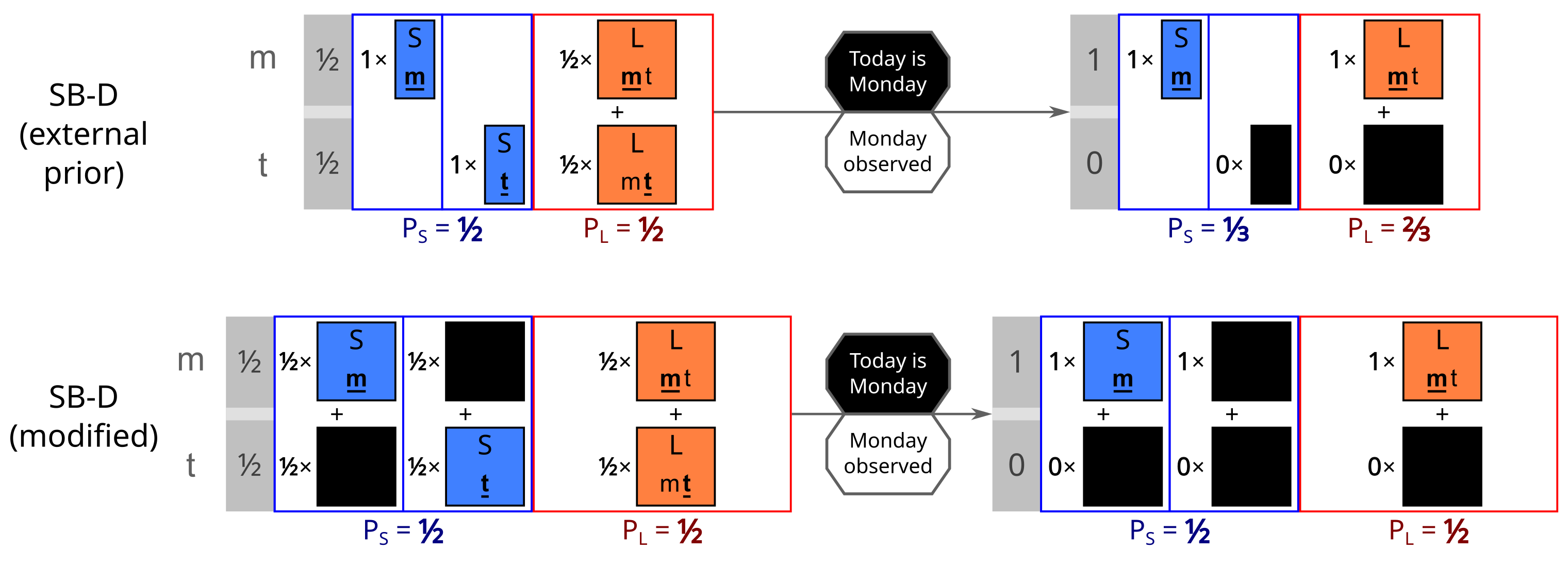}}
\caption{Illustration of an indication paradox resulting from SB-D. Whether we are told it is Monday or Tuesday, we seemingly favor the conclusion we are in the Long experiment (top). Virtual observers can be introduced to prevent this paradox (bottom).\label{fig:WeightedSBD}}
\end{figure*}

Learning that it is Monday, however, rules out the Short-Tuesday microhypothesis but the weights shift so that the Long theory is unaffected. As a result, we now favor Long 2:1 (Figure~\ref{fig:WeightedSBD}). Crucially, unlike SB-C, this also happens if one learns it is Tuesday. Any data about one's location favors the Long theory, a clear absurdity. Even if one initially favored Short 2:1, the Bayesian shift still occurs, so the problem remains.

Clearly, we cannot use this ``external'' credence distribution. For  $X$, the statement ``it is Monday'' has no relevance. Presumably, $X$ asks something like ``is the participant awake on Monday?'', but an affirmative reply is informative because it is possible for the answer to be ``No'', whereas there is a selection effect for the participant. Or, if $X$ asks ``what is one day the participant is awake on?'', then there is equal chances of the result being Monday or Tuesday in the Long case, so the credence of the Long theory is reduced accordingly (essentially converting SB-D into \SBBCompound).

But in fact the provisional prior distributions are the prior distributions one would have used if one was entirely certain of one's location. For each location, the Short and Long prior probabilities should be equal. This is supported by the analogy between AB-B and AB-D: surely our present belief in the diversity of ETIs should not depend on whether we learned of Earth's environment before or after the question was first posed. One's situation after learning ``today is Monday'' in SB-D is the same as if we had learned ``today is Monday'' in SB-B.

How, then, are we to avoid the Bayesian shift that occurs? I propose that we add \emph{virtual observers} to each Short microhypothesis, expanding $\OSet_{S,M}$ and $\OSet_{S,T}$ to $\{\mathrm{M}, \mathrm{T}\}$. The observational model is extended to allow virtual observers to ``observe'' the same data that a real observer at the same location would. The provisional prior credence of a microhypothesis for a virtual observer is zero. The indexical distributions {emerge from} the provisional credences, so they still give the correct result. The virtual observes dilute the Short microhypotheses' credences by spreading the indexical weights. Furthermore, now Short's weights are as responsive as Long's, allowing the remaining Short microhypothesis to remain on an even level with Long (Figure~\ref{fig:WeightedSBD}).

Virtual observers are placed to prevent Bayesian shifts upon learning any data (as in SB-D), but not to inhibit valid inferences when there are more outcomes in one theory than another (as in SB-C) -- a distinction dependent on symmetry. The nature of these virtual observers is unclear, and a full theory of how they are placed must be developed. This is clear when we consider a hybrid thought experiment, where there is a Short variant with a wakening on Monday only, a Long variant with awakenings on Monday and Tuesday, and an Intermediate variant with awakenings on one day chosen at random. If one then learns the experiment is not-Short, not-Intermediate, or not-Long, one would expect the situation to reduce SB-D, SB-B, or SB-C, respectively, but this is impossible if there is only one provisional distribution for each location. The reductions are possible if there are \emph{two} Monday and Tuesday provisional distributions, one for Short-Intermediate and one for Long-Intermediate, but this is admittedly contrived. This hybrid thought experiment itself is unlike the types of problems that arise in astrobiology, though, and itself may be contrived.

\subsection{{Variants on weighting}}
{The indexical weights are based on sums of the provisional credences for each position. Our evaluated credence in microhypotheses according to equation~\ref{eqn:WeightedFG} therefore depends on the provisional credences in separate theories. This can lead to some odd results. Consider \SBBCompound{}, after one learns that the experimenter reports a Heads outcome. Only one microhypothesis in each theory for each day survives this observation (Short-A and Long-A for Monday, and Long-B for Tuesday), thus neither the weights, nor the relative credence in Short and Long change. However, because the weights favor Monday two-to-one, that means that $\ProbPos(\mathrm{Long+A}) = 2 \ProbPos(\mathrm{Long+B})$ (c.f., Figure~\ref{fig:IndicationParadox}). Of course, if the experimenters had announced Tails, the relative credences of the microhypotheses would be reversed; it is not as though any data leads to us favoring Long+A over Long+B.

One ultimately problematic way to address this issue is to adopt a weight distribution that starts out not favoring any particular position: ${\WeightPriorI}^{\prime} = 1/|\OSet|$. The weights then are updated by multiplying with an indexical likelihood,
$$\frac{\sum_{\MicrohKJ} \ProvKJI \Prod(o @ i \to D | \MicrohKJ)}{\sum_{\MicrohKJ} \ProvKJI}$$
and normalizing. In many situations this gives the same result as the standard weights. Unfortunately, the weights now retain a ``memory'' of theories that are no longer viable, with two negative consequences. First, the weights can no longer directly serve as an effective indexical distribution. This can be seen in SB-B if the participant learns only that they are in the Long experiment without learning the day of the week. Then because the Short theory is ruled out, the Monday weight has an indexical likelihood of $1/2$ while the Tuesday weight has an indexical likelihood of $1$, leading the weights to favor Tuesday two-to-one. Second, this memory effect can still lead to undesirable asymmetry in credences. For suppose the participant in \SBBCompound{} learns they are in a Long experiment and the otherwise irrelevant detail that the coin flip ``outcome'' for the day is Heads. Now, because the weights favor Tuesday, the participant would conclude that $\ProbPos(\mathrm{Long+B}) = 2 \ProbPos(\mathrm{Long+A})$. In contrast, in the standard weights, the now-unviable Short theory no longer influences $\WeightPos$ or $\ProbPos$.

A different solution is to abandon universal sets of weights. Instead the weights are defined only for a subset of microhypotheses, most naturally individual macrotheories. Defining a separate set of weights for each macrotheory $k$,
\begin{equation}
\WeightKI = \frac{\sum_{\MicrohKJ \in \MacrotK} \ProvKJI}{\sum_{z \in \OSetK} \sum_{\MicrohKJ \in \MacrotK} \ProvKJZ},
\end{equation}
still allows for an averaging over microhypotheses, but now credence shifts in one theory has no effect on the weights in another. This eliminates the asymmetric microhypothesis credences in \SBBCompound{}.} The main issue with macrotheory-level weights is their ad hoc nature. It leaves open the question of what even counts as a macrotheory, and why that level should be ``correct'' other than it giving seemingly correct results.

Using separate sets of indexical weights for every single microhypothesis {may seem the most natural solution, with $\WeightKJI = \IndexDistKJI$.} Each microhypothesis’ credence is determined by the likelihood for the locations most compatible with observations. Essentially, microhypothesis-level weighting implements the simplest forms of naive FGAI, without implicit indexing. And therein lies the problem: as long as there is any observer anywhere who observes one’s data in the microhypothesis, that microhypothesis cannot be constrained. Thus in the Proximan life thought experiment, if life on Proxima b is independent of our existence, then L-B cannot be ruled out if we observe life there: it simply means we live on one of the Earths in an endless universe that just so happens to be around the ultrarare inhabited Proxima b, whether that is Earth 1 or Earth 492,155. Implicit indexing in naive FGAI gets around this by fixing one position as the only possible location, preventing indexical weight shifts. A similar restriction of the reference class (e.g., only Earth 1 has nonzero indexical weight) is necessary if microhypothesis level weighting is to be practical.

\subsection{Boltzmann Brains in WFG}
\label{sec:Boltzmann}
The Boltzmann brain problem is deeply related to the notion of typicality and has guided previous treatments of the matter. Cosmological theories with large thermal baths -- including the event horizons of black holes and de Sitter universes -- predict the appearance of Boltzmann brains, observers with your memories and beliefs that assemble out of thermal fluctuations and flicker into activity briefly \citep[e.g.,][]{Dyson02,Albrecht04,Carroll17}. Not only are they predicted to exist with probability $1$, but current cosmological theories have the disturbing tendency to predict that Boltzmann brains with your memories vastly outnumber those of you who have evolved naturally. Furthermore, even in universes very unlike ours, there should exist observers with our memories who remember our $\Lambda$CDM cosmology with probability $1$. So what justification do we have in concluding that the universe is actually anything like we observe it to be?

{Boltzmann brains and similar problems} motivated \citet{Bostrom02} to introduce the SSA and argue for the use of a wider reference class than observers identical to us. The \emph{vast} majority of Boltzmann brains have experiences completely unlike ours; even those that momentarily have our memories almost always find their sensoria dissolving immediately afterwards as they are exposed to their environment (say, all but $1$ in $\gg e^{10^{25}}$). We might hope that the persistence of our observed world favors us being evolved observers viewing a real universe. In the strictest interpretation of FGAI, this is of no help for inference because there always exist Boltzmann brains that observe a persisting universe (living on Boltzmann Earths, for example).

Let us now apply {WFG} to a simplified Boltzmann brain problem. In false cosmologies, only the Boltzmann brains ({b}) exist, while the true cosmology also contains evolved brains ({e}). We start with a set of indexical weights assigning equal credence to each possible location, Boltzmann or evolved. Every evolved observer sees a long-lived persistent universe, but only a small fraction $\eta$ of the Boltzmann brains observes one. As the observed cosmology persists in our sensory data, we update both the indexical weights and each cosmology's provisional physical credence. Then we will find ourselves both favoring the true cosmology and, if we use provisional-prior updating for the weights, our being an evolved observer within that cosmology.

For example, suppose we consider two cosmologies, a False one (F) and a True (T) one. The False one contains only {$N_e$ locations with} Boltzmann brains, while the True one contains {those same $N_b$} locations with Boltzmann brains and {an additional $N_e$} locations with evolved observers. The {provisional credences of each microhypothesis start equal for each occupied position in each theory. This results in initial weights of $1/(N_e + 2 N_b)$ for each evolved location and $2 / (N_e + 2 N_b)$ for each Boltzmann brain location.}

\begin{figure*}
\includegraphics[width=17.2cm]{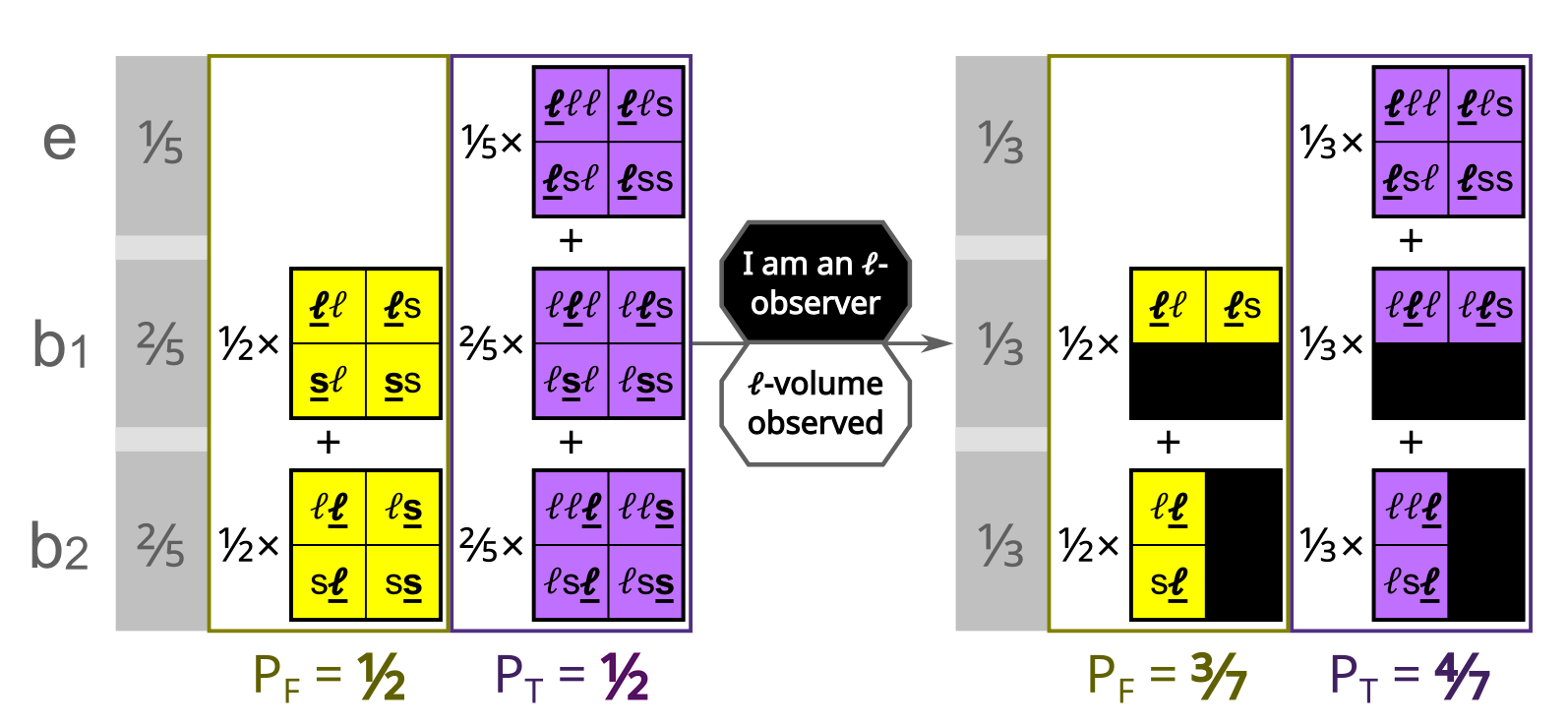}
\caption{Minimal example of Boltzmann brain problem treated in {WFG} {with $N_b = 2$, and $N_e = 1$}. Both the false and true cosmologies have two Boltzmann observers identical to ourselves; in this case, each with an equal probability of being long-lived ($\ell$) and decaying in a short time ($s$). The true cosmology also has an evolved observer $E$ that always observes it correctly. An observation of T leaves the {$e$} provisional physical distribution untouched and also shifts indexical weight to {$e$}, favoring T.\label{fig:Boltzmann}}
\end{figure*}

Our not instantly dissipating is certain if we are evolved observers but unlikely if we are Boltzmann brains. {It can be shown that the effective likelihood for each evolved location is then $1$ but $\eta$ for Boltzmann brain locations. When the provisional credences are updated, the weight for each evolved location shifts to $1/(N_E + 2 \eta N_B)$ and the weight for each Boltzmann brain location falls to $2 \eta / (N_e + 2 \eta N_b)$.} If we are an evolved individual, all microhypotheses {in the T theory are consistent with} the observation, while if we are any given Boltzmann individual, {only a fraction $\eta$ of the provisional credence in each theory remains}. Using equation~\ref{eqn:WeightedFG}, we now evaluate a posterior credence in cosmology T of: {
\begin{equation}
\ProbPos_T = \frac{N_E + 2 \eta^2 N_B}{N_E (1 + \eta) + 4 \eta^2 N_B} .
\end{equation}}
As long as $\eta \ll 1$ and $N_E \gg {2} \eta^2 N_B$, we now almost entirely favor cosmology T ($\ProbPos_T \approx 1 - \eta - 2 \eta^2 N_B/N_E$) (see Figure~\ref{fig:Boltzmann} for a minimal example). {When these criteria are fulfilled, the provisional credence for the evolved locations dominates the evaluated credence in T, even if $\WeightPos_E \ll \WeightPos_B$.}\footnote{This does not solve the Boltzmann brain problem in general. First, Boltzmann brains may so outnumber us ($2 \eta^2 N_B \gg N_E$ in this example) that the credence shift is negligible \citep{Dyson02}. The measure problem is the lack of consensus on how these probabilities should be calculated in multiverses \citep[e.g.,][]{Bousso08,Freivogel11}. Second, our memories of even a few moments ago are probably unreliable if we are Boltzmann brains. Boltzmann brains who ``remember'' observing the world a few seconds ago and thus conclude it is stable may outnumber the evolved observers \citep{Carroll17}. The implication is that we should reset our cosmological priors every moment. All we can say is that if some distribution, apparently dating back to some past date, is valid, it is updated to another distribution given our current observations.}

In the above example, we consider only observers with exactly our memories at the start of the observation \citep[as in][]{Neal06}. There are two reasons this is allowed despite \citet{Bostrom13}'s objections. First, the indexical weights are averaged instead of being defined on the microhypothesis level, and they describe a location reference class instead of an observer reference class. This lets the vast majority of Boltzmann brains that later experience a dissipating universe to influence the final credence: although a given microhypothesis has some locations with long-lived Boltzmann brains, these are no more favored than locations with dissipating Boltzmann brains. Second, unlike the SSA, {WFG} results in an uneven indexical distribution: we place much more weight in each evolved brain location than the Boltzmann brain locations. This actually results in a faster convergence to the realism of the T cosmology than the SSA.

\section{\texorpdfstring{FGAI and \xday arguments}{FGAI and \textit{x}-day arguments}}
\label{sec:FGAI_xDay}

\subsection{\texorpdfstring{Classes of \xday models in FGAI}{Classes of \textit{x}-day models in FGAI}}
\label{sec:xDay_FineGrainings}
One advantage of underpinning typicality arguments with fine-grained hypotheses is that doing so forces one to use a well-specified model that makes the assumptions explicit. In an \xday Argument, each microhypothesis corresponds to a possible permutation of people born as well as a complete set of possible observations by each observer (whether a human or an effectively independent external observer like an alien).  This entails an observation model, a set of constraints on who can observe whom. In particular, realistic theories of observation are \emph{causal} -- one cannot ``observe'' people living in the future -- and \emph{local} -- one cannot ``observe'' another person without some physical mechanism linking them. 

\begin{figure*}
\centerline{\includegraphics[width=17.2cm]{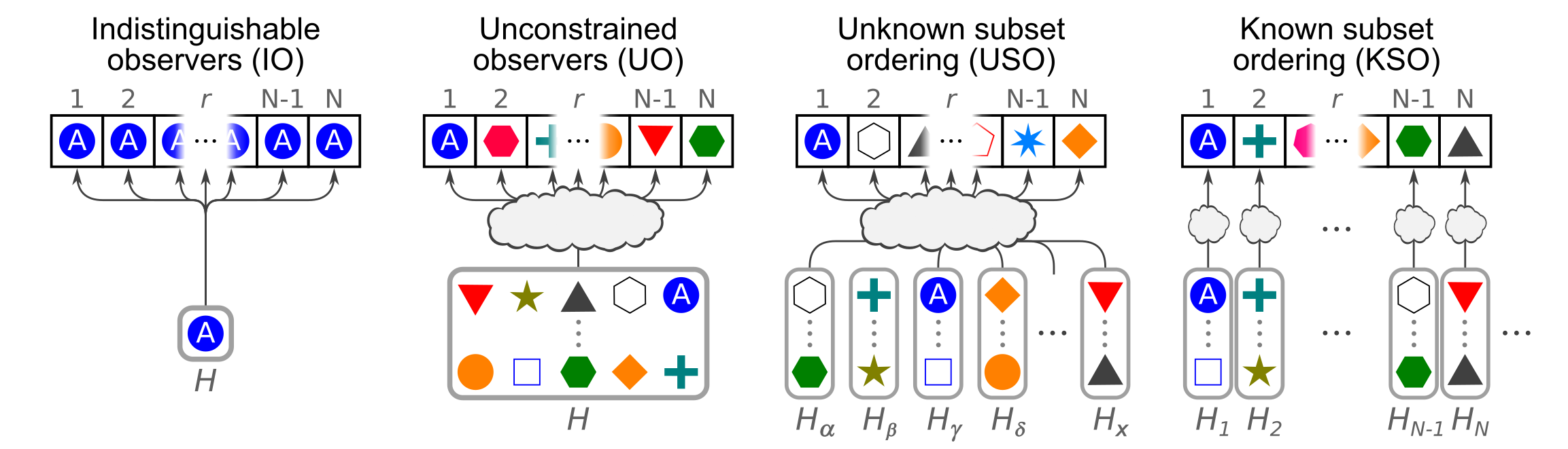}}
\caption{Different fine-grainings of the \xday Argument have different assumptions about the interchangeability of humans. The clouds represent an unknown or random selection from a set. The shown configurations are chosen so that observer $A$ (blue circle) exists with rank $1$, but this is a contingent selection in the models. \label{fig:FGAI_xDay}}
\end{figure*}

I present four general schema of fine-grained \xday theories, resulting in microhypotheses with different combinatorial properties. These fine-grainings are illustrated by a simplified model where we consider only two theories \citep[c.f.,][]{Leslie96}: a Short/Small theory {where all humanities have a} final population $\Nsmall$ and a Long/Large theory with final population $\Nlarge$ {in all humanities}, where $\Nlarge > \Nsmall$. Each human at \xrankText{} $r$ is drawn from a set of possible humans $\HumanSet$, and measures their \xrankText{} to be $r$, specifying the observation model. The possible humans in $\HumanSet$ correspond to different genetic makeups, microbiomes, life histories, memories encoded in the brain, and so on. The details of how humans at rank $r$ are selected from $\HumanSet$ form the basis for each of the different model classes (Figure~\ref{fig:FGAI_xDay}).

The four schemas differ by their restrictions on the possible permutations:
\begin{itemize}
\item[] \textbf{Indistinguishable Observers (IO)} -- Every possible human is treated as identical, with $\HumanSet = \{A\}$. There is only one possible microhypothesis for each theory because the permutations are indistinguishable. Any information about \xrankText{} is treated as purely indexical. IO is essentially the same kind of scenario as SB-B.
\item[] \textbf{Unconstrained Observers (UO)} -- Humans are drawn from a very large set $\HumanSet$ (with $|\HumanSet| = {\cal N}$) of distinguishable observers, and any $h \in \HumanSet$ can be born at any rank $r$. In UO, people with different names, identities, and memories could be treated as distinct members of $\HumanSet$, but these details would no correlation with historical era.
\item[] \textbf{Unconstrained Subset Ordering (USO)} -- The set of possible humans $\HumanSet$ is partitioned into ${\cal R}$ ($\ge \Nlarge$) mutually disjoint subsets $\HumanSetR$, with one for each $r$. Humans with rank $r$ can only be drawn from $\HumanSetR$. This reflects the fact that individuals are the result of a vast constellation of historical circumstances that should never be repeated again. In USO, we can specify the contents of these subsets, but we do not know which subset is assigned to each $r$. In an infinite Universe, every ``copy'' of you will have the same $r$ as you.
\item[] \textbf{Known Subset Ordering (KSO)} -- As in USO, the set of possible humans is partitioned into disjoint subsets $\HumanSetR$, with humans at rank $r$ drawn only from $\HumanSetR$. Unlike USO, we already know beforehand the ordering of these subsets -- which one is $\HumanSetONE$, $\HumanSetTWO$, and so on.  All that remains to be discovered is the final human population $N$ and which humans, in fact, are selected out of each $\HumanSetR$. KSO models emphasize how we already know, by virtue of our historical knowledge, our place in history before applying an \xday Argument.\footnote{The Doomsday Argument yields a ``correct'' result for most people in history, leading \citet{Leslie96} (and with more cavaets, \citealt{Bostrom13}) to conclude that we should use it as well, as should have early humans. This logic fails in the KSO model. Of course, the Doomsday Argument works for most people by construction, but in KSO, it always fails for early humans in their Large future theories. We are not interested in whether Doomsday ``works'' for most people in history, but whether it applies to us \emph{specifically}.}
\end{itemize} 

None of these models exactly corresponds to how we would approach the Doomsday Argument, since we do not actually know the set $\HumanSet$ or exactly how it is partitioned -- these details are implicit for us. But by making explicit models, they illustrate when the Doomsday Argument is appropriate. Of these, KSO arguably is most analogous to our situation when we apply the Doomsday Argument, since humans who know they live in the year 2022 cannot be born in the Paleolithic or an interstellar future (to the extent our memories are reliable). Our Short and Long theories start with our known history, and then propose an additional $\Nfuture$ future people after us, with the likelihood of us existing at birthrank $10^{11}$ being $1$ by assumption. This is not just a tautology because who you are is shaped by your place in history; realistic Small and Large models both predict that you, with all your personality, memories, and beliefs, could only appear this point in history \citep[as in][]{Korb98}.\footnote{All measurements are physical events. When someone learns an indexical fact, they are actually interacting with a physical environment that is location-dependent and changing as a result. Therefore, a truly physical theory cannot regard observers as identical if they have different indexical knowledge; observers with different (and reliable) indexical knowledge are necessarily found at different locations, which makes USO and KSO more physically grounded. For example, hypotheses about individual $A$ of Figure~\ref{fig:FGAI_xDay} really should be fine-grained into hypotheses about $A_1$, who can only exist at rank $1$, and $A_2$, who can only exist at rank $2$, and so on. UO or IO may be regarded as coarse-grainings of these hypotheses.}

\subsection{\texorpdfstring{Self-applied \xday Arguments in FGAI}{Self-applied \textit{x}-day Arguments in FGAI}}
\label{sec:NaiveFGAI_Self}
\label{sec:BigUniverse_Self}
An \xday Argument to constrain final population may be self-applied by a member of the population being constrained (as in the Bayesian Doomsday Argument, or astrobiological Copernican arguments), or applied by an external observer who happens upon the population at some time (analogous to some of \citealt{Gott93}'s examples). Self-applied \xday Arguments are by far more problematic. 

\subsubsection{Fine-graining the self-applied Doomsday Argument when our rank is $1$}
Starting with a ``single-world'' assumption that there is only one humanity in the cosmos, it is relatively simple to calculate the number of microhypotheses in IO, UO, USO, and KSO with combinatorial arguments. These are listed in Table~\ref{table:xdayMicro}, where the observer in question is labeled $A$ and is located at $\xrank = 1$ without loss of generality. We can also consider ``many worlds'' models in which there are many copies of humanity out the larger cosmos, each with different rank permutations of human individuals. This increases the number of microhypotheses, but as in the urn problem, our ignorance of which world we are in and the symmetry between the worlds yields the same results {when our rank is $1$}.

Each macrotheory about $\Ntotal$ has its own set of indexical weights. In the single world versions, the locations (or trajectories for these indexical weights are simply each possible \xrankText{} $1$ to $\Ntotal$. In multiple world versions, there is an location for each \xrankText{} and each world. These weights are initialized to be equal for each possible location. Every human can learn their \xrankText{} but not which of the worlds they live in.

\begin{table*}[!ht]
\tabcolsep4pt
\processtable{Microhypotheses counts for self-applied \xday Arguments in single world models \label{table:xdayMicro}}
{\begin{tabular}{lccccc}
\rowcolor{Theadcolor}
  & IO & \multicolumn{2}{c}{UO} & USO & KSO \\ 
\rowcolor{Theadcolor}
 & & No replacement & With replacement & All $|\HumanSetR|$ equal  & \\
\hline
Total permutations                 & $1$       & $\displaystyle \frac{{\cal N}!}{({\cal N} - N)!}$         & ${\cal N}^N$                    & $\displaystyle {\cal N}_r^N \frac{\MaxRanks!}{(\MaxRanks - N)!}$         & $\displaystyle {\cal N}_r^N$\\
Permutations where $A$ selected    & $1$       & $\displaystyle N \frac{({\cal N} - 1)!}{({\cal N} - N)!}$ & ${\cal N}^N - ({\cal N} - 1)^N$ & $\displaystyle N {\cal N}_r^{N-1} \frac{(\MaxRanks-1)!}{(\MaxRanks-N)!}$ & $\displaystyle {\cal N}_r^{N-1}$\\
Permutations with $A$ at rank $1$  & $1$       & $\displaystyle \frac{({\cal N} - 1)!}{({\cal N} - N)!}$   & ${\cal N}^{N - 1}$              & $\displaystyle {\cal N}_r^{N-1} \frac{(\MaxRanks-1)!}{(\MaxRanks-N)!}$   & $\displaystyle {\cal N}_r^{N-1}$\\
Fraction where $A$ exists          & $1$       & $N / {\cal N}$                                            & $1 - (1 - 1/{\cal N})^N$        & $\displaystyle \frac{N}{{\cal N}_r {\cal R}}$                                & $1/{\cal N}_r$\\
Fraction with $A$ at rank $1$      & $1$       & $1/{\cal N}$                                              & $1/{\cal N}$                    & $1/(\MaxRanks {\cal N}_r)$                                       & $1/{\cal N}_r$
\end{tabular}}
{\begin{tablenotes}
\end{tablenotes}}
\end{table*}

The four schemas all arrive at the ultimate conclusion that the surviving microhypothesis fraction if $A$ measures their rank to be $1$ is independent of $N$. The self-applied \xday Argument is nullified in all of them, although they take different routes to reach this conclusion.

IO is qualitatively equivalent to SB-B. $A$ is the only possible individual, and there only ever is one microhypothesis per theory, which can never be ruled out -- $A$ merely learns indexical information. Learning one has an \xrankText{} of 1 shifts the indexical weights in the Long theory and the credences do not change. In a many world case where all worlds have the same $\Ntotal$, there is uncertainty about which world one is in, but the symmetry results in the same conclusion. If there are many worlds and we consider theories where $\Ntotal$ is assigned randomly, we get a series of microhypotheses about which worlds are Small and which Large. A self-observed \xrankText{} of 1 zeroes out the indexical weights for all positions with $\xrank > 1$ in every world, inhibiting the Doomsday Argument just as for the single-world case.

For UO and USO, there are more locations for $A$ to be born in the Large theory, and thus more microhypotheses $A$ observes themself to exist.\footnote{$A$ having more locations to be born is a \objectiveName{} statement about the number of actual physically occurring trials, with each outcome being a physically distinct history \citep[contra the ``souls waiting to be born'' characterization of SIA in][]{Leslie96,Bostrom13}.} In {WFG}, indexical weights are initialized evenly for all observers, and the observation of one's existence provides no indexical information to change that. Thus, although the amount of prior credence that survives the self-observation is greater by $\Ntotal$ for a single world model, it is weighted by a factor of $1/\Ntotal$, as only that fraction of individuals has that specific self-data. Upon learning your \xrankText{}, large $\Ntotal$ theories are penalized by having fewer microhypotheses survive, but this is compensated by the indexical weights shifting to locate you at your actual \xrankText{}. Thus you can conclude nothing from the self-applied \xday Argument. A worked example is given in the Appendix.\footnote{Actually, USO can be much more complicated if the different $\HumanSetR$ have different sizes.  Then the fraction of surviving microhypotheses (and posterior probabilities) is $N$-dependent if each possible permutation of humans is given equal initial credence within each macrotheory (e.g., if $N = 1$, $\HumanSetA = \{A_a\}$, and $\HumanSetB = \{A_b, B_b\}$, then the microhypotheses of $A_a$ only, $A_b$ only, and $B_b$ only are given equal credence). A more natural weighting is to first divide each macrotheory into ``mesohypotheses'' about the ordering of the $H_r$, and  each permutation microhypothesis inherits an equal fraction of its mesohypothesis's credence (e.g., $A_a$ only would get twice as much initial credence as $A_b$ only).}

\begin{figure*}
\centerline{\includegraphics[width=12.9cm]{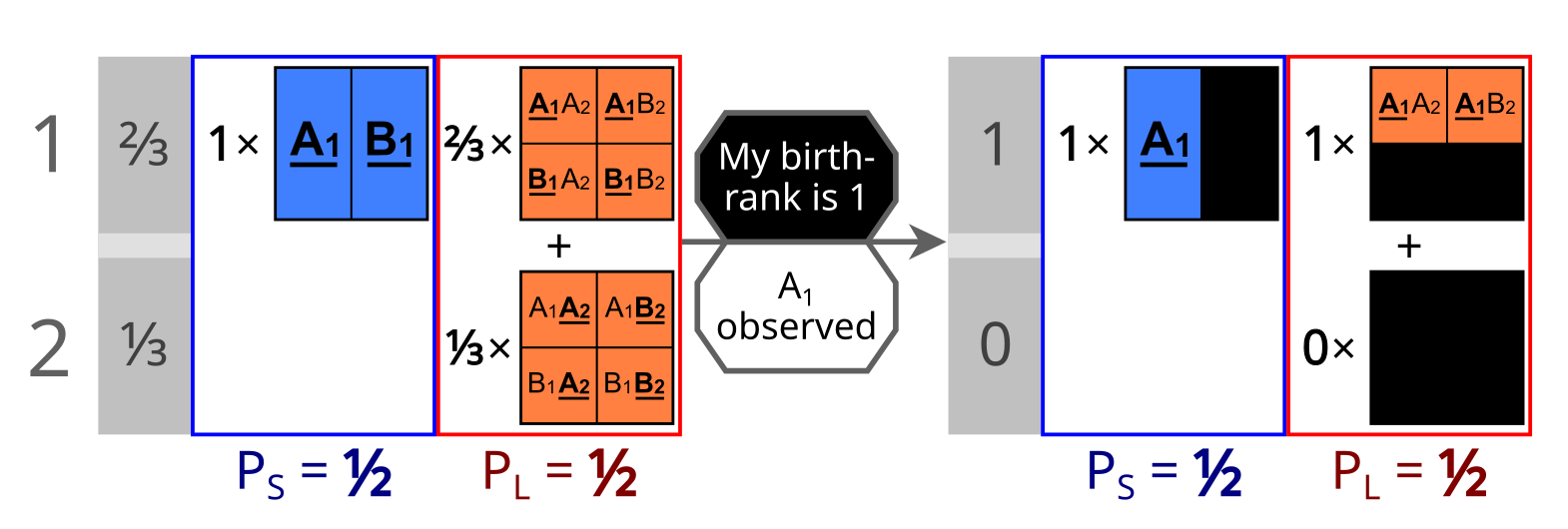}}
\caption{Illustration of how a self-applied Doomsday Argument fails in the KSO schema according to {WFG}. Two models are compared, with $N_S = 1$, $N_L = 2$, $\HumanSetONE = \{A_1, B_1\}$, and $\HumanSetTWO = \{A_2, B_2\}$. The combinatorial properties of the microhypotheses, aided by the shifting indexical weights, ensure the prior is unaffected by an observer learning they are $A_1$ at rank $1$.\label{fig:WeightedFG_KSO}}
\end{figure*}

KSO proceeds more simply because learning your individual identity also allows one to assign your \xrankText{} with certainty in all microhypotheses. The self-observation eliminates the same fraction of prior weight in all theories for observers with your \xrankText{} (Figure~\ref{fig:WeightedFG_KSO}). The self-applied \xday Argument is again rendered powerless.  

\subsubsection{{The anthropic shadow: when our rank is bigger than $1$}}
{According to Large theories about possible futures where we are likely to survive for a long time on Earth, or spread across the stars, we are among the earliest people in our society, but our birthranks are not actually $1$. This can become crucial when judging Small theories. Perhaps it was nigh inevitable that humanity would be annihilated shortly after the invention of nuclear weapons and we are one of the rare exceptions, for example. Does the birthrank of today's newborns serve as evidence against $\Ntotal$ distributions where we are an outlier?

{WFG} enforces a high level of agnosticism about the $\Ntotal$ distribution below our current $\Npresent$ when we first apply the argument. Our initial application of the self-observation of a current \xrankText{} fixes our indexical position in each world -- really it selects a class of observer-trajectories that \emph{start} at $\xrank$. The result is that any microhypothesis where there is at least one world that is large enough for $\Ntotal \ge \xrank$ survives unscathed. The only nonzero indexical weights are for our \xrankText{} in those worlds that have them, all of which are completely compatible with our \xrankText{}. This is the ``anthropic shadow'' effect of \citet{Cirkovic10}. Curiously, this leads to the credence in the distribution depending on the number of worlds, because with more worlds there is more likely to be one that survives long enough for our self-observation. In the limit that there are an infinite number of worlds, any distribution with nonzero probability of a world surviving as long as ours has a likelihood of $1$.

After that initial self-observation, however, there is no more freedom to shift the indexical weights. Our observer-trajectory has a fixed beginning. As the world continues to survive, only some fraction of worlds last. The provisional credence for worlds that are destroyed in a microhypothesis is zeroed out by our continued survival, reducing our overall credence in the microhypothesis and its parent distribution theory. {WFG} conditionalizes the credences on our initial \xrankText{}, still allowing us to constrain Small theories by their large $\Ntotal$ tails.

This raises the question of which \xrankText{}-trajectory we should use. Science is a collective enterprise, so when engaging in it we should not all be using our individual \xrankText{} \citep[c.f. the discussion in][]{Garriga08}. Future generations can use our \xrankText{} because the indexical weights are for trajectories which can start before they were born. We might ``start the clock'' at some date corresponding to anywhere from the earliest time in cultural memory to the invention of the Doomsday Argument, or perhaps the beginning of humanity itself (or even earlier!). This is an open issue.}

\subsubsection{{Why the self-applied Doomsday Argument fails}}
{WFG}, with its emphasis on observations as physical events, requires exactly specified (if simplified) models instead of simply relying on analogies. Purely indexical problems, like SB-A or Leslie's ``emerald'' thought experiment, do not accurately model the Doomsday Argument.  Nor do purely physical problems like SB-C, urn problems, or the external Doomsday Argument (Figure~\ref{fig:WeightedSB}).  The self-applied Doomsday Argument never yields new information within {WFG} because:
\begin{itemize}
\item \emph{Indexical facts do not directly constrain \objectiveName{} distributions} -- {In {WFG},} the {indexical distributions emerge from the independently updating provisional credences}, shrink{ing the} reference class {and} the power of the $1/\Ntotal$ likelihoods, averting solipsism.
\item \emph{Large worlds have more positions for specific individuals to be born, balanced by the uninformative weights} -- In UO and USO, a greater fraction of physical microhypotheses is consistent with the existence of any specific individual in a Large world. Yet the smaller weights accorded to each position prevent any ``Presumptuous Philosopher'' problems.
\item \emph{Physical observations obey causality, limiting outcomes} -- KSO provides an explicit model where one \emph{cannot} treat the birthrank of yourself, a specific individual, as a uniform random variable. Physical distributions in themselves are updated solely by \objectiveName{} observations, which are constrained by causality. If an individual like you can only be born at birthrank $10^{11}$, there is only one possible outcome, with likelihood $1$.
\end{itemize}
The rejection of the Doomsday Argument itself does not negate the many potential threats to ourselves and the Earth that we know about, especially when considering anthropic effects \citep{Leslie96,Cirkovic10}. Evaluating these threats could itself shift our beliefs to a short future, but this must be done through normal scientific investigation of their natures.

\subsubsection{Implications for Copernican arguments in astrobiology}
If we accept {WFG} as a theory of typicality, then the same arguments against self-applied \xday Arguments work against self-applied Copernican Arguments in astrobiology. Instead of birthranks, we may have habitats, and we are interested in constraining the existence of as-yet unobservable beings in other locations. But the number of microhypotheses where one individual specifically exists on Earth will be independent of the number of inhabited locations. The existence of exotic forms of alien life multiply the number of microhypotheses, but do not change the probability ratios.

Thus in {WFG}, we can conclude nothing about the existence of alien life in un-Earthlike locations from our own existence on Earth, as required by \citet{Hartle07}. We must either constrain them based on physical plausibility or by observation, just like any other astrophysical phenomenon.

\subsection{\texorpdfstring{Externally-applied \xday arguments in FGAI}{Externally-applied \textit{x}-day arguments in FGAI}}
\label{sec:FGAI_Extern}

Externally-applied \xday Arguments are structurally similar to ``urn'' thought experiments, while seeming to also justify the self-applied Doomsday Argument by analogy. Under some circumstances, one can infer something about the lifespan distribution of an external phenomenon by drawing a sample of it at random (as in the most popularized examples of \citealt{Gott93}). But the key difference with self-applied Doomsday arguments is the existence of an additional ``observer'' that is the one carrying out the experiment, who is completely outside the population being constrained. This adds another physical degree of freedom, which changes the microhypothesis counts in FGAI to allow for a Doomsday Argument.

Suppose an observer $X$ happens upon humanity, with individuals drawn from $\HumanSet$ according to IO, UO, USO, or KSO.  All of the possible sequences of human individuals listed by the microhypotheses still are valid, and would have the same relative prior probabilities. Yet there is a final observable quantity, which is actually what $X$ measures: the relative time $t$ between humanity's start and $X$ happening upon us. Even if the evolution of $X$ is deterministic enough to demand they arrive at a specific time, $X$ does not have enough information beforehand to determine a particular $t$; presumably such factors are microscopic and chaotic. Therefore, every single one of the microhypotheses of the previous section must now be divided into $T$ even finer microhypotheses, one for each possible $t$ interval $X$ can arrive in.   

Whether a Doomsday Argument applies depends on the selection process, in particular how $X$ found humanity. If $X$ could plausibly have arrived before or after humanity's existence, then $T$ is both much larger than humanity's lifespan $\ttotal$ (or its equivalent) and also independent of $\ttotal$. This selection method is quite common: if one looks for examples of a rare phenomenon at a particular location, then most likely one will be searching in the vast epochs before or after it passes. It is in fact less of a coincidence for $X$ to arrive at the right time to observe humanity if $N$ is large, favoring a Large history -- the fraction of surviving microhypotheses scales as $\sim \ttotal/T$. This is well known in SETI, where the abundance of currently observable extraterrestrial intelligences scales directly with their mean lifespan in the Drake equation. Now suppose $X$ measures the age of humanity. The fraction of microhypotheses consistent with $X$ arriving at this particular epoch of human history ($\sim 1/T$) is the same in Small and Large models. Thus there is no Doomsday Argument as such; $X$ cannot constrain humanity's lifespan.\footnote{For this reason, if we specifically observe Proxima b and discover a very young technological society, we are not justified in concluding they typically live for a short period before intelligence goes extinct on a planet forever, because of the sheer improbability of making that discovery in \emph{any} theory. We could legitimately suspect that every planet gives rise to a long procession of many such short-lived societies, however.}

But what if there is no possibility of $X$ failing to make an observation of humanity? One could suppose that rather than searching a particular location, $X$ seeks out the nearest technological society, scouring the infinite cosmos if they have to until they find one. (We will also assume all such societies have exactly the same lifespan for the sake of argument.) This would be more like picking someone on the street to ask them their age. The only possible values of $t$ are values within humanity's lifespan -- each Large history theory has a larger number of possible outcomes. The microhypothesis counts in Table~\ref{table:xdayMicro} are multiplied by $N$ to account for the extra degree of freedom in $t$. Only $\sim 1/\ttotal$ of them are consistent with any particular measurement of humanity's current age.

$X$ may assume there are several possible locations they can arrive at, one for each moment in human history, and create a provisional physical distribution for each. These provisional physical distributions should be constructed as in SB-C: they should favor Small theories at a ratio proportional to $1/\ttotal$. This is initially compensated by additional sets of microhypotheses in Large theories where $X$ arrives at later times. Note that in each microhypothesis there is only one possible location for $X$, thus an indexical weight of $1$ is applied to all standing microhypothesis. $X$ learning that they have arrived at an earlier time rules out microhypotheses in Long theories where they arrive late in history, but the indexical weights do not shift. Thus, $X$ now favors Small histories, and the externally-applied Bayesian Doomsday argument is in effect. Alternatively, $X$ may work under the assumption that their location is fixed, and that it is the start of human history that is the free quantity, to get the same result.

And what if $X$ then announces their finding to Earth -- should we agree that humanity's lifespan is short?  Actually, we do not have enough information to tell. It would be strong evidence if $X$ is the only entity carrying out a survey like this, or if there is one or more target society per survey.  If there are many such surveys/observers, however, then we expect an observer like $X$ to show up frequently, even in the early epochs of a Large history.  In fact, $X$ being the first such observer implies our history is Long -- otherwise, the multitudes of observers like $X$ destined to choose our Earth would have had to squeeze into our short historical epoch before we went extinct.

These are not the only possible models for the time of $X$'s arrival. Actual observations may deliberately be carried out at a specific time during its history, limiting the microhypotheses about $t$ to a narrow range independent of $\ttotal$ \citep[as in the wedding example of][]{Gott97-Grim}. Furthermore, the age of the Universe sets an upper limit on the relative $|t|$.  We cannot rule out astrophysical objects having lifespans of trillions of years -- as we suspect most do \citep{Adams97}. Finally, the observational outcomes will be biased if the population varies with time. If we survey an exponentially growing population of human artifacts, for example, \citet{Gott93}'s Doomsday Argument necessarily underpredicts the lifespan.

Thus, an observation of a single ETI, for example, does give us information about the distribution of ETI properties {\citep{Madore10}}. We must be mindful of selection biases; in a volume-limited survey, we are more likely to observe a long-lived society, and we will only find life that we can recognize in places we look. In an unbiased survey, however, we can favor theories where a small sample is more typical.

\section{Conclusion}
\label{sec:Conclusion}
\subsection{Summary of WFG}
I have developed {WFG} as a framework for interpreting typicality arguments. It shares elements with other frameworks: the auxiliary information about reference classes in \citet{Bostrom13}, the use of extremely fine-grained data about observers in \citet{Neal06}, and the {distinction between indexical and \objectiveName{} questions from} \citet{Srednicki10} {and \citet{Garisto20}}. It is based on two principles.

The first principle I have argued is that purely indexical facts cannot directly alter \objectiveName{} credence distributions. Indexical facts and physical facts are different types of data, and mixing them together into a joint prior leads to ``Presumptuous Philosopher'' problems, biasing us towards Large worlds or Small worlds. The difference between indexicals and physical facts is elaborated by different variants of the Sleeping Beauty thought experiment, where the probability refers to different distributions. Yet both types of distributions always exist, even when perfect knowledge about the world or our location in it hides one. {In {WFG}, the provisional credences are evaluated according to the likelihood of an observer at a specified position making a specific observation, a physical event. The indexical distribution emerges as a kind of ``projection'' of these provisional credences, and the evaluated \objectiveName{} credences are a different projection using an averaged indexical distribution as weights.} 

The second principle is that high-level macrotheories about the world can be resolved into a multitude of physically-distinct microhypotheses. The consistency of evidence with each microhypothesis is calculated through physical distributions. In {WFG}, we do not necessarily know which observer we are, so there are provisional physical distributions for every possible trajectory we could be located along in a theory. These are averaged by the indexical weights, summarizing our current beliefs about our location. By defining reference classes by location rather than observer, observations in a large universe can nonetheless be constraining. Although updated separately, they are combined together to form a single physical credence and single indexical distribution for each microhypothesis. Our third-person credence in a macrotheory is then found by summing over all microhypotheses.

I have showed how {WFG} handles several problems involving ``typicality'', avoiding fallacious Doomsday Arguments. Several issues remain. First, it is not immune to Bayesian shifts favoring Large theories (for example, in SB-D) unless ``virtual observers'' are inserted into theories, trading physical credence with additional indexical weights. Further issues stem from the need to understand how indexical information is defined and handled. In particular, the indexical weights must be {averaged} between multiple microhypotheses, {nominally all of them}. Finally, the formulation in this paper assumes a finite number of possible observer locations. Nonetheless, it avoids solipsism and reduces to simple Bayesian updating when either indexical or physical data is fixed.

\subsection{The role of the Copernican Principle in WFG}
\label{sec:CopernicanRole}

The Copernican insight that we are not unique miracles is deeply ingrained, but not all applications of the idea that ``we are typical'' are defensible. The parable of the Noonday Argument describes a fallacious application of the principle, demonstrating the problem with the frequentist Doomsday Argument or trying to pick out ``special'' points of history like its beginning. Other Copernican arguments can have more subtle flaws.

Bayesian typicality arguments {about the distribution of observers} rely on $1/\Ntotal$ likelihoods: the more possible outcomes there are, the less likely any specific one is observed. Although valid in some circumstances, applying it to rule out Large theories {in general} is unacceptable -- it can lead to solipsism, in which you are the most typical being because you are the only being{, negating science as an endeavour}. Whereas the Copernican principle motivates us to consider the possibilities of a cosmos where we are living in but one world of many, unrestricted application leads to an epistemic arrogance where you decide your world is the only kind of thing that exists. Yet the mere existence of the environments unlike ours already invalidates the strongest formulations of the Copernican argument: we are either atypical because we are not the only observers that exist, or we live in an atypical place because it contains observers unlike everywhere else in the Universe.

{WFG} is my attempt to deal with how to make inferences in a vast cosmos, and it provides two manifestations of the Copernican Principle. First is an indexical role as a prior, choosing the initial {provisional credences for each microhypothesis}. This role is not trivial because the indexical weights are {averages of these} between microhypotheses, which penalizes theories where our location is both rare and random.

The second manifestation of the Copernican Principle is embodied by our \objectiveName{} distributions. {WFG} fine-grains physical theories into microhypotheses describing the exact physical details of every possible outcome. These microhypotheses make no reference to indexical notions like ``me'' or ``us'', only specific physical observers who make specific physical observations at specific physical positions. Observer-relative typicality emerges from the combinatorial properties of these microhypotheses when there is symmetry with respect to position. In other cases, when specific individuals can only exist in certain locations, typicality can fail -- this is the anthropic principle at work. Ultimately, the reason why we may often assume ourselves to be typical is because of the uniformity of the physical laws of the cosmos, manifesting both in the symmetry and the panoply of microhypotheses. 

\section*{Conflict of Interest}
The author reports no conflicts of interest.

\ack[Acknowledgements]{I thank S. Jay Olson for comments on an earlier version of this paper. {I also thank the referees for their comments.} I acknowledge the Breakthrough Listen program for their support. Funding for Breakthrough Listen research is sponsored by the Breakthrough Prize Foundation (\url{https://breakthroughprize.org/}). }

\appendix

\section{Explicit Doomsday problems in WFG}
\label{sec:DoomsdayExample}

\setcounter{table}{1}

In this {appendix}, I work out explicit example{s} of Doomsday-type problem{s} in {WFG}. Although extremely simple {models are presented}, {they} nonetheless illustrates several points: Doomsday failing for internal observers while working for an external observer, how to handle multiple ``worlds'' of humanity, and how to treat hierarchical models in USO with unequal prior credence in the microhypotheses.

\subsection{{The Doomday Argument in a one-world USO model}}
{This} model consists of {only one world (or, alternatively, a series of worlds that are exactly identical due to deterministic evolution). The} world contains a short sequence of human observers. {We wish to compare two extreme hypotheses for the distribution of $\Ntotal$.} In the Small model, $\Ntotal = 1$ for both, while in the Large model, $\Ntotal = 2$ for both. These humans are drawn from two subsets, $\HumanSetA = \{A_a\}$ and $\HumanSetB = \{A_b, B_b\}$, of unequal size. This is a USO model: due to unknown historical contingencies, $\HumanSetA$ humans can only be born at one of birthrank $1$ or $2$, and likewise for $H_b$. In addition, an external alien observer $X$ who knows all this deliberately seeks out {this unique instance of} humanity, but not knowing {their arrival time} beforehand.  Each human observes which member of $\HumanSet \equiv \HumanSetA \cup \HumanSetB$ they are and measures their birthrank accurately, but cannot observe humans at {the} other birthrank. Likewise, if $X$ is present at that time, they measure which human is present and what their birthrank is.

The microhypotheses in this problem are listed in Table~\ref{table:IODoomsdayExample}. These microhypotheses indicate the birthrank order of $\HumanSetA$ and $\HumanSetB$, the exact sequence of {individuals in the} world, and the human observed by $X$, denoted by $\obsd{Y}$. The arrival of $X$ is a physical event and thus different locations of $X$ are properly treated as physical microhypotheses. This example is somewhat complicated because $\HumanSetA$ and $\HumanSetB$ are different size. Thus permutations with only $\HumanSetB$ are more numerous than those with $\HumanSetA$. The correct approach is to regard the ordering of $a$ and $b$ as intermediate-level parameters: it is just as likely that $\HumanSetA$ has rank 1 as $\HumanSetB$ has rank 1. Thus Small microhypotheses with only $b$ humans have lower $\ProbPriorKJ$ than those with only $a$ humans for a balanced credence. In the Table, quantities referring to humans at different birthranks within a world are separated by a comma. 

\begin{table*}[!ht]
\tabcolsep4pt
\processtable{Microhypotheses for {one}-world USO Doomsday problem \label{table:USODoomsdayExample}}
{\footnotesize \begin{tabular}{ccc|c|cccc|cccc}
\rowcolor{Theadcolor}
$\Ntotal$ & Order       & $\MicrohCore$               & $\ProbPriorKJ$ & \multicolumn{4}{c}{Human observation likelihoods} & \multicolumn{4}{c}{External observation likelihoods}\\
\rowcolor{Theadcolor}
          &             &                             &           & $o = A_a$     & $o = A_a @ 1$ & $o = A_b$     & $o = A_b @ 1$ & $A_a$ & $A_a @ 1$ & $A_b$ & $A_b @ 1$\\
\hline       
$1$       & $a$         & $\obsd{A_a}$                & $1/4$     & $1$       & $1$       & $0$       & $0$       & $1$ & $1$ & $0$ & $0$\\
          & $b$         & $\obsd{A_b}$                & $1/8$     & $0$       & $0$       & $1$       & $1$       & $0$ & $0$ & $1$ & $1$\\
          &             & $\obsd{B_b}$                & $1/8$     & $0$       & $0$       & $1$       & $1$       & $0$ & $0$ & $1$ & $1$\\
\hline
$2$       & $ab$        & $\obsd{A_a}, A_b$           & $1/16$    & $1, 0$    & $1, 0$    & $0, 1$    & $0, 0$    & $1$ & $1$ & $0$ & $0$\\
					&             & $A_a, \obsd{A_b}$           & $1/16$    & $1, 0$    & $1, 0$    & $0, 1$    & $0, 0$    & $0$ & $0$ & $1$ & $0$\\
					&             & $\obsd{A_a}, B_b$           & $1/16$    & $1, 0$    & $1, 0$    & $0, 1$    & $0, 0$    & $1$ & $1$ & $0$ & $0$\\
					&             & $A_a, \obsd{B_b}$           & $1/16$    & $1, 0$    & $1, 0$    & $0, 1$    & $0, 0$    & $0$ & $0$ & $0$ & $0$\\
					& $ba$        & $\obsd{A_b}, A_a$           & $1/16$    & $0, 1$    & $0, 0$    & $1, 0$    & $1, 0$    & $0$ & $0$ & $1$ & $1$\\
					&             & $A_b, \obsd{A_a}$           & $1/16$    & $0, 1$    & $0, 0$    & $1, 0$    & $1, 0$    & $1$ & $0$ & $0$ & $0$\\
				  &             & $\obsd{B_b}, A_a$           & $1/16$    & $0, 1$    & $0, 0$    & $0, 0$    & $0, 0$    & $0$ & $0$ & $0$ & $0$\\
					&             & $B_b, \obsd{A_a}$           & $1/16$    & $0, 1$    & $0, 0$    & $0, 0$    & $0, 0$    & $1$ & $0$ & $0$ & $0$\\
\end{tabular}}
{\begin{tablenotes}
{The order column indicates lists the relative order of $\HumanSetA$ and $\HumanSetB$, with $ab$ meaning $\HumanSetA = \HumanSetONE$ and $\HumanSetB = \HumanSetTWO$ and $ba$ meaning the reverse. The $\MicrohCore$ column lists the microhypotheses: each possible sequence of humans drawn from $\HumanSetA$ and/or $\HumanSetB$ along with which is observed by the external observer $X$. The human observation likelihoods indicate the probability that an observer $o$ observes themself to be a particular human ($A_a$ or $A_b$), possibly at a specified rank, according to each microhypotheses. The external observation likelihoods indicate the probability that $X$ observes a particular human when they arrive, and possibly measure their birthrank to be $1$.} 
\end{tablenotes}}
\end{table*}

\emph{Internal observers} -- You are a human and you have worked out this model exactly. You wish to use a Doomsday Argument by determining which human in $H$ you are and also your birthrank. To do this in {WFG}, you {must establish provisional prior credences for each microhypothesis at each possible location or trajectory.} Since you know you are human and not the alien, {these locations are birthrank $1$ and birthrank $2$ in the Large model and only birthrank $1$ in the Small model. Now, if you were already certain of which microhypothesis is true, you would want to have an uninformative indexical prior. Therefore,} $\ProvPriorKJI = \ProbPriorKJ$. {With the provisional prior credences in hand, you can calculate your indexical weights as $\WeightPriorONE = 2/3$ and $\WeightPriorTWO = 1/3$ (equation~\ref{eqn:WFGAI_IndexicalDist})}. Columns 5--8 {then} give likelihoods that a human at each position will make a given observation.

\begin{figure*}
\includegraphics[width=17.2cm]{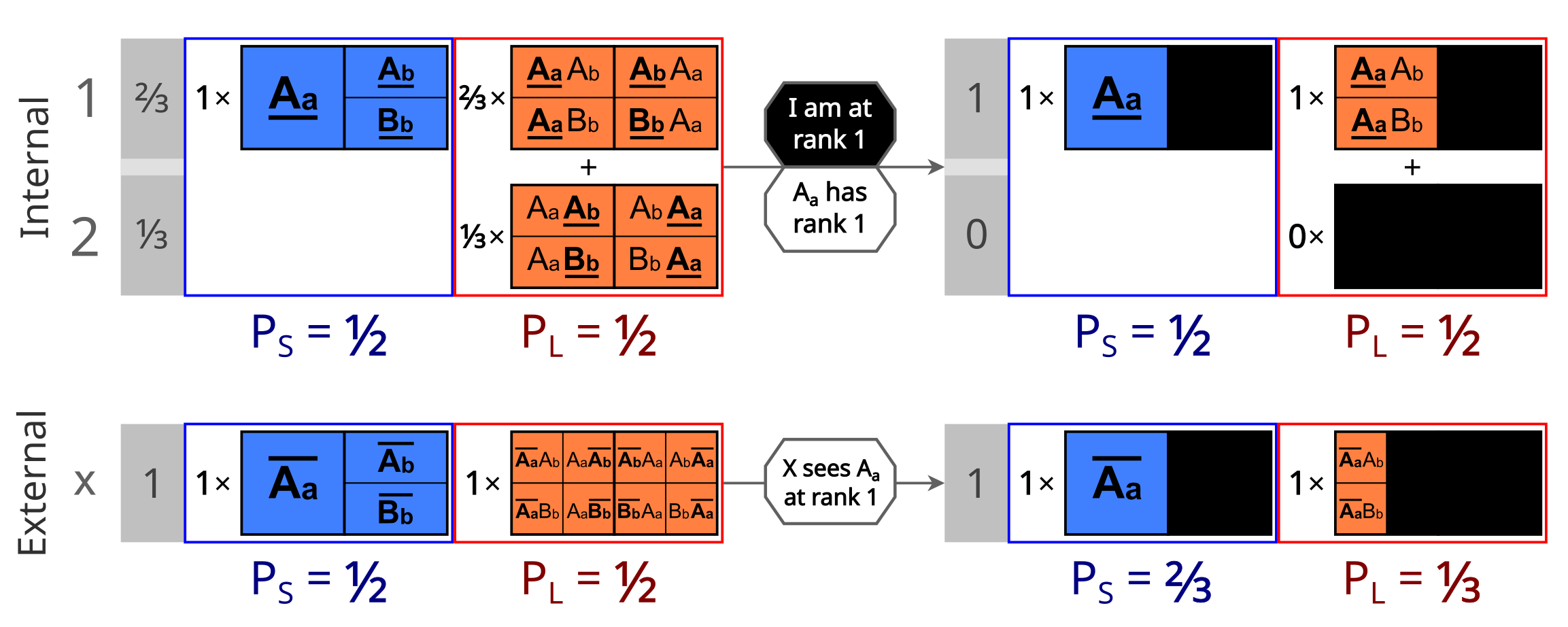}
\caption{Illustration of the self-applied {(top) and external (bottom)} Doomsday Argument {USO} example in {WFG} when $A_a$ is observed at rank $1$. The {observed} human is bold {and underlined (self-observation) or overlined (external observation)}. {For the self-applied argument,} the distinctions about $X$'s position are ignored, where each shown microhypothesis stands for {one} (Short) to {two} (Long) from Table~\ref{table:USODoomsdayExample}. {For the external argument}, $X$'s location is fixed. The case where the {epoch of humanity's start is} fixed proceeds similarly, with the {two} locations splitting the microhypotheses between them for their provisional distributions. This is an example where $X$ can apply the Copernican Argument to make inferences -- note that $X$ has no possibility of failing to observe humanity in this example.\label{fig:DoomsdayExample_WFGAI_External}\label{fig:DoomsdayExample_WFGAI_USO}}
\end{figure*}

Now suppose you observe that you are $A_a$. Half of the {provisional credence at each position is zeroed out because of incompatible microhypotheses. For the other half, the provisional credence is updated to 0. The weights remain unchanged ($\WeightPosONE = 2/3$, $\WeightPosTWO = 1/3$).} The credence in the Small theory updates to {$(1 \times 1/4 + 2 \times 1 \times 0) / [(1 \times 1/4 + 2 \times 1 \times 0) + (4 \times (2/3 \times 1/16 + 1/3 \times 0) + 4 \times (2/3 \times 0 + 1/3 \times 1/16))] = (1/4) / [(1/4) + (4/16)] = 1/2$.} Next, you happen to measure your birthrank to be $1$. In all microhypotheses in both theories, this is only consistent with {the observer} having a birthrank of $1$, {so all provisional credences for birthrank $2$ are zeroed out. The indexical weights update to $\WeightPosONE = 1$ and $\WeightPosTWO = 0$.} Using the likelihoods in the Table, the credence in the Small theory updates to {$(1 \times 1/4 + 2 \times 1 \times 0)/[(1 \times 1/4 + 2 \times 1 \times 0) + (4 \times (1 \times 1/16 + 0 \times 0) + 4 \times (1 \times 0 + 0 \times 0))] = (1/4) / [(1/4) + (4/16)] = 1/2$.} The Doomsday Argument has failed for $A_a$ because of the shifting indexical weights (Figure~\ref{fig:DoomsdayExample_WFGAI_USO}).

The Doomsday Argument also fails for $A_b$ (and $B_b$ by symmetry) if they attempt it. Upon discovering they are $A_b$, {the weights remain unchanged ($\WeightPosONE = 2/3$ and $\WeightPosTWO = 1/3$) as does} the credence in the Small theory{, $(1 \times 0 + 1 \times 0 + 1 \times 1/8) / [(1 \times 0 + 1 \times 0 + 1 \times 1/8) + (2 \times (2/3 \times 0 + 1/3 \times 0) + 2 \times (2/3 \times 0 + 1/3 \times 1/16) + 2 \times (2/3 \times 1/16 + 1/3 \times 0) + 2 \times (2/3 \times 0 + 1/3 \times 0)] = (1/8) / [(1/8) + (2/16)] = 1/2$.} If $A_b$ finds themself at birthrank 1, the Large theory indexical weights {shift as the provisional credences update}, and the credence in the Small theory is {$(1 \times 0 + 1 \times 0 + 1 \times 1/8) / [(1 \times 0 + 1 \times 0 + 1 \times 1/8) + (2 \times (1 \times 0 + 0 \times 0) + 2 \times (1 \times 0 + 0 \times 1/16) + 2 \times (1 \times 1/16 + 0 \times 0) + 2 \times (1 \times 0 + 0 \times 0)] = (1/8) / [(1/8) + (2/16)] = 1/2$.}

\emph{External observers} -- You are the alien observer and you have worked out this model exactly. You wish to use a Doomsday Argument based on your observations of who is currently living and their birthrank. You are the only external observer. There are {two} possible relative locations where you may observe humanity{, the two possible birthranks.} You may use Earth-relative locations, in which case there {are two} possible positions. Alternatively, you may assume that your location is fixed at $x$, and these positions correspond to hypotheses about the {epoch} of humanity's origin. The different possible locations of $X$ -- a physical distinction -- also correspond to different groups of microhypotheses, each with only one indexical weight (Figure~\ref{fig:DoomsdayExample_WFGAI_USO}). With only one possible position per microhypothesis, $\ProvPriorKJI = \ProbPriorKJ$ in each case. Columns 9--12 give likelihoods for the observations you may make.

Suppose you {treat humanity's origin as uncertain while your own location is fixed}, and your sample reveals $A_a$ from $\HumanSetA$. There is only one indexical weight, and one provisional physical distribution following the $\ProbPriorKJ$ listed in the Table. {{WFG} reduces to Bayesian updating in this case. The observation of $A_a$ in the sample} does not by itself affect your credence in the Small theory: {$(1/4 + 2 \times 0) / [(1/4 + 2 \times 0) + (4 \times 1/16 + 12 \times 0)] = (1/4) / [(1/4) + (4/16)] = 1/2$.} But if you measure $A_a$'s birthrank to be $1$, that credence updates to {$(1/4 + 2 \times 0) / [(1/4 + 2 \times 0) + (2 \times 1/16 + 14 \times 0)] = (1/4) / [(1/4) + (2/16)] = 2/3$.} The same results are found if you observe $A_b$ at rank $1$: first, a Small credence of {$(0 + 1/8 + 0) / [(0 + 1/8 + 0) + (2 \times 1/16 + 14 \times 0)] = (1/8) / [(1/8) + (2/16)] = 1/2$.} when you observe $A_b$, {updated to $(0 + 1/8 + 0) / [(0 + 1/8 + 0) + (1 \times 1/16 + 15 \times 0)] = (1/8) / [(1/8) + (1/16)] = 2/3$ when measuring $A_b$'s birthrank of $1$.} For you, the Doomsday Argument is validated by {WFG}.

If you instead treat humanity's origin as {fixed} while your own location is {uncertain}, these same conclusions are reached. There {are two indexical weights, but only one of them is in $\OSetKJ$ for each microhypothesis, since each microhypothesis specifies which location you observe. Thus all the relevant weights remain $1$, and} an observation of $A_a$ at rank 1 leads to a credence in Small of $2/3$.

The difference between the internal and external cases results from two factors. First, there is an extra temporal degree of freedom for $X$'s arrival. Second, $X$'s position is entirely fixed in each microhypothesis -- knowing they are the external observer, $X$ can predict exactly who they will observe simply based on time of arrival, since the time of arrival is entirely a physical statement. In contrast, there is no single ``time of observation'' for the humans in Large model, since one observes early and one observes late. For the internal case, all microhypotheses are present in the Long provisional physical distributions, because there are observers at all locations. \emph{A priori}, the internal observer places equal provisional prior credence in the Short and Long theories for the rank $1$ positions. But in the external case, the rank $1$ and rank $2$ positions constrain completely different Long microhypotheses, splitting the Long prior credence between them. \emph{A priori}, the external observer places twice as much provisional prior credence in Short than Long for the rank $1$ positions. This is illustrated in Figure~\ref{fig:DoomsdayExample_WFGAI_External}.

\subsection{{A case with two worlds}}
{In a large enough universe, there is likely to be multiple instances of humanity (or technological societies in general). These may have varying properties, increasing the number of microhypotheses exponentially. A larger number of worlds makes it more likely that any particular viable world in a theory is realized, but otherwise the worlds may not be in contact with each other. To demonstrate how to handle cases with more than one world, I present a simple indistinguishable observer (IO) model with two worlds. These are labeled $\WorldONE$ and $\WorldTWO$, with positions/trajectories consisting of a world and a birthrank. This time there will be three possible distributions for $\Ntotal$ that the observers will try to compare: a Small theory in which all worlds have $\Ntotal = 1$; a Large theory in which all worlds have $\Ntotal = 2$; and an Intermediate ($I$) theory in which each world has even probability of being small or large independent of the other. All three macrotheories start with equal credence ($\ProbPrior(S) = \ProbPrior(I) = \ProbPrior(L) = 1/3$). Because of the indistinguishability of the observers, there is only one possible pair of sequences each in the Small and Large theories, and four in the Intermediate theory corresponding to the four possible choices for $\Ntotal(\WorldONE)$ and $\Ntotal(\WorldTWO)$. Table~\ref{table:IODoomsdayExample} lists all the microhypotheses in this scenario, with $\|$ separating observable and likelihoods of worlds $\WorldONE$ and $\WorldTWO$.}

\begin{table*}[!ht]
\tabcolsep4pt
\processtable{Microhypotheses for two-world IO Doomsday problem \label{table:IODoomsdayExample}}
{\footnotesize \begin{tabular}{lcc|c|ccc|cc}
\rowcolor{Theadcolor}
Theory & $\Ntotal$ & $\MicrohCore$                                     & $\ProbPriorKJ$ & \multicolumn{3}{c}{Human observation likelihoods} & \multicolumn{2}{c}{External observation likelihoods}\\
\rowcolor{Theadcolor}
             &           &                                           &                & $o @ (w, 1)$        & $o @ (w, 2)$        & $o @ (w,1) \to 2$ & Rank $1$ & Rank $2$\\
\hline
Small        & $1 \| 1$  & $\obsd{o_{\WorldONE,1}}\|o_{\WorldTWO,1}$                 & $1/6$          & $1 \| 1$       & $0 \| 0$       & $0 \| 0$      & $1$      & $0$    \\
             &           & $o_{\WorldONE,1}\|\obsd{o_{\WorldTWO,1}}$                 & $1/6$          & $1 \| 1$       & $0 \| 0$       & $0 \| 0$      & $0$      & $1$    \\
Intermediate & $1 \| 1$  & $\obsd{o_{\WorldONE,1}}\|o_{\WorldTWO,1}$                 & $1/24$         & $1 \| 1$       & $0 \| 0$       & $0 \| 0$      & $1$      & $0$    \\
             &           & $o_{\WorldONE,1}\|\obsd{o_{\WorldTWO,1}}$                 & $1/24$         & $1 \| 1$       & $0 \| 0$       & $0 \| 0$      & $0$      & $1$    \\
			       & $1 \| 2$  & $\obsd{o_{\WorldONE,1}}\|o_{\WorldTWO,1},o_{\WorldTWO,2}$ & $1/36$         & $1 \| 1, 0$    & $0 \| 0, 1$    & $0 \| 1$      & $1$      & $0$    \\
			       &           & $o_{\WorldONE,1}\|\obsd{o_{\WorldTWO,1}},o_{\WorldTWO,2}$ & $1/36$         & $1 \| 1, 0$    & $0 \| 0, 1$    & $0 \| 1$      & $1$      & $0$    \\
			       &           & $o_{\WorldONE,1}\|o_{\WorldTWO,1},\obsd{o_{\WorldTWO,2}}$ & $1/36$         & $1 \| 1, 0$    & $0 \| 0, 1$    & $0 \| 1$      & $0$      & $1$    \\									
			       & $2 \| 1$  & $\obsd{o_{\WorldONE,1}},o_{\WorldONE,2}\|o_{\WorldTWO,1}$ & $1/36$         & $1, 0 \| 1$    & $0, 1 \| 0$    & $1 \| 0$      & $1$      & $0$    \\
			       &           & $o_{\WorldONE,1},\obsd{o_{\WorldONE,2}}\|o_{\WorldTWO,1}$ & $1/36$         & $1, 0 \| 1$    & $0, 1 \| 0$    & $1 \| 0$      & $0$      & $1$    \\
			       &           & $o_{\WorldONE,1},o_{\WorldONE,2}\|\obsd{o_{\WorldTWO,1}}$ & $1/36$         & $1, 0 \| 1$    & $0, 1 \| 0$    & $1 \| 0$      & $1$      & $0$    \\
						 & $2 \| 2$  & $\obsd{o_{\WorldONE,1}},o_{\WorldONE,2}\|o_{\WorldTWO,1},o_{\WorldTWO,2}$ & $1/48$         & $1, 0 \| 1, 0$ & $0, 1 \| 0, 1$ & $1 \| 1$      & $1$      & $0$    \\
						 &           & $o_{\WorldONE,1},\obsd{o_{\WorldONE,2}}\|o_{\WorldTWO,1},o_{\WorldTWO,2}$ & $1/48$         & $1, 0 \| 1, 0$ & $0, 1 \| 0, 1$ & $1 \| 1$      & $0$      & $1$    \\
						 &           & $o_{\WorldONE,1},o_{\WorldONE,2}\|\obsd{o_{\WorldTWO,1}},o_{\WorldTWO,2}$ & $1/48$         & $1, 0 \| 1, 0$ & $0, 1 \| 0, 1$ & $1 \| 1$      & $1$      & $0$    \\
						 &           & $o_{\WorldONE,1},o_{\WorldONE,2}\|o_{\WorldTWO,1},\obsd{o_{\WorldTWO,2}}$ & $1/48$         & $1, 0 \| 1, 0$ & $0, 1 \| 0, 1$ & $1 \| 1$      & $0$      & $1$    \\
Large 			 & $2 \| 2$  & $\obsd{o_{\WorldONE,1}},o_{\WorldONE,2}\|o_{\WorldTWO,1},o_{\WorldTWO,2}$ & $1/12$         & $1, 0 \| 1, 0$ & $0, 1 \| 0, 1$ & $1 \| 1$      & $1$      & $0$    \\
      			 &           & $o_{\WorldONE,1},\obsd{o_{\WorldONE,2}}\|o_{\WorldTWO,1},o_{\WorldTWO,2}$ & $1/12$         & $1, 0 \| 1, 0$ & $0, 1 \| 0, 1$ & $1 \| 1$      & $0$      & $1$    \\
     		     &           & $o_{\WorldONE,1},o_{\WorldONE,2}\|\obsd{o_{\WorldTWO,1}},o_{\WorldTWO,2}$ & $1/12$         & $1, 0 \| 1, 0$ & $0, 1 \| 0, 1$ & $1 \| 1$      & $1$      & $0$    \\
      			 &           & $o_{\WorldONE,1},o_{\WorldONE,2}\|o_{\WorldTWO,1},\obsd{o_{\WorldTWO,2}}$ & $1/12$         & $1, 0 \| 1, 0$ & $0, 1 \| 0, 1$ & $1 \| 1$      & $0$      & $1$    
\end{tabular}}
{\begin{tablenotes}
Each microhypothesis (listed under $\MicrohCore$) consists of two sequences of observers, one for each world, separated by a $\|$. The worlds are indexed $\WorldONE$ and $\WorldTWO$. Under human observation likelihoods, $o @ (w,1)$ indicates the probability that each human in the sequences observe themself to have rank 1, while $o@(w,2)$ indicates the probability they observe themself to have rank 2. $o @ (w,1) \to 2$ is the probability that an immortal rank 1 human observes the world surviving long enough for rank 2 to be born.
\end{tablenotes}}
\end{table*}

{\emph{Internal observers} -- If you are a human who knows all of this but are otherwise ignorant of your location and $\Ntotal$ for your society, there are now four possible starting locations you could be: world $\WorldONE$ at ranks $1$ $(\WorldONE, 1)$ and $2$ $(\WorldONE, 2)$, and world $\WorldTWO$ at ranks $1$ $(\WorldTWO, 1)$ and $2$ $(\WorldTWO, 2)$. In the Small theory, only the rank $1$ starting locations are occupied, and in the Large, all are occupied. In the Intermediate theory, none, one, or both of the rank $2$ starting locations may be occupied. Within each microhypothesis, you start with an uninformative indexical distribution, with all $\ProvPriorKJI = \ProbPriorKJ$. By summing the provisional credences, you find prior weights $\WeightPrior_{(\WorldONE,1)} = \WeightPrior_{(\WorldTWO,1)} = 1/3$ and $\WeightPrior_{(\WorldONE,2)} = \WeightPrior_{(\WorldTWO,2)} = 1/6$.

Suppose you learn your birthrank is $1$. Now all provisional credences for $(\WorldONE, 2)$ and $(\WorldTWO, 2)$ are zero, because observers at these positions necessarily have birthranks of $2$. However, the provisional credences for the $(\WorldONE, 1)$ and $(\WorldTWO, 1)$ are untouched. There remains a complete symmetry between worlds $\WorldONE$ and $\WorldTWO$, with $\WeightPos_{(\WorldONE,1)} = \WeightPos_{(\WorldTWO,1)} = 1/2$, as nothing you have learned leads you to favor one over the other. Furthermore, $(\WorldONE, 1)$ and $(\WorldTWO, 1)$ are always within $\OSetKJ$ for every microhypothesis, since every inhabited world necessarily has at least one observer. Therefore, the sum of $\WeightPosI \ProvKJI$ over all trajectories is unchanged. There is no Bayesian shift; $\ProbPos(S) = \ProbPos(I) = \ProbPos(L) = 1/3$ (Figure~\ref{fig:DoomsdayExample_WFGAI_IO}). 

\begin{figure*}
\includegraphics[width=17.2cm]{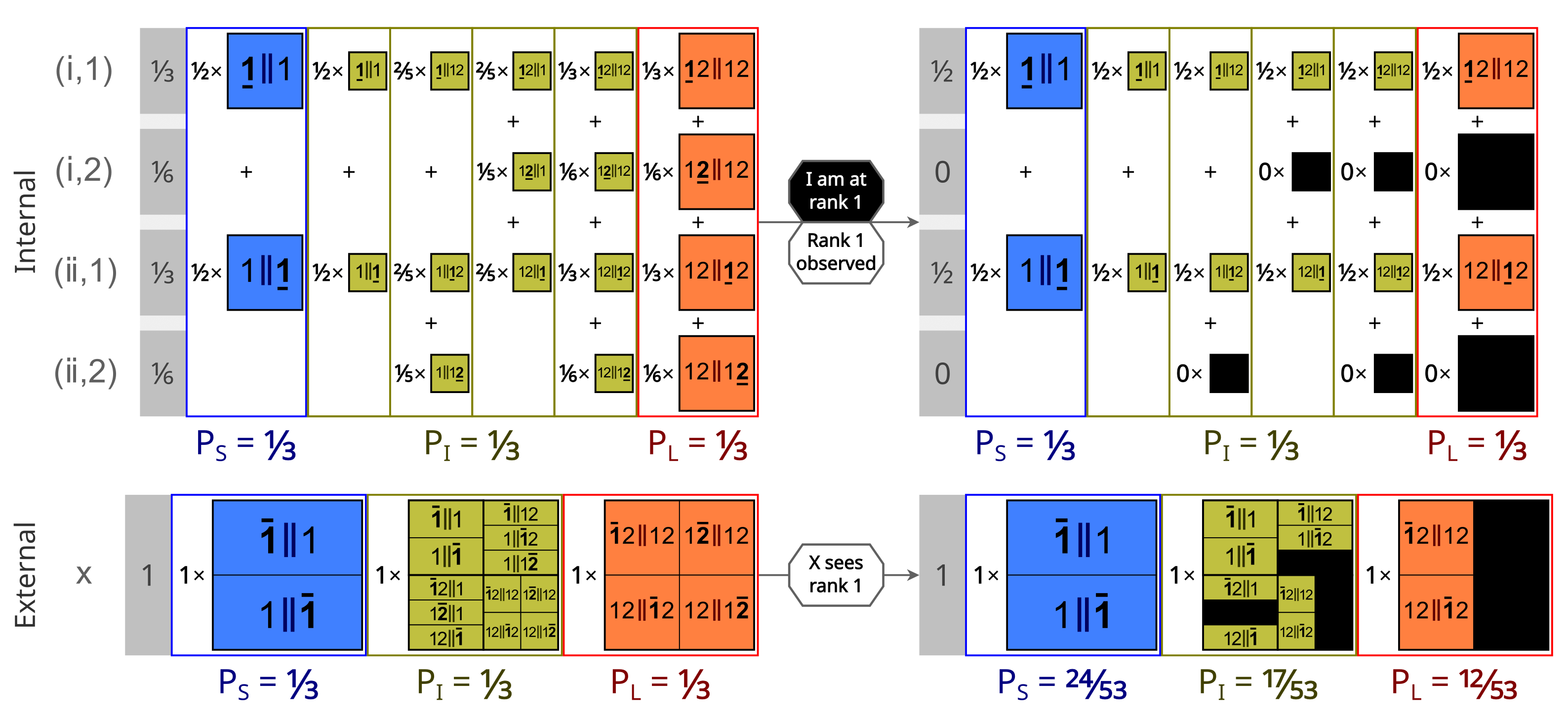}
\caption{{Two world example of the O doomsday problem. Each microhypothesis lists the sequence of realized birthranks, with a double vertical bar separating the sequences for worlds $\WorldONE$ and $\WorldTWO$. The Small, Intermediate, and Large theories are compared after observing a human at rank $1$.}\label{fig:DoomsdayExample_WFGAI_IO}}
\end{figure*}

\emph{External observers} -- Each microhypothesis specifies the location that $X$ observes, and thus has a single provisional credence associated with it. When $X$ samples an observer and finds their birthrank to be $1$, all microhypotheses where $X$ observes rank $2$ have zero evaluated credence. This removes half the total credence in the Large theory, and $1/4 (0 + 1/3 + 1/3 + 1/2) = 7/24$ of the credence in the Intermediate theory. As a result, the Short theory has twice the credence of the Long credence. 

The credences depend slightly on the number of worlds, because of the variance in $\Ntotal$ among the different Intermediate microhypotheses. Leaving all other model parameters fixed, the effective likelihood of the Intermediate distribution is
$$\frac{1}{2^{\Nworld}} \sum_{n = 0}^{\Nworld} {\Nworld \choose n} \frac{\Nworld}{\Nworld + n},$$
which decreases from $3/4$ when $\Nworld = 1$ to $2/3$ as $\Nworld \to \infty$. }

\subsection{{The anthropic shadow and the number of worlds}}
{Suppose the rank $1$ humans in the previous model live long enough to see the next generation being born, if it exists. Now, because they already have measured their birthranks, and because the weights actually parameterize temporal trajectories, the weights cannot shift any further without learning which world they are in. If such an observer witnesses the birth of rank $2$, then the provisional credences of those microhypotheses where $\Ntotal = 1$ for the world the observer is located in update to $0$. This results in an effective likelihood that equals the fraction of worlds that survive to $\Ntotal = 2$, which is $0$ in the Small theory, $1/2$ in the Intermediate theory, and $1$ in the Large theory, with $\ProbPos(I) = 1/3$ and $\ProbPos(L) = 2/3$ (Figure~\ref{fig:WFGAI_AnthropicShadow}). Indeed, this holds regardless of the number of worlds due to the underlying symmetry, with all worlds having a rank $1$ human.

\begin{figure*}
\includegraphics[width=17.2cm]{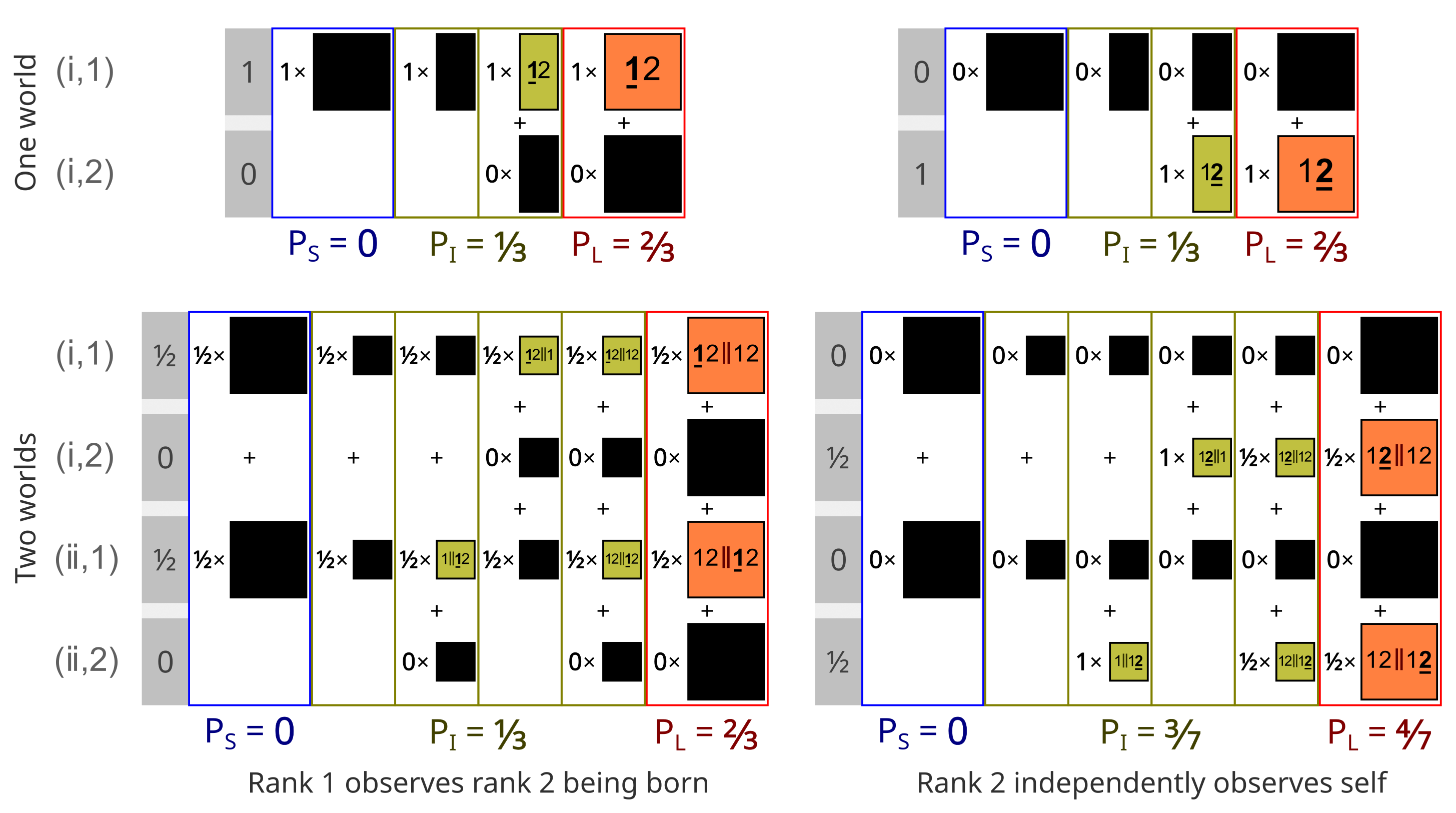}
\caption{{The anthropic shadow effect in {WFG}, demonstrated by the IO model, with one (top) and two (bottom) worlds. When an early human's posterior is adopted, the survival of their world to rank $2$ has a relatively strong update (left) compared to if the later human uses an independent posterior (right). Furthermore, the later human's independent posterior constraints on the Intermediate theory weaken as the number of worlds increases.}\label{fig:WFGAI_AnthropicShadow}}
\end{figure*}

Now, perhaps a rank $2$ human adopts the rank $1$ prior, if they see themselves as continuing a scientific program initiated by their ancestors. But it is also possible that a rank $2$ human starts from scratch, ignoring or ignorant of the ``result'' of the predecessor in their world. If they begin with uncertainty about their starting location but otherwise aware of the setup of the model, they too begin with $\WeightPrior_{(\WorldONE,1)} = \WeightPrior_{(\WorldTWO,1)} = 1/3$ and $\WeightPrior_{(\WorldONE,2)} = \WeightPrior_{(\WorldTWO,2)} = 1/6$. Upon measuring their birthrank of $2$, these update to $\WeightPos_{(\WorldONE,1)} = \WeightPos_{(\WorldTWO,1)} = 0$ and $\WeightPrior_{(\WorldONE,2)} = \WeightPrior_{(\WorldTWO,2)} = 1/2$. All of the indexical weight is in rank $2$ starting locations. In microhypotheses where there is at least one large world, all of the weight is placed in provisional credences that are unscathed by the birthrank measurement. Thus, the effective likelihood for any microhypothesis with at least one long-lived world is $1$. The effective likelihood of the Intermediate theory is $1 - 1/2^{\Nworld}$, which approaches $1$ as $\Nworld \to \infty$ (Figure~\ref{fig:WFGAI_AnthropicShadow}).

This is a demonstration of the anthropic principle in {WFG}. Later observers necessarily only are born in worlds long-lived enough to host them, even if such worlds are rare. Unless they adopt their ancestor's posterior, they draw no conclusions in an infinite universe from their birthrank except that the probability of their own existence in any given world is nonzero. 

Indeed, we can imagine this scenario playing out in humanity's future if we regard the ``observers'' not as individuals but entire species. Suppose our current society collapses in the next few centuries, with humanity reduced to a few thousand survivors, all records of our science vanishing. A million years from now, an intelligent posthuman species evolves from our scattered descendants. They come to realize that their ancestors built a technological society comparable to or exceeding their own. Should they conclude that humanity was destined to survive the catastrophe after all instead of going extinct? Not if they believe in a large universe. As per the anthropic shadow of \citet{Cirkovic10}, the remnants of humanity surviving is a necessary condition for their existence. In a large universe, for every Earth where humanity hangs on after such a catastrophe, there could be a billion where it perished or {practically} none -- to the posthumans descendents of the survivors, the result will look much the same.}

\end{document}